\documentclass[preprint,3p]{elsarticle}

\usepackage{amsfonts}
\usepackage{amsmath}
\usepackage{amssymb}
\usepackage{amsthm}
\usepackage{pifont}
\usepackage{mathtools}
\usepackage{booktabs}
\usepackage{mathrsfs}
\usepackage{graphicx}
\usepackage[normalem]{ulem}
\usepackage{wrapfig}
\usepackage{xcolor}
\usepackage{times}
\usepackage{url}
\usepackage{hyperref}
\usepackage{lineno}
\usepackage{yhmath}
\usepackage{natbib}
\usepackage{cleveref}
\usepackage{subcaption}
\usepackage{caption}
\usepackage{./definitions}
\hypersetup{
  bookmarksnumbered = true,
  bookmarksopen=false,
  pdfborder=0 0 0,         % make all links invisible, so the pdf looks good when printed
  pdffitwindow=true,      % window fit to page when opened
  pdfnewwindow=true, % links in new window
  colorlinks=true,           % false: boxed links; true: colored links
  linkcolor=blue,            % color of internal links
  citecolor=magenta,    % color of links to bibliography
  filecolor=magenta,     % color of file links
  urlcolor=cyan              % color of external links
}

\newcommand{\vertex}{Vertex-CFD}

\newcommand{\Mach}{\mathsf{Ma}}
\newcommand{\CFL}{\mathsf{CFL}}
\newcommand{\Reynolds}{\mathsf{Re}}
\newcommand{\ReynoldsM}{\mathsf{Re}_{\rm m}}
\newcommand{\Ha}{\mathsf{Ha}}
\newcommand{\tolN}{\texttt{tol}_{\rm N}}
\newcommand{\tolL}{\texttt{tol}_{\rm L}}
\newcommand{\rtolN}{\texttt{rtol}_{\rm N}}
\newcommand{\atolN}{\texttt{atol}_{\rm N}}

\newcommand{\cmark}{\ding{51}}
\newcommand{\xmark}{\ding{55}}

\defcitealias{smolentsev_2025}{S25}

\begin{document}

\begin{frontmatter}

\title{A Full-Induction Magnetohydrodynamics Solver for Liquid Metal Fusion Blankets in \vertex\tnoteref{t1}}

\author[ornl-CSMD,utk-phys]{Eirik Endeve\corref{cor1}}
\ead{endevee@ornl.gov}

\author[ornl-CSED]{Doug Stefanski}
\ead{stefanskidl@ornl.gov}

\author[ornl-RNSD]{Marc-Olivier G. Delchini}
\ead{delchinimg@ornl.gov}

\author[ornl-CSED]{Stuart Slattery}
\ead{slatterysr@ornl.gov}

\author[ornl-CSMD,utk-math]{Cory D. Hauck}
\ead{hauckc@ornl.gov}

\author[ornl-CSED]{Bruno Turcksin}
\ead{turcksinbr@ornl.gov}

\author[ornl-FED]{Sergey Smolentsev}
\ead{smolentsevsy@ornl.gov}

\tnotetext[t1]{This manuscript has been authored by UT-Battelle, LLC under Contract No. DE-AC05-00OR22725 with the U.S. Department of Energy. The United States Government retains and the publisher, by accepting the article for publication, acknowledges that the United States Government retains a non-exclusive, paid-up, irrevocable, world-wide license to publish or reproduce the published form of this manuscript, or allow others to do so, for United States Government purposes. The Department of Energy will provide public access to these results of federally sponsored research in accordance with the DOE Public Access Plan (http://energy.gov/downloads/doe-public-access-plan).}

\cortext[cor1]{Corresponding author}
\address[ornl-CSMD]{Computer Science and Mathematics Division, Oak Ridge National Laboratory, Oak Ridge, TN 37831, USA}
\address[ornl-CSED]{Computational Science and Engineering Division, Oak Ridge National Laboratory, Oak Ridge, TN 37831, USA}
\address[ornl-RNSD]{Reactor and Nuclear Systems Division, Oak Ridge National Laboratory, Oak Ridge, TN 37831, USA}
\address[ornl-FED]{Fusion Energy Division, Oak Ridge National Laboratory, Oak Ridge, TN 37831, USA}
\address[utk-phys]{Department of Physics and Astronomy, University of Tennessee Knoxville, TN 37996-1200, USA}
\address[utk-math]{Department of Mathematics, University of Tennessee Knoxville, TN 37996-1320, USA}

\begin{abstract}
Multiphysics modeling of liquid metal fusion blankets, which produce tritium and convert energy of neutrons created via fusion reactions into heat, is crucial for predicting performance, ensuring structural integrity, and optimizing energy production.  
While traditional blanket modeling of liquid metal flows during normal steady operating conditions commonly employs the inductionless approximation of the magnetohydrodynamics (MHD) equations, transient scenarios, when the plasma-confining magnetic field varies on millisecond time scales, require a full-induction MHD approach that dynamically evolves the magnetic field via the time-dependent induction equation.  
This paper presents the formulation, implementation, and initial verification of a full-induction MHD solver integrated within the open-source \vertex\ framework, which aims to achieve tight multiphysics coupling, a flexible software design enabling easy extension and addition of physics models, and performance portability across computing platforms.  
The solver utilizes finite element spatial discretization, implicit Runge--Kutta time integration, and an inexact Newton method to solve the resulting discrete nonlinear system, leveraging Trilinos packages for efficient computation.  
Verification against selected benchmark problems demonstrates accuracy and robustness of the solver.  
Furthermore, when the solver is applied to an idealized blanket model in 2.5D and full 3D, results obtained with \vertex\ are in good agreement with recently published quasi-2D simulations.  
These findings establish a computational foundation for future simulations of transient MHD phenomena in liquid metal blankets with \vertex, and open avenues for future extensions and performance optimizations.  
\end{abstract}


\end{frontmatter}

\section{Introduction}
\label{sec:intro}

This paper presents the formulation, implementation, and initial verification of a full-induction magnetohydrodynamics (MHD) solver for electrically conducting fluids, such as liquid metals, in a magnetic field, developed within the open-source \vertex\ framework~\cite{VERTEX-CFD-v100}.  
The solver is a computational tool to support the assessment of fusion blanket designs for emerging reactor concepts.  
The numerical approach employs finite element methods (FEMs) for spatial discretization, singly diagonally implicit Runge--Kutta (SDIRK) methods for time integration, and an inexact Newton method to solve the nonlinear system of the fully discrete MHD equations.  
For performance portability across computing platforms, including GPU-accelerated architectures, \vertex\ is built on Trilinos, a suite of packages for large-scale multiphysics simulations with device-level abstraction and optimization provided through Kokkos.  

Fusion reactor concepts under development for future energy production such as tokamaks and stellarators confine hot plasmas using strong magnetic fields, subjecting surrounding structures to intense thermal and electromagnetic loads.  
The fusion blanket is a critical reactor subsystem that is designed to serve multiple functions, including heat extraction for power generation, breeding tritium for fuel recycling, and radiation shielding (e.g., \cite{abdou_etal_2015,kessel_etal_2018}).  
Various blanket concepts have been proposed to fulfill these roles, including solid breeder blankets and liquid metal systems that leverage the favorable thermal and nuclear properties of materials like lead-lithium (PbLi).  
For example, in the dual-coolant lead-lithium (DCLL) blanket, a PbLi alloy circulates at low velocities ($\sim0.1$~m~s$^{-1}$) through ducts to absorb energy of neutrons produced by fusion reactions and produce tritium, while pressurized helium removes surface heat to cool the inner surface of the reactor chamber and other structures \cite{smolentsev_etal_2008,smolentsev_etal_2015}.  
Modeling of liquid metal blanket systems is essential for predicting performance under normal and off-normal operating conditions, and for guiding material selection, optimizing energy production, and ensuring structural integrity.  

Due to their high electrical conductivity, liquid metal flows in fusion blankets are governed by the MHD equations.  
Many existing computational models employ the inductionless approximation (e.g., \cite{smolentsev_etal_2015b,smolentsev_etal_2020}), which assumes a low magnetic Reynolds number (i.e., magnetic field diffusion dominates over advection) so that flow-induced magnetic fields can be neglected.  
Under this assumption, the magnetic field is treated as externally imposed and steady, and the primary electromagnetic variable is the scalar electric potential obtained by solving an elliptic equation.  
The inductionless approximation is widely used for modeling quasi-steady, normal operating conditions in blanket systems.  
See, for example, \cite{ni_etal_2007,zhou_2010,planas_etal_2011,he_etal_2015,eardley-Brunt_etal_2024} for implementations of the inductionless approximation for numerical modeling of liquid metal blanket flows, and \cite{smolentsev_2021} for a review of blanket-relevant computations using this approach.  

Magnetic field configurations in fusion devices, particularly tokamaks, are prone to instabilities such as those associated with vertical displacement events, magnetic island locking, and resistive wall modes \cite{hender_etal_2007,boozer_2012}.  
Instability-induced disruptions, which involve a thermal quench followed by a current quench, release large amounts of thermal and magnetic energy \cite{hender_etal_2007}.
During these rapid transients, thermal energy and electromagnetic loads can far exceed those during steady operation, and the resulting rapidly changing magnetic fields generate high eddy currents in both solid and liquid blanket components. 
The associated electromagnetic forces are known to accelerate the flowing liquid metal in the blanket to velocities that may significantly exceed the blanket limits.  
Such scenarios motivate the use of \emph{full-induction MHD}, which evolves the magnetic field dynamically using the full time-dependent induction equation.  
This formulation captures both the advection and diffusion of magnetic fields and is necessary when the magnetic Reynolds number is moderate or high, or when induced magnetic fields and transient electromagnetic effects are non-negligible.  
While some prior studies have applied full-induction MHD to simplified blanket geometries \cite{huang_etal_2017,kawczynski_thesis,kawczynski_etal_2018}, broader use remains limited.  

By nature, computational models for blanket analysis are both multiphysics and multiscale.  
They must be capable of handling multi-material domains (e.g., coupling liquid metal, solid structures, and plasma), complex three-dimensional geometries, and far-field boundary conditions that connect to reactor-scale electromagnetic environments.  
Furthermore, the accurate resolution of boundary layers (e.g., Hartmann layers) requires high spatial resolution near walls, and models are stiff due to disparate timescales, necessitating implicit time integration to stably capture evolution from milliseconds to seconds without resolving the fastest time scales of the system.  
At the numerical level, the design of robust and accurate solution methods is demanding, as solvers should preserve key structural properties, such as the divergence-free condition on the magnetic field \cite{brackbillBarnes_1980,toth_2000}, conservation principles, and positivity of density and temperature \cite{liuOsher_1996,zhangShu_2011,guermondPopov_2014}.  
Robust preconditioning strategies that account for multiphysics coupling across the range of relevant time scales remain a central challenge in achieving performance on high-performance computing (HPC) systems \cite{cyr_etal_2016,ohm_etal_2024}, while also ensuring software extensibility so that new physics capabilities can be added without major code refactoring \cite{pawlowski2012automating,pawlowski2012automating2}.  

As an initial step towards addressing some of these challenges, we have implemented a full-induction liquid metal MHD solver within the open-source framework \vertex~\cite{VERTEX-CFD-v100}.  
This solver is based on a FEM for spatial discretization, SDIRK methods for time integration, and inexact Newton methods for handling nonlinearities of the discrete system.  
Linear systems arising within the Newton iterations are solved using iterative methods. 
To handle the divergence-free constraints on the velocity and magnetic field, the incompressibility condition is relaxed with the artificial compressibility method due to Chorin~\cite{chorin_1967}, and the divergence-free condition is relaxed with the generalized Lagrange multiplier method introduced to MHD by Dedner et al.~\cite{dedner_etal_2002}.  
\vertex\ derives capabilities from Trilinos \cite{trilinos-website}, a suite of packages for large-scale multiphysics simulations, which provides finite-element assembly, time integrators, nonlinear solvers, linear solvers and preconditioners.  
\vertex\ aims for performance on heterogeneous architectures, including GPU-accelerated systems, via its use of Kokkos for performance portability.
Software extensibility is facilitated through C++ templates, which provide a mechanism for adding, modifying, or coupling physics models without extensive code refactoring.  

The remainder of the paper is organized as follows: Section~\ref{sec:model} provides the theoretical formulation of the full-induction MHD model, followed by Section~\ref{sec:methods}, which summarizes the numerical methods used to solve the MHD equations.  
Section~\ref{sec:tests} presents numerical verification results on standard test cases, and Section~\ref{sec:applications} presents results from the application of the MHD solver to an idealized model of blanket flow driven by a transient external magnetic field.  
Section~\ref{sec:conclusions} summarizes our findings, draws some conclusions, and discusses future directions for \vertex\ development.  
\ref{sec:wavespeeds} presents wave speeds associated with the full-induction MHD model.  
\ref{sec:implementations} provides additional details on the implementation of the numerical methods in \vertex. 

\section{AC-GLM-MHD Model}
\label{sec:model}

We employ the artificial compressibility (AC) model as an approximation to the incompressible Navier--Stokes (NS) equations.  
The formulation is a coupled hyperbolic-parabolic system of partial differential equations (PDEs) after introducing a time evolution equation for pressure that approximates incompressible dynamics (e.g., \cite{chorin_1967,rogers_1987,kwakKiris_2013,clausen_2013})
\begin{subequations}\label{eq:navierStokes}
	\begin{align}
		\pd{}{t}(\rho_{0}\,\bv)+\nabla\cdot\big(\,\rho_{0}\,\bv\otimes\bv+P\,\bI-\bstau\,\big) 
		&= \bJ\times\bB, \label{eq:momentumEquation} \\
		\pd{P}{t}+\rho_{0}\,c_{0}^{2}\,\nabla\cdot\bv
		&=0, \label{eq:pressureEquation}
	\end{align}
\end{subequations}
where $\rho_{0}$ is the constant mass density, $\bv$ the velocity field, $P$ the pressure, $c_{0}$ the (constant) artificial sound speed, and $\bstau$ the viscous stress tensor, defined as
\begin{equation}
	\bstau=\rho_{0}\,\nu\,\big(\,(\nabla\bv)+(\nabla\bv)^{\intercal}\,\big),
\end{equation}
where $\nu$ is the kinematic viscosity.  
The Lorentz force on the right-hand side of Eq.~\eqref{eq:momentumEquation} couples the NS equations to the electromagnetic fields, where $\bB$ denotes the magnetic field and the current density is $\bJ=\nabla\times\bB/\mu_{0}$, where $\mu_{0}$ is the vacuum magnetic permeability.  
The evolution of the magnetic field, which completes the MHD coupling, is discussed in more detail below.  

The AC model transforms the incompressible NS system by replacing the incompressibility constraint $\nabla\cdot\bv=0$ with the evolution equation in Eq.~\eqref{eq:pressureEquation}.  
This introduces artificial, finite-speed acoustic waves into the system, in contrast to the instantaneous pressure response of the incompressible formulation, which involves solving an elliptic equation at each time step.  
The model aims to approximate low Mach number flows, where the Mach number $\Mach=|\bv|/c_{0}\ll1$.  
Within this framework, density fluctuations are neglected, i.e., $\delta\rho/\rho_{0}\sim\cO(\Mach^{2})$ \cite{clausen_2013}, and $\rho_{0}$ is treated as a constant.  
The AC model can be derived as the isentropic limit of the first law of thermodynamics \cite{clausen_2013}, with the pressure equation commonly formulated in terms of the material derivative, but the approximation $\pd{P}{t}+\bv\cdot\nabla P\approx\pd{P}{t}$ can be made under the assumption that the acoustic wave speed is much greater than the characteristic flow speed.  
While the AC model can be derived from the first law of thermodynamics, we emphasize that the sound speed $c_{0}$ is a user-specified artificial computational parameter, not necessarily intended to reflect the true physical compressibility of the flow.  
Its value may be tuned for consistency with experimental or numerical reference data.  

The low Mach number assumption is expected to be well justified in models of liquid PbLi blankets for fusion applications.  
In these systems, the speed of sound is typically of order $10^{3}$~m~s$^{-1}$ (e.g., \cite{humrickhouseMerrill_2018,martelli_etal_2019}), while characteristic flow velocities under normal operating conditions are much lower.  
Even during transient events such as plasma disruptions, flow speeds may rise considerably but are still expected to remain within $\cO(10~{\rm m}~{\rm s}^{-1})$ (e.g., \cite{smolentsev_2025}).  

For the evolution of the magnetic field we adopt the generalized Lagrange multiplier (GLM) form of the magnetic induction equation (e.g., \cite{dedner_etal_2002}).  
The GLM formulation introduces a coupled system for the magnetic field $\bB$ and a ``divergence-correcting'' scalar field $\psi$
\begin{subequations}
	\begin{align}
		\pd{\bB}{t} + \nabla\times\bE
		&=-\bv\,(\nabla\cdot\bB) - c_{h}\,\nabla\psi, \label{eq:inductonEquation} \\
		\pd{\psi}{t} + c_{h}\,\nabla\cdot\bB
		&=-\alpha\,\psi, \label{eq:magneticCorrectionPotentialEquation}
	\end{align}
	\label{eq:maxwellMHD}
\end{subequations}
where the electric field $\bE$ is defined by Ohm's law,
\begin{equation}
	\bE=-(\bv\times\bB)+\eta\,\bJ,
	\label{eq:ohm}
\end{equation}
and $\eta$ is the electrical resistivity, here assumed to be constant in space and time.  
The GLM system in Eq.~\eqref{eq:maxwellMHD} is designed to mitigate numerically induced violations of the divergence-free condition on the magnetic field $\nabla\cdot\bB=0$, and the coupling to the scalar field $\psi$ facilitates propagation and damping of divergence errors.  
In Eq.~\eqref{eq:maxwellMHD}, $c_{h}$ is the hyperbolic cleaning speed controlling the rate at which divergence errors propagate, and $\alpha\ge0$ is a damping coefficient that controls the decay of these errors.  
The first term on the right-hand side of Eq.~\eqref{eq:inductonEquation}, which arises from a generalized form of Maxwell's equations that admits magnetic monopoles \cite{jackson_ClassicalElectroDynamics}, facilitates advection of divergence errors with the fluid flow \cite{powell_etal_1999}.  
As one of the so-called Godunov--Powell source terms it plays a role in rendering the MHD system symmetrizable \cite{godunov_1972}, which is important for ensuring stability of numerical schemes in the presence of $\nabla\cdot\bB$ errors (e.g., \cite{derigs_etal_2018,dao_etal_2024}).  

The coupled system defined by Eqs.~\eqref{eq:navierStokes} and \eqref{eq:maxwellMHD} --- referred to here as the AC-GLM-MHD model --- approximates viscous, resistive, incompressible MHD flows governed by the constraints $\nabla\cdot\bv=0$ and $\nabla\cdot\bB=0$.  
In this formulation, these solenoidal constraints are not enforced exactly but are instead relaxed through hyperbolic evolution equations that allow constraint violations to propagate with finite, user-specified speeds $c_{0}$ (velocity) and $c_{h}$ (magnetic field).  
This is reflected in the structural similarity of the pressure and magnetic correction equations, Eqs~\eqref{eq:pressureEquation} and \eqref{eq:magneticCorrectionPotentialEquation}, and their respective coupling to the momentum and induction equations via gradients of $P$ and $\psi$.  
By incorporating divergence mitigation directly into the model, rather than, e.g., enforcing the constraints through discrete projection \cite{brackbillBarnes_1980} or specialized approximation spaces \cite{hu_etal_2021}, this approach offers greater flexibility in the choice of numerical methods.  
It enables numerically induced divergence errors to propagate out of the domain, avoiding the need for computationally expensive elliptic solves at each time step, as required by traditional projection-based methods (e.g., \cite{brackbillBarnes_1980}).  
We refer to \cite{donatelli_2013} for a theoretical analysis of an artificial-compressibility-based MHD system closely related to the AC-GLM-MHD model.

\paragraph{Conservation Form of the AC-GLM-MHD Equations}
For numerical solutions using \vertex, we express the AC-GLM-MHD system in a compact hyperbolic-parabolic balance-law form that is suitable for finite element discretization
\begin{equation}
	\pd{\bU}{t} + \nabla\cdot\big(\,\bF(\bU)-\bG(\bU,\nabla\bU)\,\big) = \bS(\bU,\nabla\bU).
	\label{eq:mhd.compact}
\end{equation}
Here, the evolved state vector is $\bU=[\rho_{0}\bv,P,\bB,\psi]^{\intercal}$, and the inviscid and viscous fluxes are defined as
\begin{equation}
	\bF(\bU)
	=
	\left[\begin{array}{c}
		\rho_{0}\bv\otimes\bv + (P+P_{\rm m})\bI-\f{1}{\mu_{0}}\bB\otimes\bB \\
		\rho_{0}c_{0}^{2}\,\bv \\
		\bv\otimes\bB-\bB\otimes\bv+\,c_{h}\,\psi\,\bI \\
		c_{h}\,\bB
	\end{array}\right] 
	\,\,\text{and}\,\,
	\bG(\bU,\nabla\bU)
	=
	\left[\begin{array}{c}
		\bstau \\
		0 \\
		\f{\eta}{\mu_{0}}\big(\nabla\bB-(\nabla\cdot\bB)\bI\big) \\
		0
	\end{array}\right],
	\label{eq:mhd.compact.fluxes}
\end{equation}
with magnetic pressure $P_{\rm m}=\f{1}{2\mu_{0}}|\bB|^{2}$.  
The source terms are given by
\begin{equation}
	\bS(\bU,\nabla\bU)
	=
	-
	\left[\begin{array}{c}
		\f{1}{\mu_{0}}\,\bB \\
		0 \\
		\bv \\
		0
	\end{array}\right]\,(\nabla\cdot\bB)
	-\alpha\,\psi
	\left[\begin{array}{c}
		0 \\
		0 \\
		0 \\
		1
	\end{array}\right].
	\label{eq:mhd.compact.sources}
\end{equation}
The momentum source in Eq.~\eqref{eq:mhd.compact.sources} results from rewriting the Lorentz force in divergence form and retaining the term proportional to $\nabla\cdot\bB$, which in a numerical setting may improve the orthogonality of the Lorentz force and the magnetic field in the presence of divergence errors \cite{brackbillBarnes_1980}.  
The first term on the right-hand side of Eq.~\eqref{eq:mhd.compact.sources} is commonly referred to as the Godunov--Powell (GP) source term.  

Eq.~\eqref{eq:mhd.compact} must be equipped with initial and boundary conditions, which we leave unspecified for now, but will return to when solving specific problems in Sections~\ref{sec:tests} and \ref{sec:applications}.  

\section{Numerical Methods}
\label{sec:methods}

This section provides a brief overview of the numerical methods used by \vertex\ to solve Eq.~\eqref{eq:mhd.compact}.  
The software implementations of these methods are discussed in detail in Section~\ref{sec:implementations}.  

\paragraph{Spatial Discretization}
We discretize Eq.~\eqref{eq:mhd.compact} in space over the computational domain $\Omega\subset\bbR^{3}$, with boundary $\Gamma$, using the FEM.  
To this end, let $\cT_{h}$ denote a triangulation of $\Omega$ into disjoint elements $\bK$ such that $\Omega=\cup_{\bK\in\cT_{h}}\bK$.  
We define the finite element approximation space
\begin{equation}
	\cV_{h}^{k} = \big\{\,\varphi_{h}\in\cC^{0}(\Omega)~\vcentcolon~\varphi_{h}|_{\bK}\in\bbM^{k}(\bK),\, \forall\bK\in\cT_{h}\,\big\},
	\label{eq:fem.approximation_space}
\end{equation}
where $\bbM^{k}(\bK)$ is the space of polynomials of degree $k$ used to approximate components of $\bU$ on the element $\bK$.  
(We use the subscript $h$ to denote quantities associated with the discrete, finite-element representation of the continuous problem.)  
The semi-discrete Ritz--Galerkin variational formulation for solving Eq.~\eqref{eq:mhd.compact} is then to find the approximation to $\bU$, i.e., $\bU_{h}=((\rho_{0}\bv)_{h}^{\intercal},P_{h},\bB_{h}^{\intercal},\psi_{h})^{\intercal}\in\bV_{h}\vcentcolon=[\cV_{h}^{k}]^{8}$, such that
\begin{align}
	&\int_{\Omega}\bW_{h}^{\intercal}\,\big(\,\pd{\bU_{h}}{t}-\bS(\bU_{h},\nabla\bU_{h})\,\big)\,dV
	-\int_{\Omega}\nabla\bW_{h}^{\intercal}\cdot\big(\,\bF(\bU_{h})-\bG(\bU_{h},\nabla\bU_{h})\,\big)\,dV \nonumber \\
	&\hspace{12pt}
  	+\oint_{\Gamma}\big[\bW_{h}^{\intercal}\big(\,\bF(\bU_{h})-\widehat{\bG}(\bU_{h},\nabla\bU_{h})\,\big)\big]\cdot\bn\,d\Gamma
	=0 \label{eq:mhd.compact.fem}
\end{align}
holds for all $\bW_{h}\in\bV_{h}$.  
In Eq.~\eqref{eq:mhd.compact.fem}, $\bn$ is the unit normal on the boundary $\Gamma$, and $\widehat{\bG}$ is the numerical approximation to the viscous flux on the boundary, for which we use a symmetric interior penalty method \cite{sip-wheeler,hartmannHouston_2006,HARTMANN20089670}.  

\paragraph{Temporal Discretization}
The semi-discrete scheme in Eq.~\eqref{eq:mhd.compact.fem} can be reorganized into a system of ordinary differential equations (ODEs) of the form
\begin{equation}
	\deriv{\bY}{t}=\bsf(\bY,t),
	\label{eq:mhd.compact.ode}
\end{equation}
where the global solution vector $\bY$ represents all the degrees of freedom in $\Omega$ and $\bsf$ captures the spatial discretization of the AC-GLM-MHD model.  
(In Eq.~\eqref{eq:mhd.compact.ode}, $\bsf$ also includes a left multiplication by the inverse mass matrix.)  
Explicit time-dependence in $\bsf$ is included to, e.g., accommodate time-dependent boundary conditions and/or external forcing.  

We use implicit Runge--Kutta methods to integrate the system in Eq.~\eqref{eq:mhd.compact.ode} forward in time.  
While \vertex\ can access a wide variety of time integrators (Section~\ref{sec:implementations}), in this paper we specifically consider $s$-stage singly diagonally implicit Runge--Kutta (SDIRK) methods to advance the state vector a time step $\dt$ from time $t^{n}$ to $t^{n+1}=t^{n}+\dt$, which take the general form (e.g., \cite{kennedyCarpenter_2016})
\begin{subequations}
	\label{eq:sdirk}
	\begin{align}
		\bY^{(i)}
		&= \bY^{n} + \dt\sum_{j=1}^{s}a_{ij}\,\bsf(\bY^{(j)},t^{(n,j)}), \quad (i=1,\ldots,s), \label{eq:sdirk.stages} \\
		\bY^{n+1} 
		&= \bY^{n} + \dt\sum_{i=1}^{s}b_{i}\,\bsf(\bY^{(i)},t^{(n,i)}), \label{eq:sdirk.assembly}
	\end{align}
\end{subequations}
where $t^{(n,i)}=t^{n}+c_{i}\dt$, the stage weights $a_{ij}$ are components of the matrix $\bA\in\bbR^{s\times s}$, and the scheme weights $b_{i}$ and nodes $c_{i}$ are components of $\bb,\bc\in\bbR^{s}$.  
For diagonally implicit Runge--Kutta (DIRK) methods, $\bA$ is lower triangular, so the upper limit of the sum in Eq.~\eqref{eq:sdirk.stages} can be set to $i$.  
SDIRK methods are a subclass of DIRK methods in which all the diagonal entries of $\bA$ are identical.  
The scheme is often organized in the Butcher table format
\begin{equation}
  \begin{array}{c | c}
        \bc & \bA \\ \hline
              & \bb^{\intercal}
  \end{array}.
\end{equation}
Since $\bA$ is lower triangular, each implicit solve in the stages in Eq.~\eqref{eq:sdirk.stages} can be viewed as a backward Euler solve,
\begin{equation}
	\bY^{(i)} = \bY^{(n,i)} + a_{ii}\dt\,\bsf(\bY^{(i)},t^{(n,i)}),
	\label{eq:backward_euler}
\end{equation}
with modified time step $a_{ii}\dt$ and \emph{known} initial state
\begin{equation}
	\bY^{(n,i)} = \bY^{n} + \dt\sum_{j=1}^{i-1}a_{ij}\,\bsf(\bY^{(j)},t^{(n,j)}).  
	\label{eq:backward_euler_initial_state}
\end{equation}

In this paper, we consider the two-stage second-order SDIRK method (SDIRK22) with Butcher table (e.g., \cite[][Section~4.2.1]{kennedyCarpenter_2016})
\begin{equation}
  \begin{array}{c | c}
        \bc & \bA \\ \hline
              & \bb^{\intercal}
  \end{array}
  =
  \begin{array}{c | cc}
        \gamma &    \gamma &  \\ 
                   1 & 1-\gamma & \gamma \\ \hline
                      & 1-\gamma & \gamma
  \end{array},
  \label{eq:sdirk22}
\end{equation}
where $\gamma=\f{1}{2}(2-\sqrt{2})$, and the five-stage fourth-order SDIRK method (SDIRK54) with Butcher table \cite[][Chapter~IV.6]{hairerWanner_1996}
\begin{equation}
  \begin{array}{c | c}
        \bc & \bA \\ \hline
              & \bb^{\intercal}
  \end{array}
  =
  \begin{array}{c | ccccc}
                  1/4  & 1/4    \\ 
                   3/4 & 1/2           & 1/4 \\
               11/20 & 17/50       & -1/25         & 1/4 \\
                   1/2 & 371/1360 & -137/2720 & 15/544 & 1/4 \\
                      1 & 25/24       & -49/48       & 125/16 & -85/12 & 1/4 \\ \hline
                         & 25/24       & -49/48       & 125/16 & -85/12 & 1/4
  \end{array}.
  \label{eq:sdirk54}
\end{equation}
Both SDIRK22 and SDIRK45 are stiffly accurate, that is $a_{si}=b_{i}$, for $i=1,\ldots,s$, which implies the ``first same as last'' (FSAL) property; i.e., $\bY^{n+1}=\bY^{(s)}$, and the assembly step in Eq.~\eqref{eq:sdirk.assembly} is not needed.  

To determine the step size $\dt$ in Eq.~\eqref{eq:sdirk}, we define the element-local Courant--Friedrichs--Lewy (CFL) time step $\dtCFL$ in a computational domain of dimension $D$ as a function of the maximum eigenvalues $\{\lambda_{\max}^{i}\}$, the mesh size $|\bK|$, and the user-input CFL number ($\CFL>0$),
\begin{equation}
	\dtCFL|_{\bK} = \CFL\times\f{|\bK|}{\sum_{i=1}^{D} \lambda_{\max}^{i}},
	\label{eq:dtCFL}
\end{equation}
where $\lambda_{\max}^{i}$ is the largest absolute eigenvalue of the AC-GLM-MHD flux Jacobian in the $i$th spatial dimension (see \ref{sec:wavespeeds}).  
Because an implicit temporal integrator is employed, the CFL number can (ideally) be set to any positive value.  
For linear stability we can take $\CFL>1$, while explicit methods typically require $\CFL\le1$.  
However, temporal accuracy, nonlinear solver convergence, and linear solver conditioning considerations typically place restrictions on $\CFL$.  

\paragraph{Nonlinear Solution Method}
In \vertex, the nonlinear system arising from the semi-discretization in Eq.~\eqref{eq:mhd.compact.fem} is solved iteratively with Newton's method by considering the residual
\begin{equation}
	\bR(\dot{\bY},\bY,t) \equiv \dot{\bY}-\bsf(\bY,t) = 0.  
	\label{eq:nonlinear.residual}
\end{equation}
Considering residuals in the form of Eq.\eqref{eq:nonlinear.residual} allows for generality in terms of both software design and algorithm choices for time integration by requiring only the construction of the algorithmic building blocks in Eq.~\eqref{eq:mhd.compact.fem} via the finite element method and the associated derivatives of the residual with respect to $\bY$ and $\dot{\bY}$.  
Considering the fully discrete system obtained after applying the SDIRK method to Eq.~\eqref{eq:mhd.compact.ode}, as given by Eq.~\eqref{eq:backward_euler}, as an example, $\bY\vcentcolon=\bY^{(i)}$ represents the unknowns to be solved for and $\dot{\bY}=\dot{\bY}(\bY^{(i)})\vcentcolon=(\bY^{(i)}-\bY^{(n,i)})/(a_{ii}\dt)$.  

Given an initial guess $\bY^{[0]}$ for $\bY^{(i)}$, the solution is iteratively refined through the sequence \cite[see, e.g.,][and references therein]{knollKeyes_2004}
\begin{equation}
	W^{[k]}\,\delta\bY^{[k]} = - \bR^{[k]}, 
	\quad 
	\bY^{[k+1]}=\bY^{[k]}+\delta\bY^{[k]},
	\quad
	k=0,1,\ldots,
	\label{eq:nonlinear.iteration}
\end{equation}
until the residual norm drops below a user-specified tolerance.  
In Eq.~\eqref{eq:nonlinear.iteration}, the iteration (Jacobian) matrix is defined as
\begin{equation}
	W^{[k]} = \pderiv{\dot{\bY}}{\bY}\Big|_{\bY^{[k]}}\pderiv{\bR}{\dot{\bY}}\Big|_{\bY^{[k]}} + \pderiv{\bR}{\bY}\Big|_{\bY^{[k]}}.
	\label{eq:nonlinear.jacobian}
\end{equation}
The update from $\bY^{[k]}$ to $\bY^{[k+1]}$ requires the solution of a linear system for $\delta\bY^{[k]}$, which is carried out iteratively using preconditioned GMRES (see Section~\ref{sec:implementations} for details).  

\paragraph{Implementations}
The implementation of the above methods in \vertex\ is guided by three primary goals central to coupled physics modeling: 
({\it i}) tight algorithmic coupling \cite[``tight coupling'' in][]{keyes_etal_2013}, achieved through a fully implicit formulation in which all physics components are solved within a single monolithic system; 
({\it ii}) extensibility, enabling the addition and coupling of new physics terms or models without substantial code refactoring; and 
({\it iii}) performance portability, ensuring efficient use of leadership-class computing systems, which are currently built by a variety of vendors that require their own programming models for execution on accelerated hardware.  
Conceptually, an implicit finite element solver has two key components: (a) assembly of vectors and matrices representing the discretized equations, and (b) numerical algorithms which use them to advance the simulation in time.  
\vertex\ is designed around these two components to achieve the primary design goals through the Trilinos scientific computing libraries \cite{trilinos-website}.  
Solver implementations, including mesh representation and partitioning, operator assembly, time integrators, and the subsequent linear solvers and preconditioners arising from linearization of the nonlinear system are described in more detail in \ref{sec:implementations}.  
We also highlight the use of Kokkos for performance portability.  

\section{Numerical Tests}
\label{sec:tests}

In this section, we verify the implementation of the AC-GLM-MHD model in \vertex\ through a series of numerical tests with known reference solutions --- either analytical solutions or well-established numerical benchmarks from the literature.  
All numerical results are reported in dimensionless units (``code units'') with the vacuum permeability $\mu_{0}$ set to unity, following common practice in MHD code-verification studies (e.g., \cite{toth_2000,gardinerStone_2008,stone_etal_2020}).

For problems where an analytic solution exists, we compare the numerical and analytical solutions in the $p$-norm.  
To this end, we denote the numerical solution of a variable $X\in\{v_{x},v_{y},v_{z},B_{x},B_{y},B_{z}\}$ by $X_{h}$, and compute the $p$-norm error as
\begin{equation}
	L^{p}(X) = \frac{\Big(\int_{\Omega}|X-X_{h}|^{p}dV\Big)^{1/p}}{\int_{\Omega}dV},
	\label{eq:pNorm}
\end{equation}
for $p\in\{1,2\}$, where the integrals are evaluated by numerical quadrature.  
When performing mesh refinement studies, we quantify the convergence rate of the numerical solution towards the analytic solution as the mesh is refined.  
Let $N_{1}$ to $N_{2}$ denote the number of degrees of freedom in two successive meshes, with corresponding $p$-norm errors $L_{1}^{p}$ and $L_{2}^{p}$.
The convergence rate is then computed as
\begin{equation}
	\text{rate}
	=\Big|\f{\log\big(L_{1}^{p}/L_{2}^{p}\big)}{\log\big(N_{1}/N_{2}\big)}\Big|.
	\label{eq:convergenceRate}
\end{equation}
Thus, a rate of $r$ indicates that the error decreases as $N^{-r}$ under mesh refinement.  

We also investigate violations of the divergence-free condition on the magnetic field, and the efficacy of the various divergence control mechanisms in the model (i.e., GP sources and GLM divergence cleaning).  
To this end, the following measures for tracking the discrete global divergence in time are considered:
\begin{align}
	\max_{\bK\in\Omega}\left(|\nabla\cdot\bB_{h}|\right) 
	& = \max_{i=1}^{n_\text{ele}}\left[\,\max_{j=1}^{n_\text{qp}}|\nabla\cdot\bB_{h}|_{ij}\,\right], \label{eq:maxAbsDivB} \\
	\int_{\Omega}|\nabla\cdot\bB_{h}|dV
	& = \sum_{i=1}^{n_\text{ele}} \sum_{j=1}^{n_\text{qp}} w_{ij}\,|\nabla\cdot\bB_{h}|_{ij}, \label{eq:totAbsDivB}
\end{align}
where $n_{\text{ele}}$ and $n_{\text{qp}}$ are the number of discrete elements in the domain and the number of quadrature points per element, respectively, and $w_{ij}$ is the quadrature weight for a given point.  

We define the artificial compressibility parameter $\beta=\rho_{0}\,c_{0}^{2}$; see Eq.~\eqref{eq:pressureEquation}.  
Below, the magnetic advection-diffusion (Section~\ref{sec:magnetic_advection_diffusion}) and circularly polarized Alfv{\'e}n wave (\ref{sec:cpAlfvenWave}) tests used $\beta=1$.  
The current sheet test (Section~\ref{sec:currentSheetTest}) used $\beta=10^{4}$.  
The divergence cleaning (Section~\ref{sec:divergenceCleaningTest}) and the lid-driven cavity (Section~\ref{sec:lidDrivenCavity}) tests used $\beta=10^{3}$.  
As a rule of thumb, we set $\beta\sim\cO(1)$ when compressibility effects are negligible, and choose larger values when compressibility is expected to matter and a nearly divergence-free velocity field is required.  
In the latter case, $\beta$ should be chosen so that the associated acoustic modes remain fast relative to other physical wave speeds, but not so large that solver performance or convergence is adversely affected.  

\subsection{Magnetic Advection-Diffusion}
\label{sec:magnetic_advection_diffusion}

To verify the MHD model implemented in \vertex, we begin by solving the induction equation with various divergence control methods, as defined in Eq.~\eqref{eq:maxwellMHD}, without coupling to the fluid model in Eq.~\eqref{eq:navierStokes}.  
That is, this test is carried out by simulating the advection and diffusion of a magnetic field in a periodic domain, subjected to a prescribed velocity field that is constant in both space and time.
While modeling different physics, the computational setup here is similar to that of the circularly polarized Alfv{\'e}n wave test (e.g., \cite{toth_2000}), which we consider in Section~\ref{sec:cpAlfvenWave}.  
For a specified $v_{\parallel}$, the velocity components are given by
\begin{equation}
	v_{x} = v_{\parallel}\,\cos\varphi, \quad
	v_{y} = v_{\parallel}\,\sin\varphi,
	\quad\text{and}\quad
	v_{z} = 0,
\end{equation}
where $\varphi$ is the angle between the velocity vector and the $x$-axis.
For $\varphi \in (0, \pi/2)$, we consider a two-dimensional periodic domain $\Omega = [0, 1/\cos\varphi] \times [0, 1/\sin\varphi]$, which renders the magnetic field prescribed below periodic. 

By substituting Eq.~\eqref{eq:ohm} in Eq.~\eqref{eq:inductonEquation}, and assuming constant resistivity $\eta$ as well as $\nabla\cdot\bB = \psi = 0$, each component of the magnetic field evolves according to an advection-diffusion equation of the form
\begin{align}
	\pd{X}{t} + \bv\cdot\nabla X = \eta\,\Delta X,
\end{align}
for $X \in \{B_{x}, B_{y}, B_{z}\}$.  
The initial magnetic field is specified as
\begin{equation}
	B_{\parallel,0} = 1, \quad
	B_{\perp,0} = A\,\sin(2\pi\,x_{\parallel}),
	\quad\text{and}\quad
	B_{z,0} = A\,\cos(2\pi\,x_{\parallel}),
\end{equation}
where $x_{\parallel} = x \cos\varphi + y \sin\varphi$ denotes the coordinate aligned with the flow direction, and the amplitude is set to $A = 0.1$.  
The initial $x$- and $y$-components of the magnetic field are obtained by rotating the $\parallel$ and $\perp$ components
\begin{equation}
	B_{x,0} = B_{\parallel,0}\cos\varphi - B_{\perp,0}\sin\varphi,
	\quad\text{and}\quad
	B_{y,0} = B_{\perp,0}\cos\varphi + B_{\parallel,0}\sin\varphi.
	\label{eq:magnetic_advection_diffusion.initial_BxBy}
\end{equation}
Given the initial and boundary conditions, the exact solution is
\begin{subequations}\label{eq:magnetic_advection_diffusion_solution}
	\begin{align}
		B_{\parallel}(x_{\parallel},t)
		&=B_{\parallel,0}, \\
		B_{\perp}(x_{\parallel},t)
		&=B_{\perp,0}(x_{\parallel}-v_{\parallel}\,t)\times\exp(-(2\pi)^{2}\,\eta\,t), \\
		B_{z}(x_{\parallel},t)
		&=B_{z,0}(x_{\parallel}-v_{\parallel}\,t)\times\exp(-(2\pi)^{2}\,\eta\,t), 
	\end{align}
\end{subequations}
with $B_x$ and $B_y$ obtained by rotation according to Eq.~\eqref{eq:magnetic_advection_diffusion.initial_BxBy}.
Thus, sinusoidal perturbations in $B_{\perp}$ and $B_{z}$ are superimposed on a constant background field $B_{\parallel}$ aligned with the flow.  
These perturbations are advected with the velocity field while diffusing due to finite resistivity, with their detailed evolution dependent on the values of $v_{\parallel}$ and $\eta$.  

In the following, for $v_{\parallel}=1$ and $\varphi=\pi/6$, we assess the accuracy of the implementation by varying the resistivity $\eta$, as well as the terms responsible for the transport and dissipation of numerically induced magnetic monopoles (i.e., divergence errors) appearing on the right-hand side of Eq.~\eqref{eq:inductonEquation}.
Although all three magnetic field components are subject to numerical errors due to spatial and temporal discretization, in this two-dimensional configuration, only $B_x$ and $B_y$ are directly affected by errors arising from violations of the divergence-free constraint.
Specifically, the solution in Eq.~\eqref{eq:magnetic_advection_diffusion_solution} is obtained from the induction equation assuming $\nabla\cdot\bB=0$ holds for all times.  
We use a quasi-two-dimensional mesh with $N\times N\times1$ uniform hexahedral elements.  
Thus, for $\varphi=\pi/6$, $\Delta x=(2/\sqrt{3})/N$ and $\Delta y=2/N$.  
We set the time step as $\dt=0.8/N$, and set linear and nonlinear tolerances for the implicit solve to $\tolL=\tolN=10^{-12}$.  

To enable the demonstration of high-order accuracy, i.e., beyond second order in both space and time, all results in this section were obtained using quadratic elements and the SDIRK54 time integration scheme.  

In this test, we consider the induction equation, Eq.~\eqref{eq:inductonEquation}, in the following form
\begin{equation}
	\pd{\bB}{t}+\nabla\cdot\big[\,\big(\bv\otimes\bB-\bB\otimes\bv\big)-\eta\,\big(\,\nabla\bB-\underbrace{(\nabla\cdot\bB)\,\bI}_{\eta(\nabla\cdot\bB)\text{-term}}\,\big)\,\big] 
	= -\underbrace{\bv\,(\nabla\cdot\bB)}_{\text{GP-term}} - \underbrace{c_{h}\,\nabla\psi}_{\text{GLM-term}},
	\label{eq:induction_equation_divergence_controls}
\end{equation}
where we have emphasized terms that are associated with divergence control or present because the divergence-free condition on the magnetic field is not strictly enforced.  
The ``$\eta(\nabla\cdot\bB)$-term'' on the left-hand side of Eq.~\eqref{eq:induction_equation_divergence_controls} arises when transforming the resistive part of the induction equation into divergence form using the identity
\begin{equation}
	\nabla\times(\nabla\times\bB)=\nabla\cdot\big(\,(\nabla\cdot\bB)\bI-\nabla\bB\,\big).  
\end{equation}
The first term on the right-hand side of Eq.~\eqref{eq:induction_equation_divergence_controls}, labeled ``GP-term'', is the Godunov--Powell source for the induction equation, which results in the transport of divergence errors with the flow \cite{powell_etal_1999}.  
The second term on the right-hand side of Eq.~\eqref{eq:induction_equation_divergence_controls}, labeled ``GLM-term'', provides coupling to Eq.~\eqref{eq:magneticCorrectionPotentialEquation} and propagation and damping of divergence errors \cite{dedner_etal_2002}.  
When this term is enabled, we set $c_{h}=1$ and $\alpha=c_{h}/0.18\approx5.56$.  

Table~\ref{tab:magnetic_advection-diffusion} lists the $L^{2}$ error norm for each of the magnetic field components and the divergence error as defined by Eq.~\eqref{eq:totAbsDivB} versus spatial resolution $N$ at $t=1$ for the resistive case $\eta=(2\pi)^{-2}$.  
At this time, the initial profile has advected once across the grid and the peak amplitude has been reduced by a factor $e$.  

\begin{table}[!htbp]
	{\footnotesize
	\begin{center}
		\begin{tabular}{lccccccccccc}
			& & \multicolumn{2}{c}{({\it i}) no $\nabla\cdot\bB$ terms} & \multicolumn{2}{c}{({\it ii}) $\eta\,(\nabla\cdot\bB)$} & \multicolumn{2}{c}{({\it iii}) GP+$\eta\,(\nabla\cdot\bB)$} & \multicolumn{2}{c}{({\it iv}) GLM+$\eta\,(\nabla\cdot\bB)$} & \multicolumn{2}{c}{({\it v}) all $\nabla\cdot\bB$ terms} \\
			\cmidrule(r){3-4} \cmidrule(r){5-6} \cmidrule(r){7-8} \cmidrule(r){9-10} \cmidrule(r){11-12} 
			& $N$ & $L^{2}$ & rate & $L^{2}$ & rate & $L^{2}$ & rate & $L^{2}$ & rate & $L^{2}$ & rate \\
			\midrule\midrule
			$B_{x}$ 
			& 8   & $5.721\times10^{-5}$ & ---    & $1.783\times10^{-4}$ & ---     & $8.356\times10^{-5}$ & ---    & $5.668\times10^{-5}$ & ---     & $4.018\times10^{-5}$ & ---    \\
			&16  & $5.279\times10^{-6}$ & 3.44 & $2.230\times10^{-5}$ & 3.00 & $2.326\times10^{-5}$ & 1.85 & $9.769\times10^{-6}$ & 2.54 & $8.132\times10^{-6}$ & 2.30 \\
			& 32 & $5.632\times10^{-7}$ & 3.23 & $4.739\times10^{-6}$ & 2.23 & $6.464\times10^{-6}$ & 1.85 & $2.589\times10^{-6}$ & 1.92 & $2.716\times10^{-6}$ & 1.58 \\
			& 64 & $6.702\times10^{-8}$ & 3.07 & $1.144\times10^{-6}$ & 2.05 & $1.619\times10^{-6}$ & 2.00 & $6.569\times10^{-7}$ & 1.98 & $7.219\times10^{-7}$ & 1.91 \\
			\midrule
		 	$B_{y}$ 
			& 8   & $6.530\times10^{-5}$ & ---    & $3.163\times10^{-4}$ & ---     & $1.345\times10^{-4}$ & ---    & $8.795\times10^{-5}$ & ---     & $8.003\times10^{-5}$ & ---    \\
			&16  & $7.535\times10^{-6}$ & 3.12 & $3.785\times10^{-5}$ & 3.06 & $3.414\times10^{-5}$ & 1.98 & $1.909\times10^{-5}$ & 2.20 & $1.874\times10^{-5}$ & 2.09 \\
			& 32 & $9.199\times10^{-7}$ & 3.03 & $6.419\times10^{-6}$ & 2.56 & $8.314\times10^{-6}$ & 2.04 & $4.144\times10^{-6}$ & 2.20 & $4.485\times10^{-6}$ & 2.06 \\
			& 64 & $1.143\times10^{-7}$ & 3.01 & $1.422\times10^{-6}$ & 2.17 & $2.059\times10^{-7}$ & 2.01 & $1.007\times10^{-6}$ & 2.04 & $1.104\times10^{-6}$ & 2.02 \\
			\midrule
		 	$B_{z}$ 
			& 8   & $7.338\times10^{-5}$ & ---    & $7.338\times10^{-5}$ & ---     & $7.338\times10^{-5}$ & ---    & $7.338\times10^{-5}$ & ---     & $7.338\times10^{-5}$ & ---    \\
			&16  & $8.707\times10^{-6}$ & 3.07 & $8.707\times10^{-6}$ & 3.07 & $8.707\times10^{-6}$ & 3.07 & $8.707\times10^{-6}$ & 3.07 & $8.707\times10^{-6}$ & 3.07 \\
			& 32 & $1.063\times10^{-6}$ & 3.03 & $1.063\times10^{-6}$ & 3.03 & $1.063\times10^{-6}$ & 3.03 & $1.063\times10^{-6}$ & 3.03 & $1.063\times10^{-6}$ & 3.03 \\
			& 64 & $1.320\times10^{-7}$ & 3.01 & $1.320\times10^{-7}$ & 3.01 & $1.320\times10^{-7}$ & 3.01 & $1.320\times10^{-7}$ & 3.01 & $1.320\times10^{-7}$ & 3.01 \\
			\midrule
		 	$\nabla\cdot\bB$ 
			& 8   & $3.527\times10^{-3}$ & ---    & $3.199\times10^{-2}$  & ---    & $1.416\times10^{-2}$ & ---    & $9.335\times10^{-3}$ & ---     & $7.140\times10^{-3}$ & ---    \\
			&16  & $8.526\times10^{-4}$ & 2.05 & $9.397\times10^{-3}$ & 1.77 & $8.830\times10^{-3}$ & 0.68 & $3.983\times10^{-3}$ & 1.23 & $3.623\times10^{-3}$ & 0.98 \\
			& 32 & $2.072\times10^{-4}$ & 2.04 & $3.742\times10^{-3}$ & 1.33 & $4.488\times10^{-3}$ & 0.98 & $1.925\times10^{-3}$ & 1.05 & $2.107\times10^{-3}$ & 0.78 \\
			& 64 & $5.106\times10^{-5}$ & 2.02 & $1.763\times10^{-3}$ & 1.09 & $2.234\times10^{-3}$ & 1.01 & $9.611\times10^{-4}$ & 1.00 & $1.084\times10^{-3}$ & 0.96 \\
			\midrule\midrule
		\end{tabular}
		\vspace{-12pt}
		\caption{Magnetic advection-diffusion $L^{2}$ error norms and convergence rates for resistivity $\eta=(2\pi)^{-2}$.
		Errors are given for $B_{x}$, $B_{y}$, $B_{z}$, and $\nabla\cdot\bB$ versus number of elements $N\in\{8,16,32,64\}$.
		Columns of errors and convergence rates are grouped from left to right for the following cases: 
		({\it i}) no $\nabla\cdot\bB$-terms included in the induction equation --- i.e., the $\eta(\nabla\cdot\bB)$-, GP-, and GLM-terms all excluded; 
		({\it ii}) only the $\eta(\nabla\cdot\bB)$-term included; 
		({\it iii}) the $\eta(\nabla\cdot\bB)$- and GP-terms included; 
		({\it iv}) the $\eta(\nabla\cdot\bB)$- and GLM-terms included; and 
		({\it v}) all $\nabla\cdot\bB$-terms (i.e., the $\eta(\nabla\cdot\bB)$-, GP-, and GLM-terms) included.}
		\label{tab:magnetic_advection-diffusion}
	\end{center}}
\end{table}

From Table~\ref{tab:magnetic_advection-diffusion}, we observe that the error convergence rates for $B_{x}$ and $B_{y}$, computed as defined in Eq.~\eqref{eq:convergenceRate}, are influenced by the inclusion of $\nabla\cdot\bB$-terms in the induction equation.  
For case ({\it i}), with no $\nabla\cdot\bB$-terms, the error convergence rates for all magnetic field components are third-order.  
The error convergence rates for $B_{x}$ and $B_{y}$ are reduced to second-order when any of the considered combinations of $\nabla\cdot\bB$-terms are included.  
Among these, cases ({\it iv}) and ({\it v}) produce somewhat smaller overall errors.  
The error in $B_{z}$ is unaffected by the $\nabla\cdot\bB$-terms, and we observe third-order convergence rate for this field component for all the cases considered.  
The magnetic field divergence decreases with increasing resolution: for any of the cases with $\nabla\cdot\bB$-terms included, the divergence of the magnetic field decreases with first-order rate, while it decreases with second-order rate when none of the $\nabla\cdot\bB$-terms are included.  

\begin{table}[!htbp]
	{\small
	\begin{center}
		\begin{tabular}{cccccccccc}
			& & \multicolumn{2}{c}{({\it i}) no $\nabla\cdot\bB$ terms} & \multicolumn{2}{c}{({\it ii}) GP} & \multicolumn{2}{c}{({\it iii}) GLM} & \multicolumn{2}{c}{({\it iv}) GLM+GP} \\
			\cmidrule(r){3-4} \cmidrule(r){5-6} \cmidrule(r){7-8} \cmidrule(r){9-10}
			& $N$ & $L^{2}$ & rate & $L^{2}$ & rate & $L^{2}$ & rate & $L^{2}$ & rate \\
			\midrule\midrule
			$B_{x}$ 
			& 8   & $2.184\times10^{-3}$ & ---    & $1.812\times10^{-4}$ & ---     & $1.906\times10^{-4}$ & ---    & $1.359\times10^{-4}$ & ---     \\
			&16  & $4.838\times10^{-4}$ & 2.17 & $5.773\times10^{-5}$ & 1.65 & $3.787\times10^{-5}$ & 2.33 & $3.462\times10^{-5}$ & 1.97 \\
			& 32 & $1.141\times10^{-4}$ & 2.08 & $1.567\times10^{-5}$ & 1.88 & $9.157\times10^{-6}$ & 2.05 & $9.001\times10^{-6}$ & 1.94 \\
			& 64 & $2.804\times10^{-5}$ & 2.02 & $3.939\times10^{-6}$ & 1.99 & $2.277\times10^{-6}$ & 2.01 & $2.280\times10^{-6}$ & 1.98 \\
			\midrule
		 	$B_{y}$ 
			& 8   & $9.898\times10^{-4}$ & ---    & $3.139\times10^{-4}$ & ---     & $1.772\times10^{-4}$ & ---    & $2.381\times10^{-4}$ & ---     \\
			&16  & $2.026\times10^{-4}$ & 2.29 & $9.999\times10^{-5}$ & 1.65 & $7.200\times10^{-5}$ & 1.30 & $7.249\times10^{-5}$ & 1.72 \\
			& 32 & $4.866\times10^{-5}$ & 2.06 & $2.714\times10^{-5}$ & 1.88 & $1.715\times10^{-5}$ & 2.07 & $2.046\times10^{-5}$ & 1.83 \\
			& 64 & $1.194\times10^{-5}$ & 2.03 & $6.822\times10^{-6}$ & 1.99 & $4.152\times10^{-6}$ & 2.05 & $5.151\times10^{-6}$ & 1.99 \\
			\midrule
		 	$B_{z}$ 
			& 8   & $3.624\times10^{-4}$ & ---    & $3.624\times10^{-4}$ & ---     & $3.624\times10^{-4}$ & ---    & $3.624\times10^{-4}$ & ---     \\
			&16  & $1.155\times10^{-4}$ & 1.65 & $1.155\times10^{-4}$ & 1.65 & $1.155\times10^{-4}$ & 1.65 & $1.155\times10^{-4}$ & 1.65 \\
			& 32 & $3.134\times10^{-5}$ & 1.88 & $3.134\times10^{-5}$ & 1.88 & $3.134\times10^{-5}$ & 1.88 & $3.134\times10^{-5}$ & 1.88 \\
			& 64 & $7.877\times10^{-6}$ & 1.99 & $7.877\times10^{-6}$ & 1.99 & $7.877\times10^{-6}$ & 1.99 & $7.877\times10^{-6}$ & 1.99 \\
			\midrule
		 	$\nabla\cdot\bB$ 
			& 8   & $6.311\times10^{-2}$ & ---     & $2.723\times10^{-2}$ & ---    & $1.968\times10^{-2}$ & ---    & $1.813\times10^{-2}$ & ---     \\
			&16  & $2.912\times10^{-2}$ & 1.12 & $1.651\times10^{-2}$ & 0.72 & $8.247\times10^{-3}$ & 1.25 & $7.863\times10^{-3}$ & 1.21 \\
			& 32 & $1.390\times10^{-2}$ & 1.07 & $8.502\times10^{-3}$ & 0.96 & $4.271\times10^{-3}$ & 0.95 & $3.955\times10^{-3}$ & 0.99 \\
			& 64 & $6.857\times10^{-3}$ & 1.02 & $4.206\times10^{-3}$ & 1.02 & $2.160\times10^{-3}$ & 0.98 & $1.992\times10^{-3}$ & 0.99 \\
			\midrule\midrule
		\end{tabular}
		\vspace{-12pt}
		\caption{Magnetic advection-diffusion $L^{2}$ error norms and convergence rates for resistivity $\eta=0$.
		Errors are given for $B_{x}$, $B_{y}$, $B_{z}$, and $\nabla\cdot\bB$ versus number of elements $N\in\{8,16,32,64\}$.
		Columns of errors and convergence rates are grouped from left to right for the following cases: 
		({\it i}) no $\nabla\cdot\bB$-terms included; ({\it ii}) only the GP-term included; ({\it iii}) only the GLM-term included; and ({\it iv}) GP- \emph{and} GLM-terms included.}
		\label{tab:magnetic_advection}
	\end{center}}
\end{table}

Table~\ref{tab:magnetic_advection} shows similar results for the ideal case ($\eta=0$).  
Here, we observe that the error for all magnetic field components converges with a second-order rate.  
The error for the $x$- and $y$-components depends on which $\nabla\cdot\bB$-terms are included.  
Interestingly, case ({\it i}) with no $\nabla\cdot\bB$-terms exhibits the largest errors.  
Suboptimal error convergence rate for unstabilized FEMs for advection-dominated problems have been reported by others (e.g., \cite{garcia-archilla_etal_2021}).  

Finally, in additional experiments not tabulated, we observed asymptotic third-order accuracy for both the $\eta=(2\pi)^{-2}$ and $\eta=0$ cases when the flow is aligned with the coordinate axes.  
Also, with $\varphi=\pi/6$, we observe second-order convergence rate for all magnetic field components when using linear elements and SDIRK22 time stepping.  

\subsection{Circularly Polarized Alfv{\'e}n Wave}
\label{sec:cpAlfvenWave}

Described in \cite{toth_2000}, this test considers the propagation of circularly polarized Alfv{\'e}n waves in a periodic domain.  
By varying the background flow velocity, we consider both traveling and standing waves.  
Similar to the magnetic advection-diffusion test in Section~\ref{sec:magnetic_advection_diffusion}, the Alfv{\'e}n wave propagates in the $xy$-plane with the propagation direction forming an angle $\varphi$ relative to the $x$-axis.  
The periodic domain is $0\le x \le1/\cos\varphi$, $0 \le y \le 1/\sin\varphi$, and $0\le z \le 1$, and, as in \cite{toth_2000}, the propagation angle is set to $\varphi=\pi/6$.  
Then, $x\in[0,2/\sqrt{3}]$, $y\in[0,2]$, and we let $z\in[0,1]$.  
The magnetic field components parallel ($B_{\parallel}$) and perpendicular ($B_{\perp}$ and $B_{z}$) to the propagation direction are initialized as
\begin{equation}
    B_{\parallel} 
    = 1, \quad
    B_{\perp} 
    = A\,\sin(2\pi\,x_{\parallel}), \quad\text{and}\quad
    B_{z}
    = A\,\cos(2\pi\,x_{\parallel}),
    \label{eq:cpAlfvenWave.magnetic_field}
\end{equation}
where $x_{\parallel}$ is the coordinate along the wave propagation direction, and we set the amplitude to $A=0.1$.  
The density is set to $\rho_{0}=1$, so that the Alfv{\'e}n speed is $v_{\rm A}=B_{\parallel}/\sqrt{\rho_{0}}=1$.  
The velocity components are set as
\begin{equation}
	v_{\parallel}
    	=\left\{\begin{array}{rc} 0 & \text{for traveling waves} \\ 1 & \text{for standing waves} \end{array}\right., \quad
    	v_{\perp}
    	=A\times\sin(2\pi\,x_{\parallel}), \quad\text{and}\quad
    	v_{z}
    	=A\times\cos(2\pi\,x_{\parallel}).
	\label{eq:cpAlfvenWave.velocity}
\end{equation}

The coordinates in the coordinate system rotated by an angle $\varphi$ about the $z$-axis, ($x_{\parallel},x_{\perp}$), are related to the corresponding Cartesian coordinates, ($x,y$), by $x_{\parallel}=x\cos\varphi+y\sin\varphi$ and $x_{\perp}=y\cos\varphi-x\sin\varphi$.  
The Cartesian $x$ and $y$ magnetic field and velocity components are obtained from Eqs.~\eqref{eq:cpAlfvenWave.magnetic_field} and \eqref{eq:cpAlfvenWave.velocity}, respectively, by
\begin{equation}
    B_{x} 
    = B_{\parallel}\,\cos\varphi-B_{\perp}\,\sin\varphi
    \quad\text{and}\quad
    B_{y} 
    = B_{\perp}\,\cos\varphi+B_{\parallel}\,\sin\varphi,
\end{equation}
and
\begin{equation}
    v_{x} 
    = v_{\parallel}\,\cos\varphi-v_{\perp}\,\sin\varphi
    \quad\text{and}\quad
    v_{y} 
    = v_{\perp}\,\cos\varphi+v_{\parallel}\,\sin\varphi.
\end{equation}

We let $N$ be the number of elements in the $x$ and $y$ dimensions.
We use a single element in the $z$ dimension.  
We set the time step as $\dt=0.8/N$, with $N\in\{8,16,32,64\}$, and evolve to $t=5$, when the traveling wave has crossed the computational domain five times.  
Linear and nonlinear tolerances for the implicit solve are set to $\tolL=\tolN=10^{-12}$.  
For this test, we include the GP sources, but do not include GLM divergence cleaning.  
To measure the order of accuracy of the implementation in \vertex, we compare the numerical solution against the analytical solution at $t=5$.  
Table~\ref{tab:cp_alfven_wave_linear_sdirk22} shows results for both traveling and standing waves, obtained using linear elements and SDIRK22 time stepping.  
We list average $L^{1}$ and $L^{2}$ error norms, where the average is computed across all the components of $\bv$ and $\bB$.  
We observe second-order accuracy across all error measures.  

\begin{table}[!htbp]
	\small
	\begin{center}
		\begin{tabular}{ccccccccc}
			& \multicolumn{4}{c}{Traveling Wave} & \multicolumn{4}{c}{Standing Wave} \\
			\cmidrule(r){2-5} \cmidrule(r){6-9} 
			$N$ & $L^{1}$ & rate & $L^{2}$ & rate & $L^{1}$ & rate & $L^{2}$ & rate \\
			\midrule\midrule
			8   & $2.651\times10^{-2}$ & ---    & $1.935\times10^{-2}$ & ---     & $4.860\times10^{-3}$ & ---    & $3.605\times10^{-3}$ & ---     \\
			16 & $6.529\times10^{-3}$ & 2.02 & $4.765\times10^{-3}$ & 2.02 & $1.271\times10^{-3}$ & 1.93 & $9.325\times10^{-4}$ & 1.95 \\
			32 & $1.616\times10^{-3}$ & 2.01 & $1.181\times10^{-3}$ & 2.01 & $3.214\times10^{-4}$ & 1.98 & $2.351\times10^{-4}$ & 1.99 \\
			64 & $4.025\times10^{-4}$ & 2.01 & $2.942\times10^{-4}$ & 2.00 & $8.058\times10^{-5}$ & 2.00 & $5.891\times10^{-5}$ & 2.00 \\
			\midrule\midrule
		\end{tabular}
		\vspace{-12pt}
		\caption{Circularly polarized Alfv{\'e}n wave average errors and convergence rates for traveling and standing waves (columns 2-5 and 6-9, respectively) using linear elements and SDIRK22 time stepping.
		Note: for the standing wave case with linear elements, the initial condition results in a sufficiently small residual that the nonlinear solver considers the solution converged.  Therefore, the listed errors for this case are due to the projection of the initial condition onto the FEM basis.}
		\label{tab:cp_alfven_wave_linear_sdirk22}
	\end{center}	
\end{table}

We repeated these tests using quadratic elements and SDIRK54 time stepping, with results listed in Table~\ref{tab:cp_alfven_wave_quadratic_sdirk54}.  
While we only observe second-order error convergence rate for this case, as is expected given the results in Section~\ref{sec:magnetic_advection_diffusion}, for a given $N$, the errors are considerably smaller than for the case with linear elements and SDIRK22 time stepping.  
Specifically, for the traveling wave case, errors obtained with quadratic elements and $N=8$ are comparable to those obtained with linear elements and $N=64$.  

\begin{table}[!htbp]
	\small
	\begin{center}
		\begin{tabular}{ccccccccc}
			& \multicolumn{4}{c}{Traveling Wave} & \multicolumn{4}{c}{Standing Wave} \\
			\cmidrule(r){2-5} \cmidrule(r){6-9} 
			$N$ & $L^{1}$ & rate & $L^{2}$ & rate & $L^{1}$ & rate & $L^{2}$ & rate \\
			\midrule\midrule
			8   & $5.409\times10^{-4}$ & ---    & $4.179\times10^{-4}$ & ---     & $4.942\times10^{-4}$ & ---    & $4.264\times10^{-4}$ & ---    \\
			16 & $1.328\times10^{-4}$ & 2.03 & $1.058\times10^{-4}$ & 1.98 & $1.199\times10^{-4}$ & 2.04 & $1.099\times10^{-4}$ & 1.96 \\
			32 & $3.802\times10^{-5}$ & 1.80 & $2.944\times10^{-5}$ & 1.85 & $3.189\times10^{-5}$ & 1.91 & $3.013\times10^{-5}$ & 1.87 \\
			64 & $9.138\times10^{-6}$ & 2.06 & $6.927\times10^{-6}$ & 2.09 & $7.787\times10^{-6}$ & 2.03 & $7.542\times10^{-6}$ & 2.00 \\
			\midrule\midrule
		\end{tabular}
		\caption{Circularly polarized Alfv{\'e}n wave test showing the same quantities as in Table~\ref{tab:cp_alfven_wave_linear_sdirk22} with quadratic elements and SDIRK54 time stepping.}
		\label{tab:cp_alfven_wave_quadratic_sdirk54}
	\end{center}	
\end{table}

\subsection{Divergence Cleaning Tests}
\label{sec:divergenceCleaningTest}

To investigate the efficacy of divergence cleaning approaches in \vertex, we consider the evolution of an initial divergence in the magnetic field over time.
Following \cite{dedner_etal_2002} and \cite{TRICCO20127214}, the initial defect in the magnetic field is advected by a background flow in a two-dimensional periodic domain $\Omega=[-0.5,1.5]^{2}$, according to the initial conditions
\begin{equation}
    \label{eq:div-adv}
        \rho_{0} = 1.0, \quad 
        p = 6.0, \quad
        \bv = \begin{pmatrix}1.0 \\ 1.0 \\ 0 \end{pmatrix}, \quad
        \bB = \frac{1}{\sqrt{4\pi}}\begin{pmatrix} \left[(r/r_0)^8 - 2(r/r_0)^4+1\right] \\ 0 \\ 1 \end{pmatrix},
\end{equation}
where $r=\sqrt{x^2+y^2}$ and $r_0 = 1 / \sqrt{8}$ prescribes the radial extent of the perturbation in $B_x$. 
This perturbation results in violation of the divergence-free condition on $\bB$ in the initial condition, and the evolution of $\nabla \cdot \bB$ in the solution is examined to determine the extent to which the cleaning method reduces the divergence in the magnetic field.  
We run this test using a diagonalized $180\times180$ linear mesh ($64800$ triangular elements) and evolve with the SDIRK22 time stepper, with $\dt=0.0025$, until $t=2$, when the divergence perturbation has had time to advect once across the domain diagonally.  
The linear and nonlinear tolerances for the implicit solve are set to $\tolL=10^{-10}$ and $\tolN=10^{-8}$, respectively.  

\begin{figure}[!ht]
    \centering
    \includegraphics[width=0.98\linewidth]{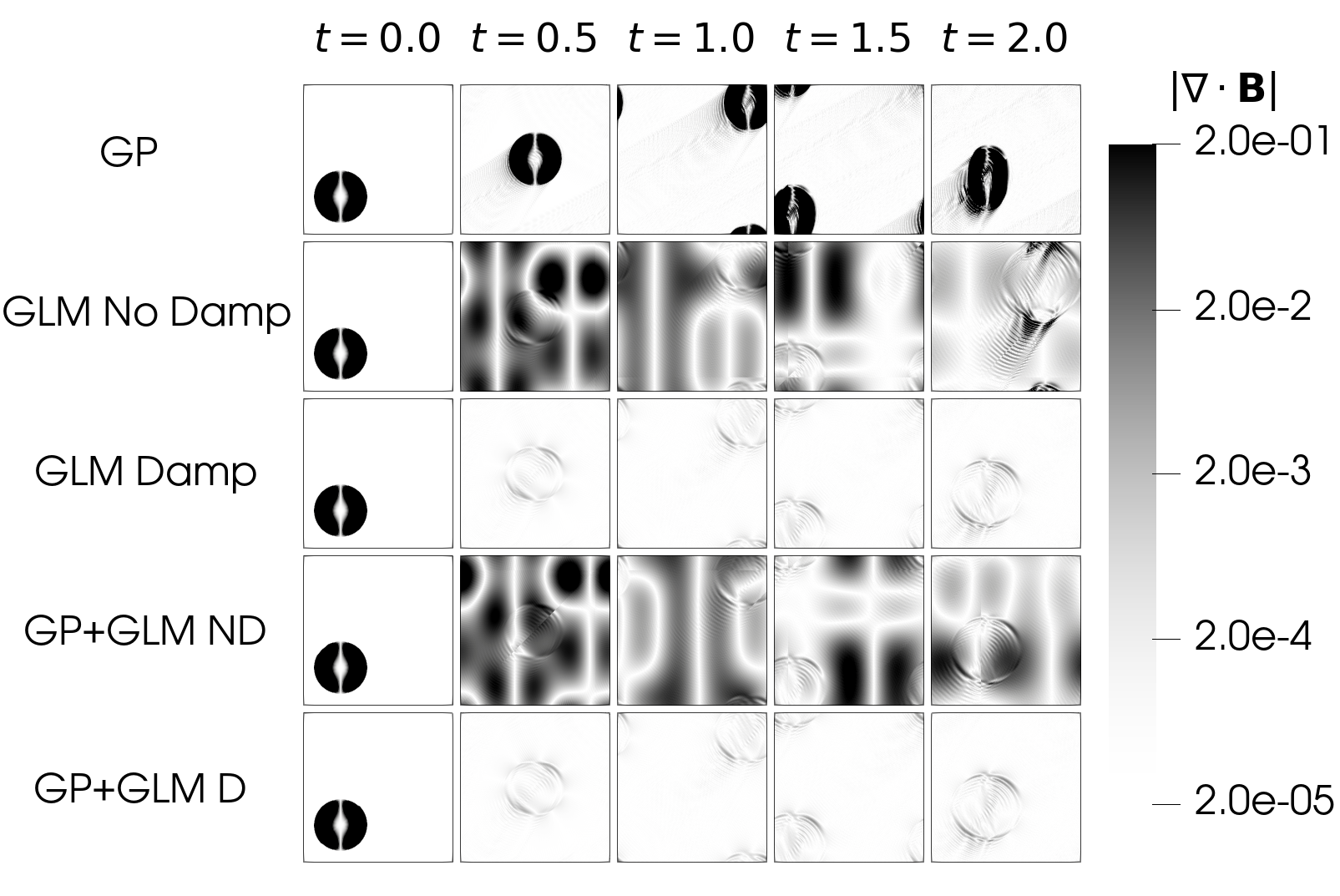}
    \caption{Divergence cleaning test: $\nabla\cdot\bB$ at several times under the effects of various cleaning terms: 
    only GP sources, (first row; GP); 
    only GLM divergence cleaning without damping, $c_{h}=50$ and $\alpha=0$ (second row; GLM No Damp); 
    only GLM divergence cleaning, $c_{h}=50$ and $\alpha=c_h/0.18$ (third row; GLM Damp); 
    GP sources and GLM divergence cleaning without damping, $c_{h}=50$ and $\alpha=0$ (fourth row; GP+GLM ND); and 
    GP sources and GLM divergence cleaning, $c_{h}=50$ and $\alpha=c_h/0.18$ (fifth row; GP+GLM D).}
    \label{fig:div-adv-terms}
\end{figure}

The impact of various divergence cleaning terms is shown in Figure~\ref{fig:div-adv-terms}, which shows results from five different models.  
With only the GP sources, the initial divergence defect is carried by the background velocity with limited dispersion and essentially no reduction in amplitude.  
With GLM divergence cleaning without damping, the divergence error spreads across domain at a rate proportional to the cleaning speed, which was set to $c_h=50$ here.  
The added damping source term (GLM Damp) facilitates active removal of divergence error.  
Comparing the second and third rows with the fourth and fifth rows, respectively, the addition of the Godunov-Powell source increases dispersion by transporting divergence errors with the flow velocity.  
For the specified time step $\dt=2.5\times10^{-3}$, Eq.~\eqref{eq:dtCFL} gives $\CFL\approx 0.73$ for the case with only GP sources, while it increases to $\CFL\approx 32.5$ for the cases with GLM divergence cleaning and $c_{h}=50$.  
The Lorentz force is active in the momentum equation for this test, which results in a velocity magnitude deviation of about 10\% relative to the initial value at the end of the simulation.  
However, within the core of the divergence bubble, both the Lorentz force and velocity deviation remain effectively zero.  

\begin{figure}[ht!]
    \centering
    \subfloat[Variation of cleaning speed $c_h$]{
        \includegraphics[width=0.48\linewidth]{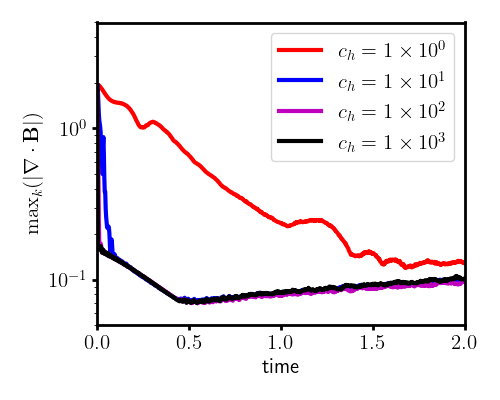}
        \label{fig:div-adv-ch}
    }
    \hfill
    \subfloat[Variation of damping parameter $\alpha$]{
    \includegraphics[width=0.48\linewidth]{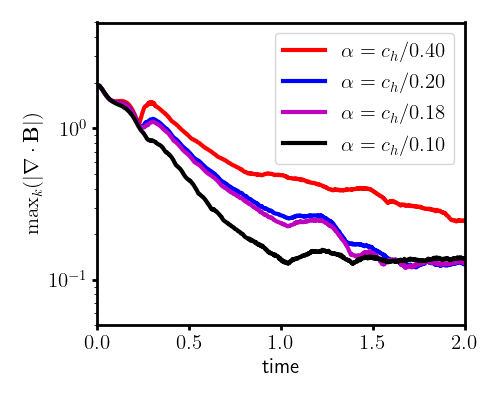}
        \label{fig:div-adv-alpha}
    }
    \caption{Impact of divergence cleaning parameters on the time evolution of maximum local divergence metrics.}
    \label{fig:div-adv-cleaning-main}
\end{figure}

The effect of altering the cleaning parameters $c_h$ and $\alpha$ is shown in Figure~\ref{fig:div-adv-cleaning-main}.  
(The GP sources are also active in these simulations.)  
From Figure~\ref{fig:div-adv-ch}, we observe that increasing the divergence cleaning speed $c_h$ with a fixed relative damping of $\alpha=c_h/0.18$ results in more rapid reduction of the maximum divergence in the domain, here computed as in Eq.~\eqref{eq:maxAbsDivB}.
However, there is a point of diminishing returns at which additional increases to $c_h$ no longer provide appreciable benefit. 
Cases with $c_h=1$, $10$, $10^{2}$, and $10^{3}$ resulted in $\CFL\approx1.3$, $7$, $64$, and $64$, respectively; for $c_{h}=10^{3}$, we reduced $\dt$ by an order of magnitude due to problems with linear solver convergence.  
The impacts of altering the damping parameter $\alpha$ for a fixed $c_h=1.0$ are shown in Figure~\ref{fig:div-adv-alpha}. 
Increased damping results in a reduction in maximum divergence; however, values of $\alpha$ larger than the $c_h/0.18 \approx 5.56$ level suggested by \cite{dedner_etal_2002} results in slightly elevated maximum divergence values later in the simulation, indicating that $\alpha=c_{h}/0.18$ may be a suitable choice for \vertex\ as well.  

\subsection{Current Sheet Problem}
\label{sec:currentSheetTest}

Here, \vertex\ is employed to model the current sheet problem described in \cite{fambri_etal_2023}.  
It consists of a computational domain $\Omega=[-1,1]\times[-0.1,0.1]$. The fluid is initially at rest with an uniform pressure $P=0$.  
The initial magnetic field is given by
\begin{align}\label{eq:current-sheet-ics}
    \begin{split}
    B_x = 0 \text{ and }
    B_y = \begin{cases} -0.1~\text{if}~x\leq0 \\ \phantom{-}0.1~\text{if}~x>0 \end{cases}
    \end{split}.
\end{align}
In this problem, the fluid remains at rest and an exact solution for the $y$-component of the magnetic field is
\begin{equation}\label{eq:sheet_exact}
    B_y(x,t) = \frac{1}{10}\text{erf}\left(\frac{x}{2\sqrt{\eta t}}\right).
\end{equation}

\begin{figure}[!htbp]
     \centering
     \begin{subfigure}[b]{0.45\textwidth}
         \centering
         \includegraphics[width=\textwidth]{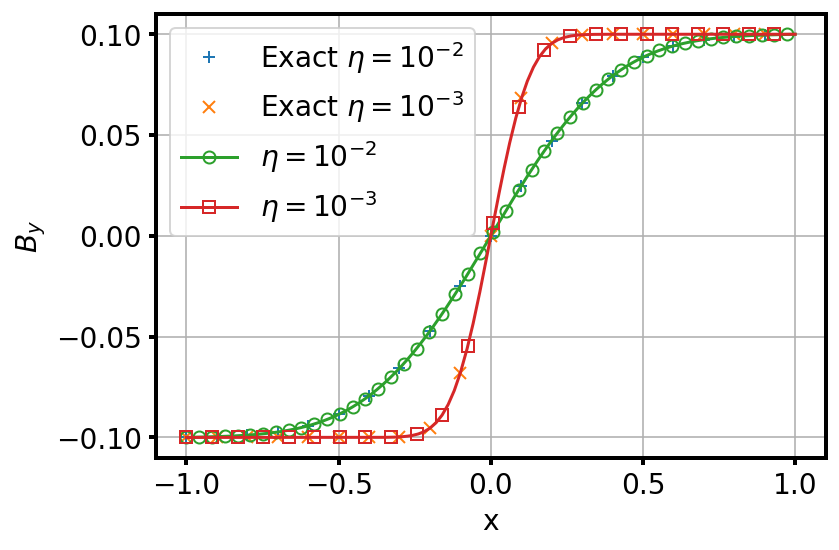}
         \caption{Quadrilateral elements.}
         \label{fig:quads_eta_0p01}
     \end{subfigure}
     \hfill
     \begin{subfigure}[b]{0.45\textwidth}
         \centering
         \includegraphics[width=\textwidth]{./current_sheet_results_t_5s.png}
         \caption{Triangle elements.}
         \label{fig:quads_eta_0p001}
     \end{subfigure}
        \caption{Comparison of the \vertex\ solution against the exact solution for the current sheet problem obtained with a quadrilateral (left) and triangle (right) mesh composed of 50 elements in each Cartesian direction. 
        Fewer markers than nodal points are plotted to reduce marker density.}
        \label{fig:quads-current-sheet-t5}
\end{figure}

\vertex\ is run until $t = 5.0$ with a $\CFL=1.0$, using the SDIRK22 temporal integrator and linear elements. 
The linear and nonlinear tolerances for the implicit solve are set to $\tolL=10^{-10}$ and $\tolN=10^{-8}$, respectively.  
The computational domain is meshed with quadrilateral elements or triangle elements. 
The domain is periodic in the $y$-direction, and Dirichlet-like boundary conditions are applied in the $x$-direction to match the initial conditions. 
Numerical results obtained with \vertex\ for two values of the resistivity $\eta$ and two mesh element types are presented in Figures~\ref{fig:quads-current-sheet-t5}.  
It is observed that the \vertex\ numerical solution matches the exact solution well, independent of resistivity values and mesh element types.  
We also verify that the other variables (pressure and velocity) remain constant during the evolution of $B_{y}$.

\subsection{Lid-Driven Cavity with Magnetic Field}
\label{sec:lidDrivenCavity}

The lid-driven cavity (LDC) problem (e.g., \cite{ghia_etal_1982}), which was simulated with magnetic fields in \cite{fambri_etal_2023}, considers a flow in a two-dimensional box, $(x,y)\in\Omega=[x_{\min},x_{\max}]\times[y_{\min},y_{\max}]$, with stationary no-slip walls at $x_{\min}$, $y_{\min}$, and $x_{\max}$, and the no-slip wall at $y_{\max}$ traveling at a constant velocity $\bv_\text{wall} = (V_{0},0)^{\intercal}$.  
The flow is initially at rest and develops over time due to the motion of the upper wall (the lid) and the viscosity of the fluid, under the influence of an initially constant magnetic field $\bB^0=(B^0_x, B^0_y)^{\intercal}$.  
The cavity walls are considered to be perfectly conducting, which requires on the boundaries
\begin{equation}
	\bB\cdot \bn = \bB^0 \cdot \bn
	\quad\text{and}\quad
        \bE\times\bn = 0,
\end{equation}
where $\bn$ is the boundary normal unit vector.  
The first condition provides the normal component of the boundary magnetic field.  
The second gives a condition on the gradient of the magnetic field.  
Using Eq.~\eqref{eq:ohm}, the condition $\bE\times\bn = 0$ can be recast from
\begin{equation}
	\bE\times\bn 
	= \bn \times \big(\, \bv \times \bB \,\big) - \bn \times \big(\, \eta\,\nabla \times \bB \,\big),
\end{equation}
so that
\begin{equation}
    \bn \times \big(\, \bv \times \bB \,\big) = \bn \times \big(\, \eta\,\nabla \times \bB \,\big).
\end{equation}
With some additional manipulation, the above can be expressed as
\begin{equation}
    (\bn\cdot\nabla)\bB = -\eta\,\bn\times(\bv\times\bB) = \eta\,\big(\,\bB(\bn\cdot\bv)-\bv(\bn\cdot\bB)\,\big).  
\end{equation}

As in \cite{fambri_etal_2023}, we consider two orientations of the initial magnetic field for the LDC: horizontal and vertical.  
For these cases, the above boundary conditions reduce to the following:
\begin{itemize}
  \item For the initially horizontal magnetic field, on the $x$-boundary faces ($x_{\rm b}\in\{x_{\min},x_{\max}\}$):
    \begin{equation}
        B_{x,\text{b}} = B^0_x
    \end{equation}
  \item For the initially vertical magnetic field, on the $y$-boundary faces ($y_{\rm b}\in\{y_{\min},y_{\max}\}$):
    \begin{equation}
        B_{y,{\rm b}} = B^0_y 
        \quad\text{and}\quad
        \frac{\partial B_x}{\partial y}|_{{\rm b}} = -\frac{V_0 B^0_y}{\eta},
    \end{equation}
\end{itemize}
where all other unspecified components are trivially zero on the boundaries.

Following \cite{fambri_etal_2023}, LDC simulations with domain boundaries at $x_{\min}=y_{\min}=-0.5$ and $x_{\max}=y_{\max}=0.5$, lid velocity $V_0 = 1$, and kinematic viscosity and resistivity of $\nu = \eta = 0.01$ were run to $t=20$ with \vertex\ for horizontal and vertical initial magnetic fields, $\bB^0 = (B_0, 0)^{\intercal}$ and $\bB^0 = (0, B_0)^{\intercal}$, respectively, with field strengths $B_0 \in \{0.10,\,0.25,\,0.50,\,1.00\}$.  
The mesh used for these simulations consists of $128\times128$ linear quadrilateral elements, with a fine spacing of about $4.2\times10^{-5}$ near the boundaries, which increases towards the center of the domain with an approximate geometric growth factor of $1.134$ near the wall.  
For all models, the time step is set by increasing $\CFL$ according to the following schedule:
For the first 200 time steps, $\CFL=10$ ($\dt\approx3.6\times10^{-6}$); for time steps 201-2000, $\CFL$ increases linearly from 10 to $10^{4}$; from time step 2001, $\CFL=10^{4}$ ($\dt\approx3.6\times10^{-3}$).  
Each model completed in about 6637 time steps using the SDIRK22 time integrator.  
Linear and nonlinear implicit solver tolerances are set to $\tolL=5\times10^{-7}$ and $\tolN=10^{-8}$, respectively.  

\begin{figure}[ht!]
    \centering
    \begin{subfigure}[t]{0.485\textwidth}
        \centering
        \includegraphics[width=0.8\linewidth]{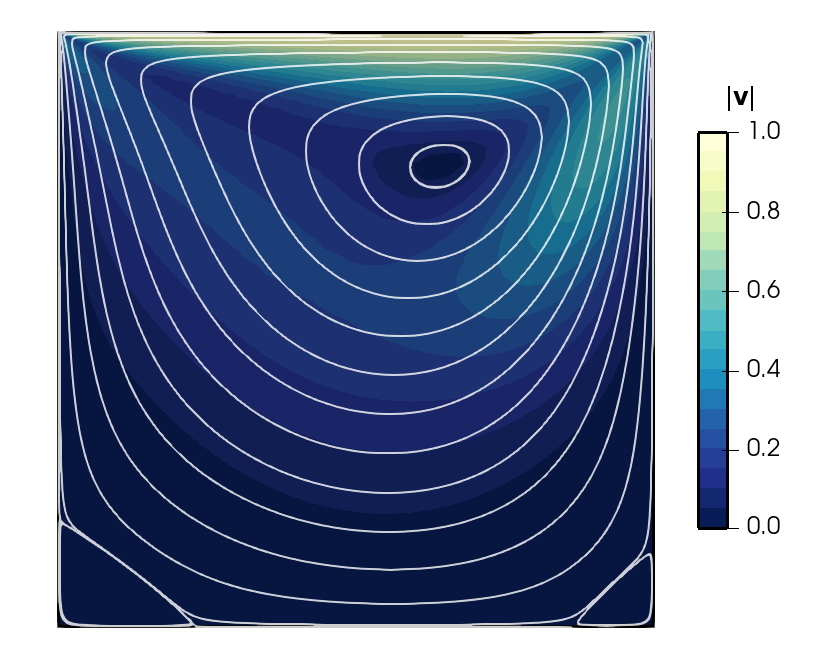}
        \caption{$B^0_x = 0.1$.}
    \end{subfigure}
    \begin{subfigure}[t]{0.485\textwidth}
        \centering
        \includegraphics[width=0.8\linewidth]{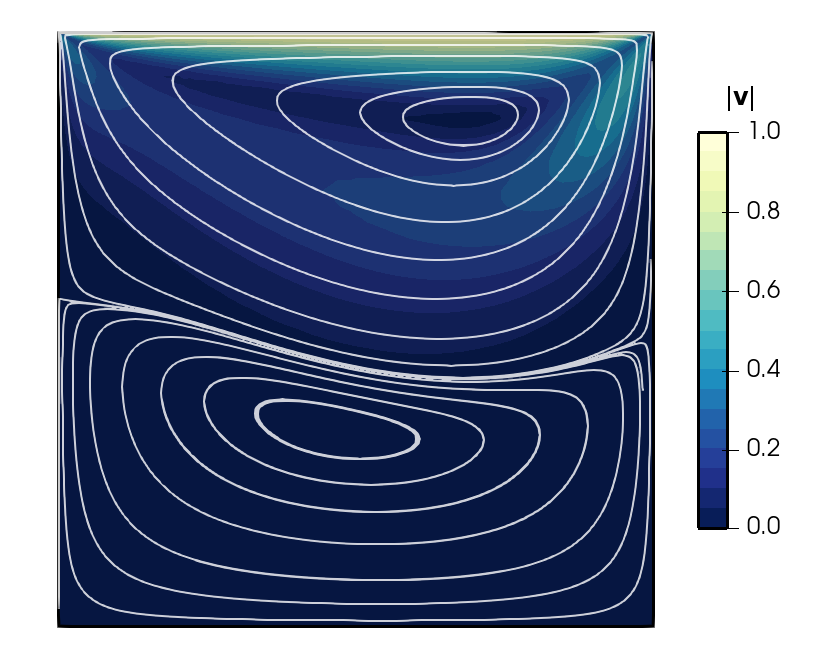}
        \caption{$B^0_x = 0.25$.}
    \end{subfigure}
    \\
    \vspace{5pt}
    \begin{subfigure}[t]{0.485\textwidth}
        \centering
        \includegraphics[width=0.8\linewidth]{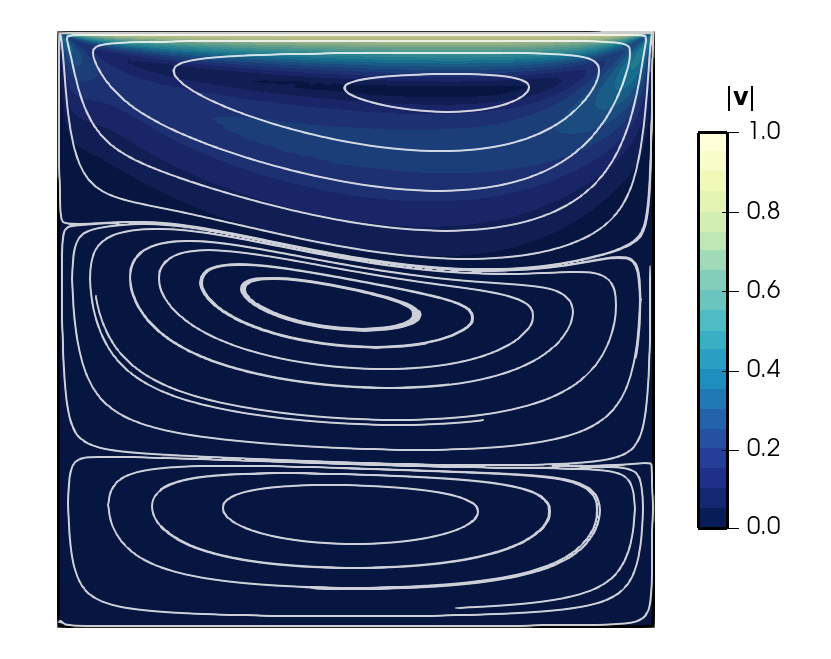}
        \caption{$B^0_x = 0.5$.}
    \end{subfigure}
    \begin{subfigure}[t]{0.485\textwidth}
        \centering
        \includegraphics[width=0.8\linewidth]{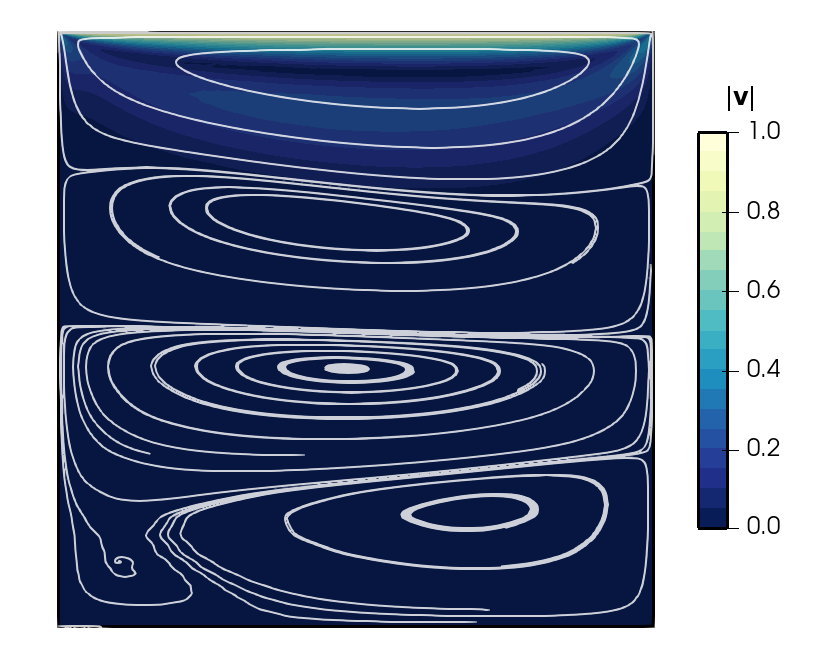}
        \caption{$B^0_x = 1.0$.}
    \end{subfigure}
    \caption{Velocity magnitude with streamlines overlaid for the MHD LDC with horizontal initial magnetic field of varying strength.}
    \label{fig:ldc_bx_vel}
\end{figure}

\begin{figure}[ht!]
    \centering
    \begin{subfigure}[t]{0.485\textwidth}
        \centering
        \includegraphics[width=0.8\linewidth]{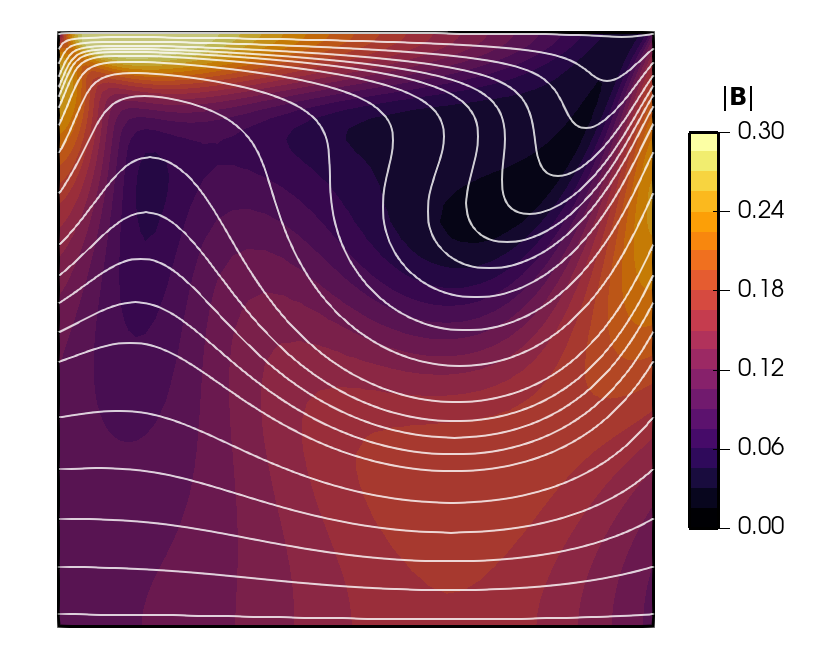}
        \caption{$B^0_x = 0.1$.}
    \end{subfigure}
    \begin{subfigure}[t]{0.485\textwidth}
        \centering
        \includegraphics[width=0.8\linewidth]{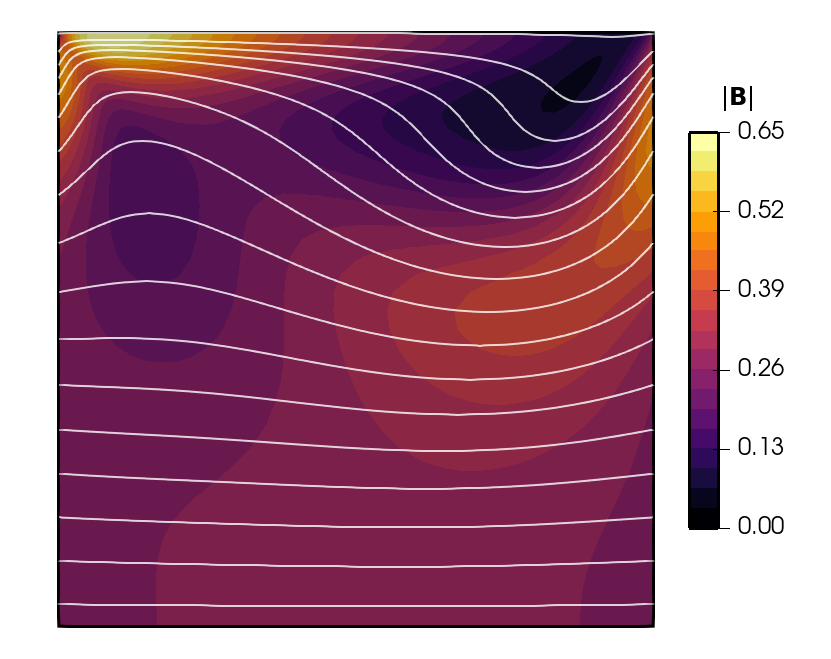}
        \caption{$B^0_x = 0.25$.}
    \end{subfigure}
    \\
    \vspace{5pt}
    \begin{subfigure}[t]{0.485\textwidth}
        \centering
        \includegraphics[width=0.8\linewidth]{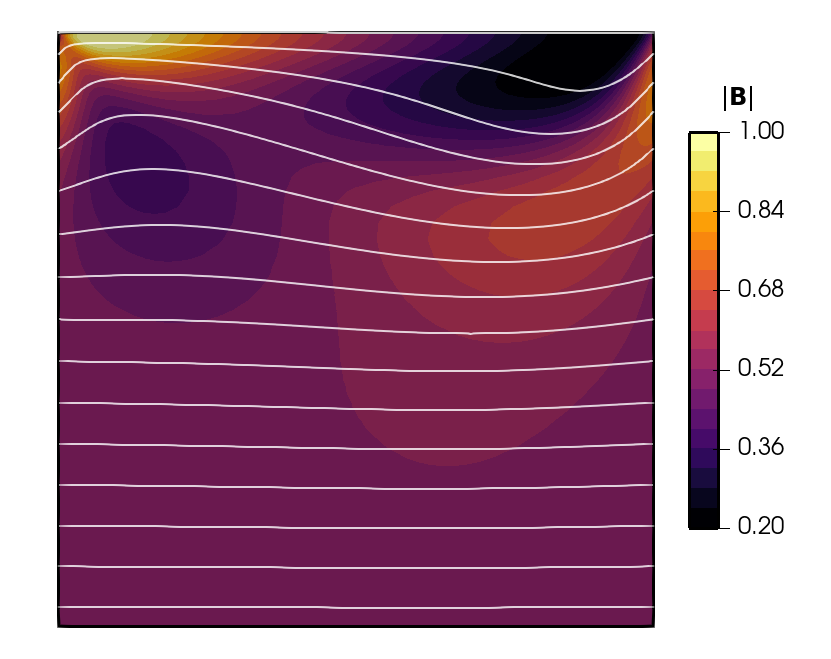}
        \caption{$B^0_x = 0.5$.}
    \end{subfigure}
    \begin{subfigure}[t]{0.485\textwidth}
        \centering
        \includegraphics[width=0.8\linewidth]{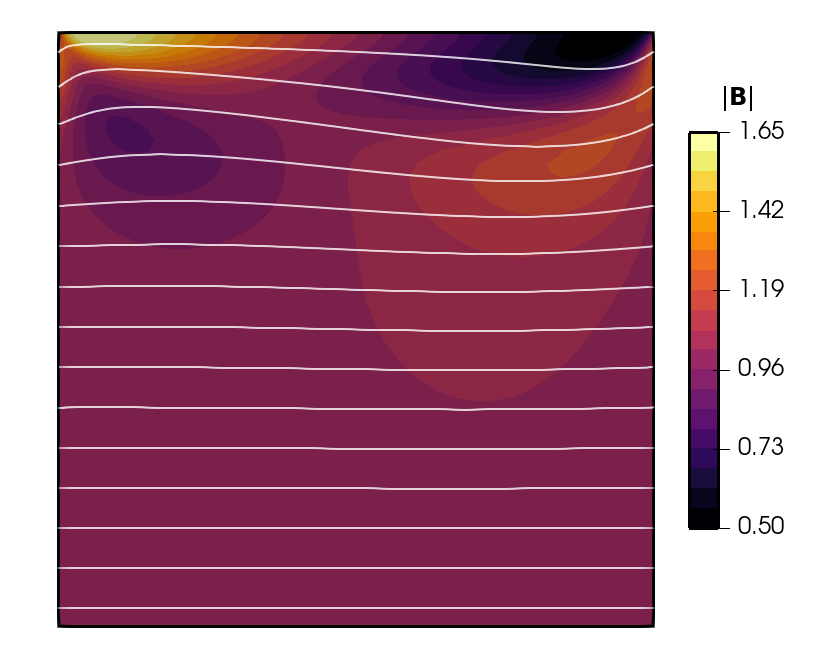}
        \caption{$B^0_x = 1.0$.}
    \end{subfigure}
    \caption{Magnetic field magnitude with field lines overlaid for the MHD LDC with horizontal initial magnetic field of varying strength.}
    \label{fig:ldc_bx_mag}
\end{figure}

\begin{figure}[ht!]
    \centering
    \begin{subfigure}[t]{0.485\textwidth}
        \centering
        \includegraphics[width=0.8\linewidth]{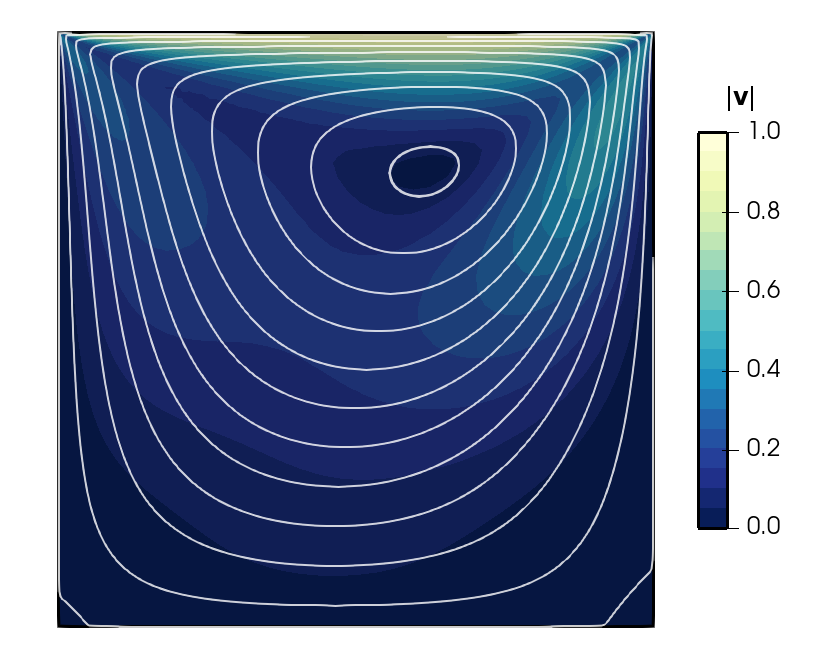}
        \caption{$B^0_y = 0.1$.}
    \end{subfigure}
    \begin{subfigure}[t]{0.485\textwidth}
        \centering
        \includegraphics[width=0.8\linewidth]{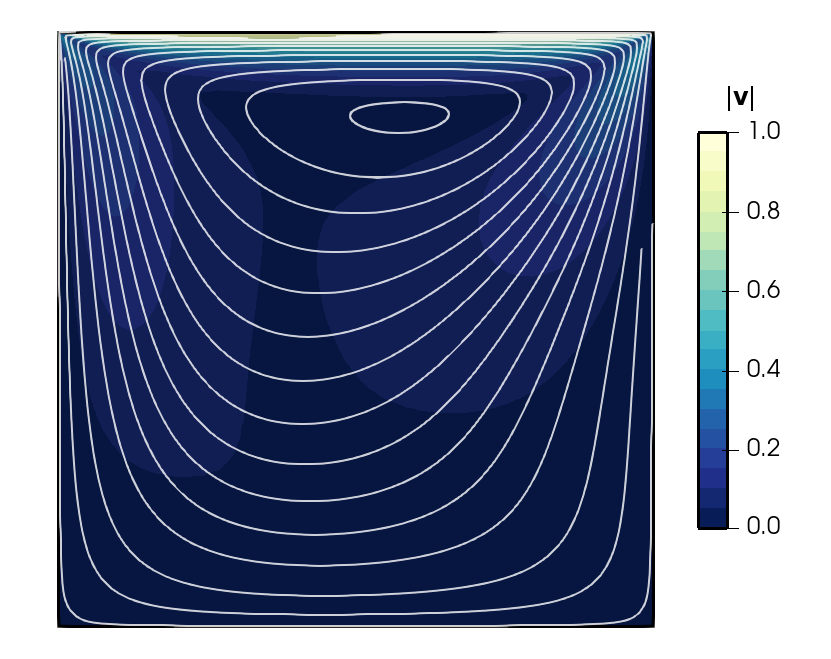}
        \caption{$B^0_y = 0.25$.}
    \end{subfigure}
    \\
    \vspace{5pt}
    \begin{subfigure}[t]{0.485\textwidth}
        \centering
        \includegraphics[width=0.8\linewidth]{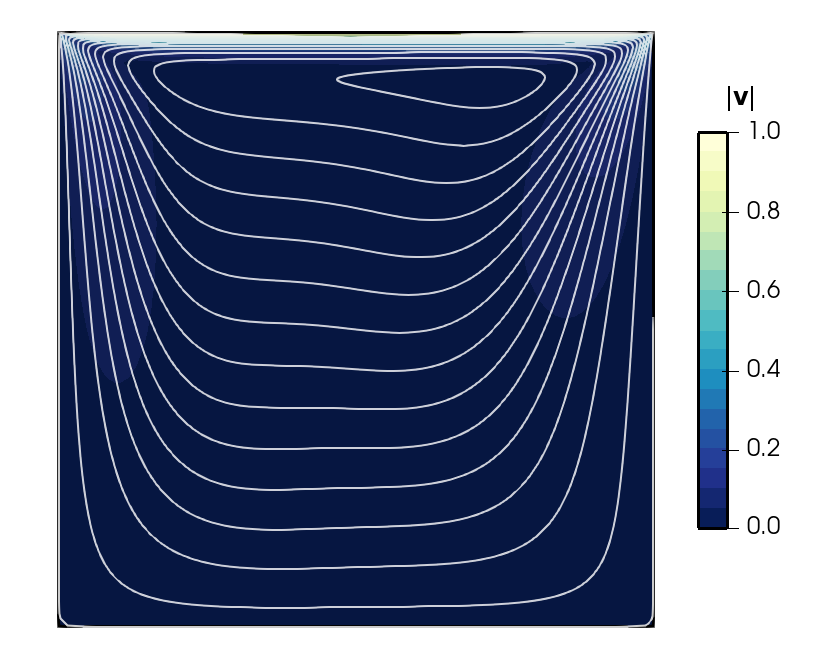}
        \caption{$B^0_y = 0.5$.}
    \end{subfigure}
    \begin{subfigure}[t]{0.485\textwidth}
        \centering
        \includegraphics[width=0.8\linewidth]{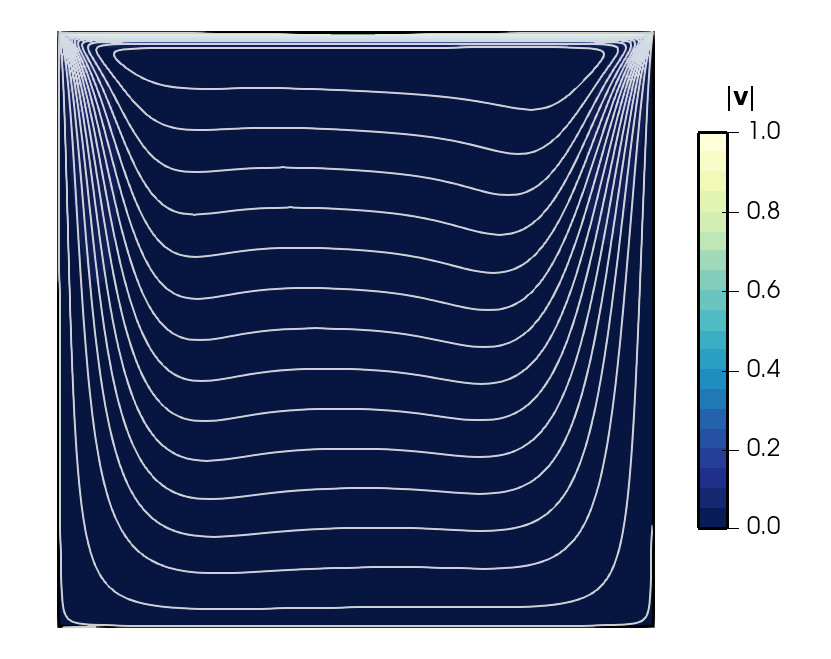}
        \caption{$B^0_y = 1.0$.}
    \end{subfigure}
    \caption{Velocity magnitude with streamlines overlaid for the MHD LDC with vertical initial magnetic field of varying strength.}
    \label{fig:ldc_by_vel}
\end{figure}

\begin{figure}[ht!]
    \centering
    \begin{subfigure}[t]{0.485\textwidth}
        \centering
        \includegraphics[width=0.8\linewidth]{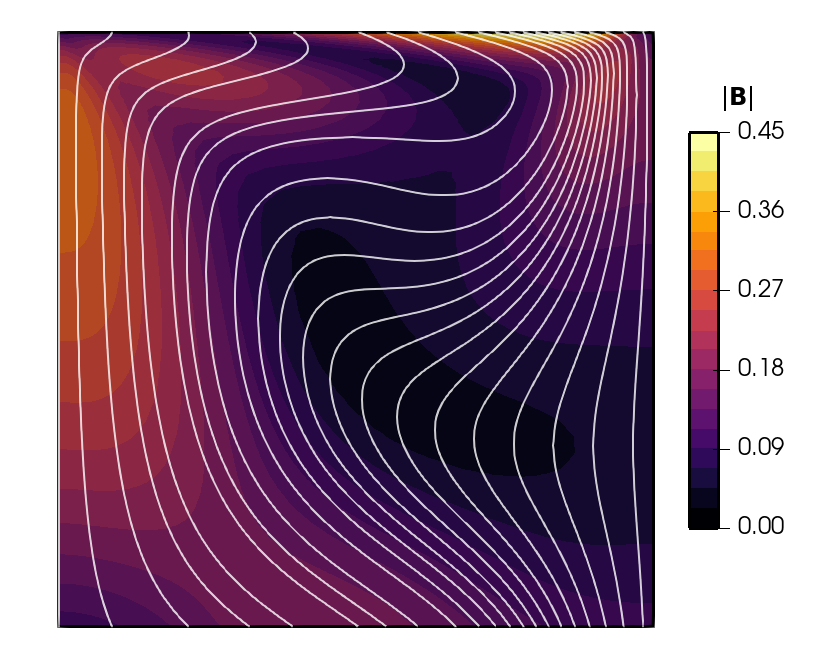}
        \caption{$B^0_y = 0.1$.}
    \end{subfigure}
    \begin{subfigure}[t]{0.485\textwidth}
        \centering
        \includegraphics[width=0.8\linewidth]{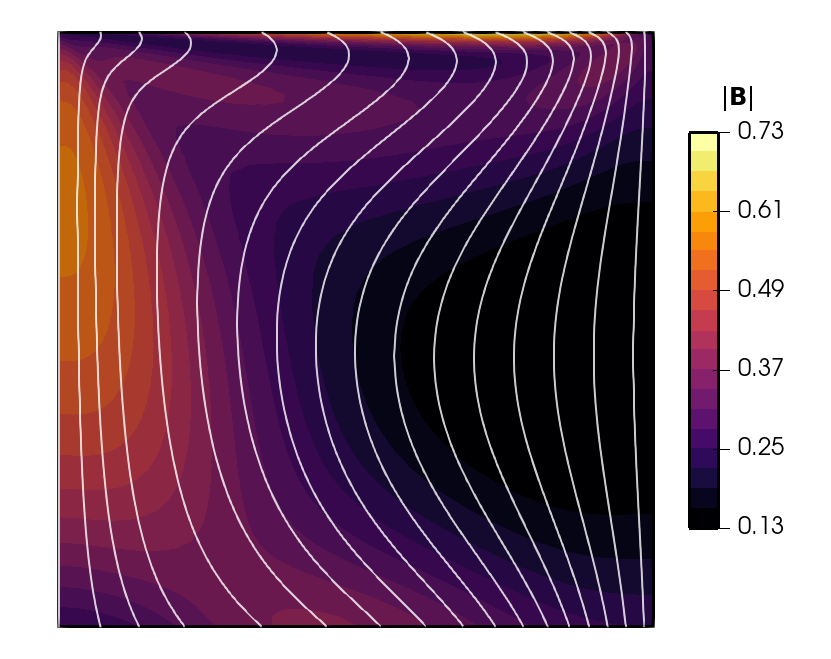}
        \caption{$B^0_y = 0.25$.}
    \end{subfigure}
    \\
    \vspace{5pt}
    \begin{subfigure}[t]{0.485\textwidth}
        \centering
        \includegraphics[width=0.8\linewidth]{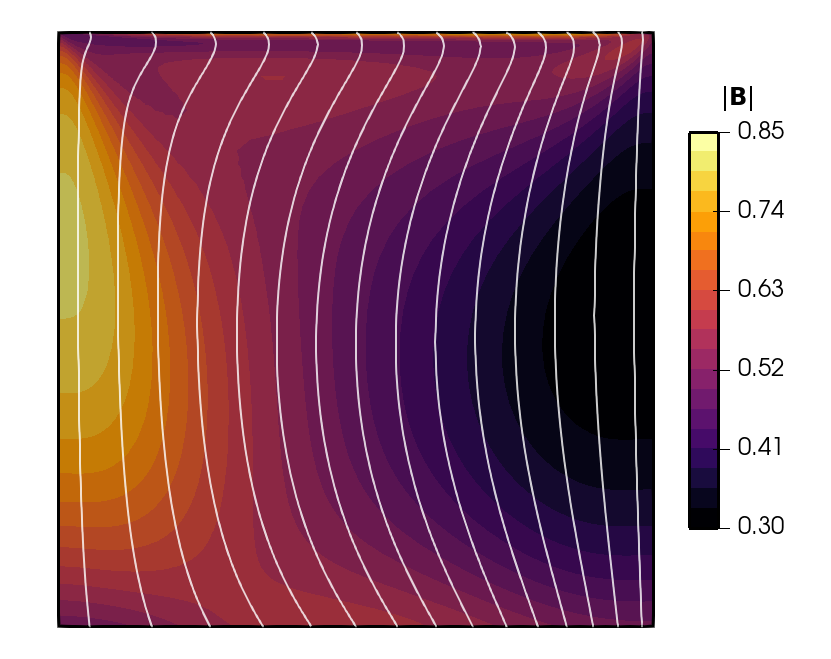}
        \caption{$B^0_y = 0.5$.}
    \end{subfigure}
    \begin{subfigure}[t]{0.485\textwidth}
        \centering
        \includegraphics[width=0.8\linewidth]{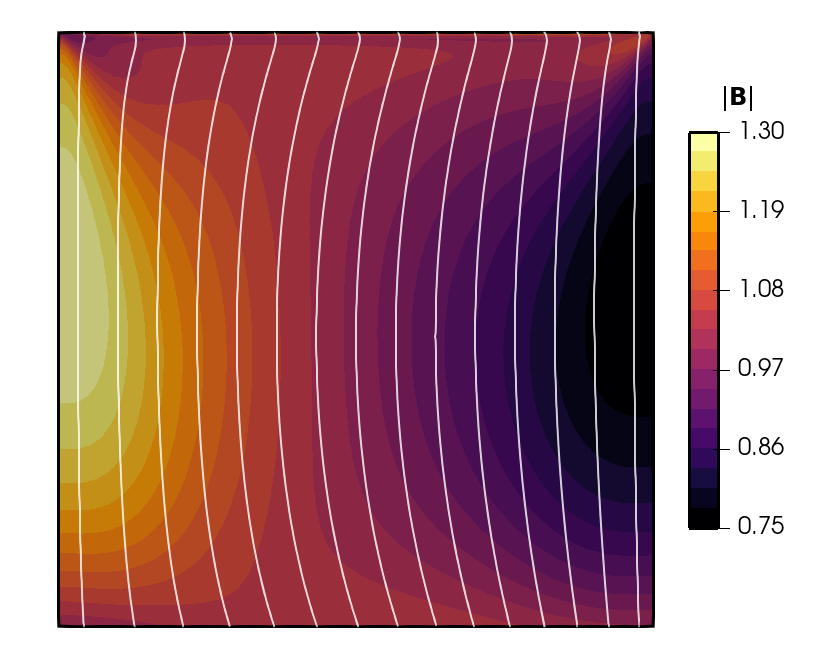}
        \caption{$B^0_y = 1.0$.}
    \end{subfigure}
    \caption{Magnetic field magnitude with field lines overlaid for the MHD LDC with vertical initial magnetic field of varying strength.}
    \label{fig:ldc_by_mag}
\end{figure}

\begin{figure}[ht!]
    \centering
    \begin{subfigure}[t]{0.85\textwidth}
        \centering
        \includegraphics[width=0.9\linewidth]{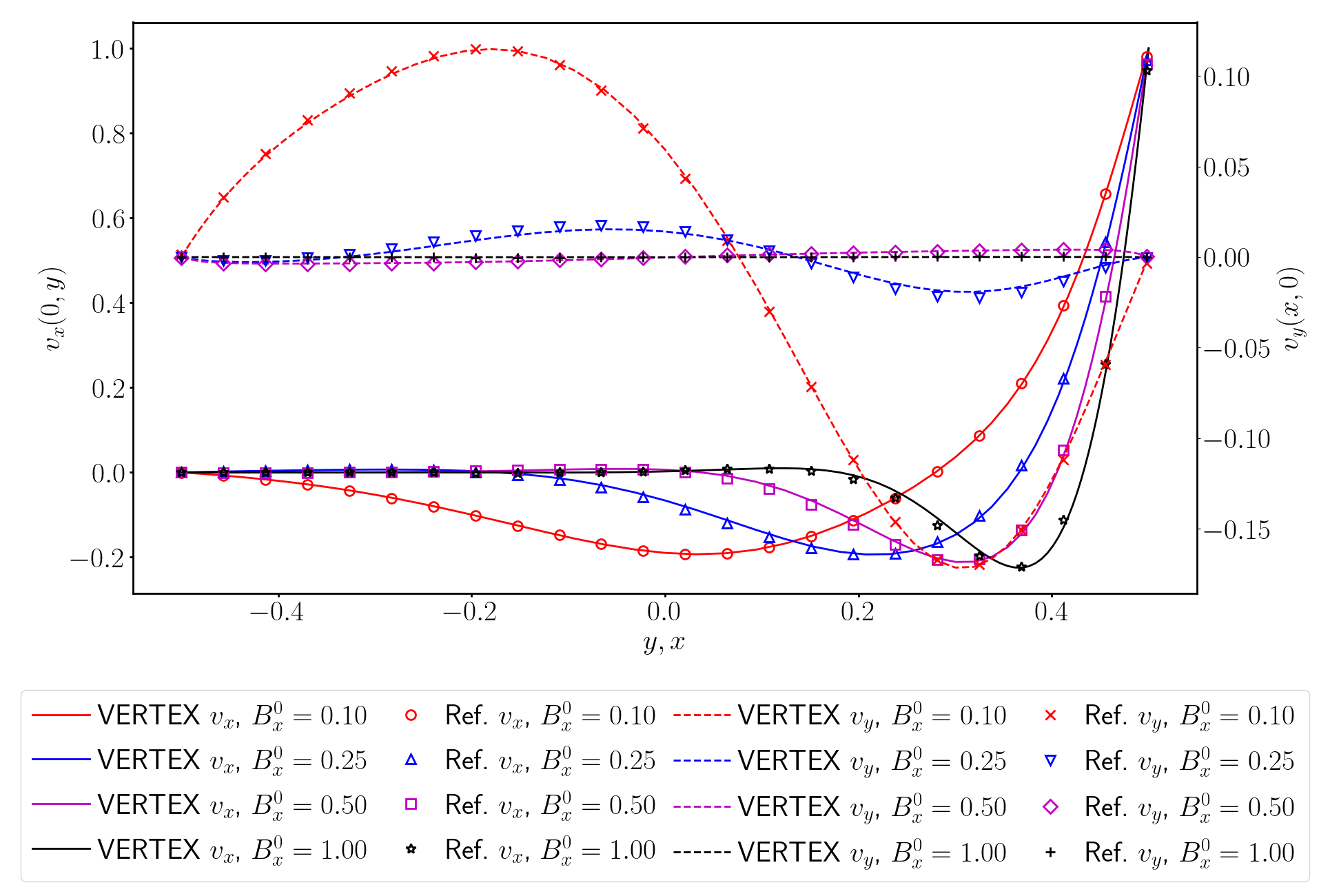}
        \caption{Velocity profiles.}
    \end{subfigure}
    \\
    \vspace{5pt}
    \begin{subfigure}[t]{0.85\textwidth}
        \centering
        \includegraphics[width=0.9\linewidth]{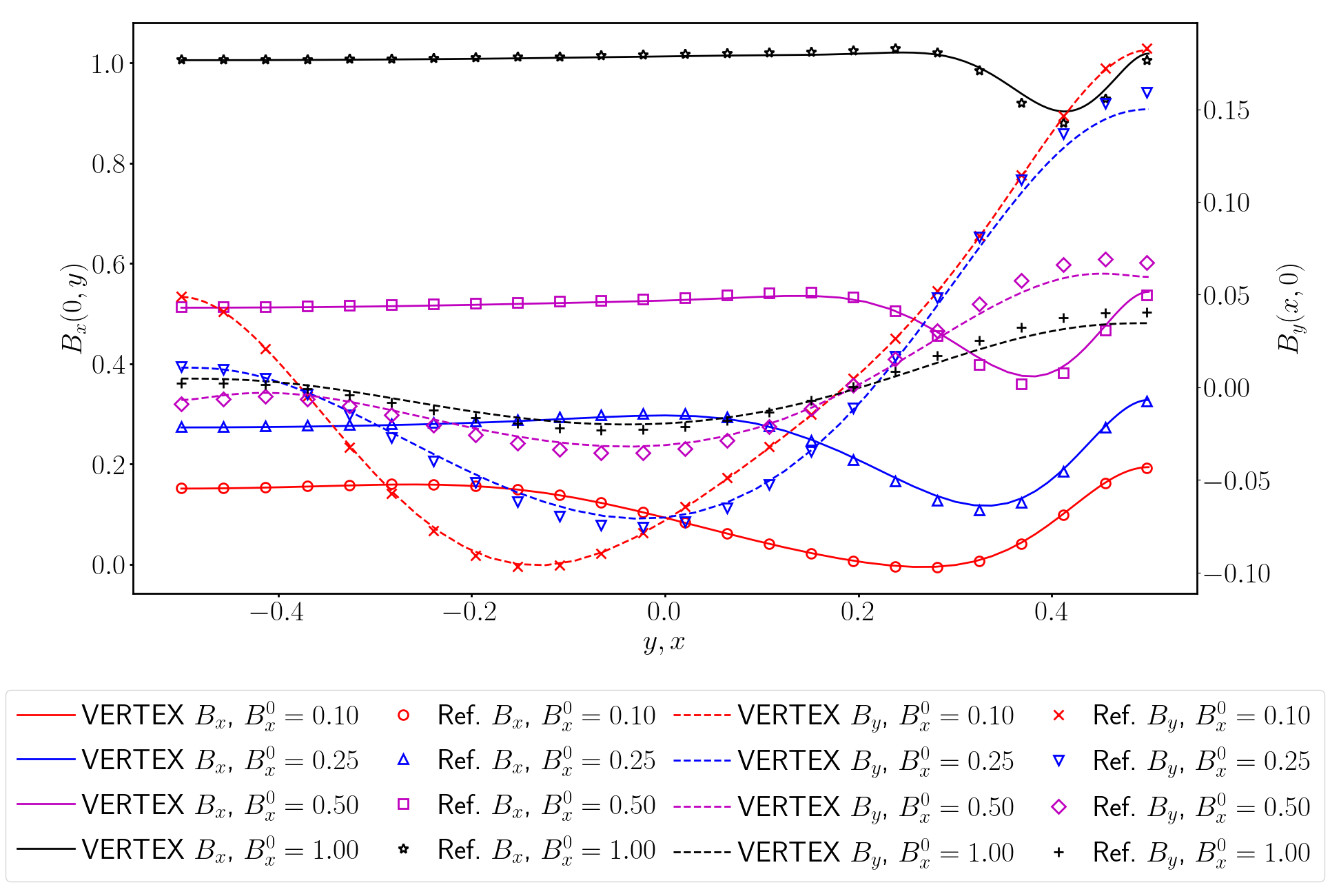}
        \caption{Magnetic field profiles.}
    \end{subfigure}
    \caption{Comparison of velocity (top) and magnetic field (bottom) component profiles along $x=0$ ($v_{x}$ and $B_{x}$) and $y=0$ ($v_{y}$ and $B_{y}$) from \vertex\ (lines) with reference data from \cite{fambri_etal_2023} (symbols) for the initially horizontal magnetic field.}
    \label{fig:ldc_bx_profiles}
\end{figure}

\begin{figure}[ht!]
    \centering
    \begin{subfigure}[t]{0.85\textwidth}
        \centering
        \includegraphics[width=0.9\linewidth]{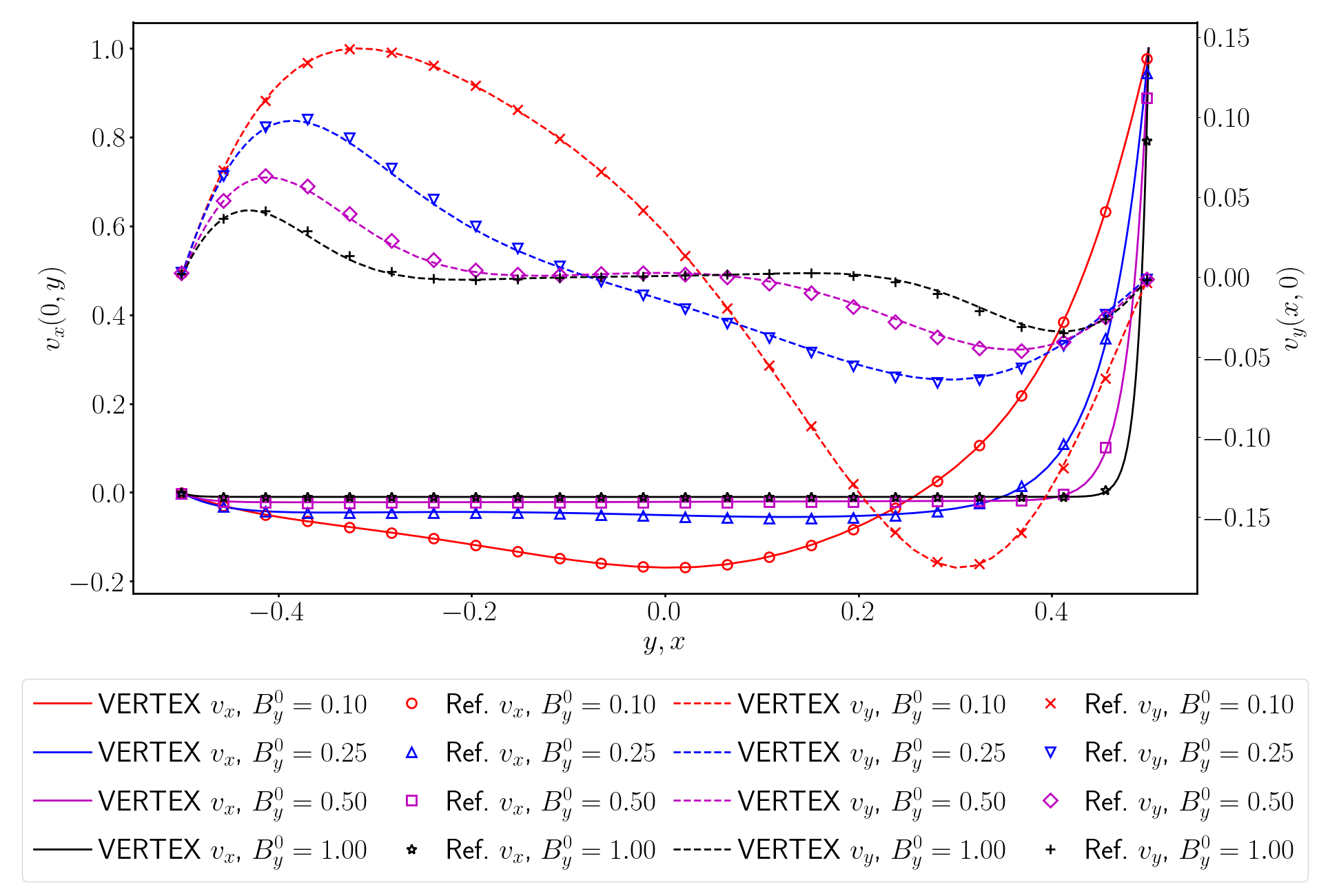}
        \caption{Velocity, $B^0_y$}
    \end{subfigure}
    \\
    \vspace{5pt}
    \begin{subfigure}[t]{0.85\textwidth}
        \centering
        \includegraphics[width=0.9\linewidth]{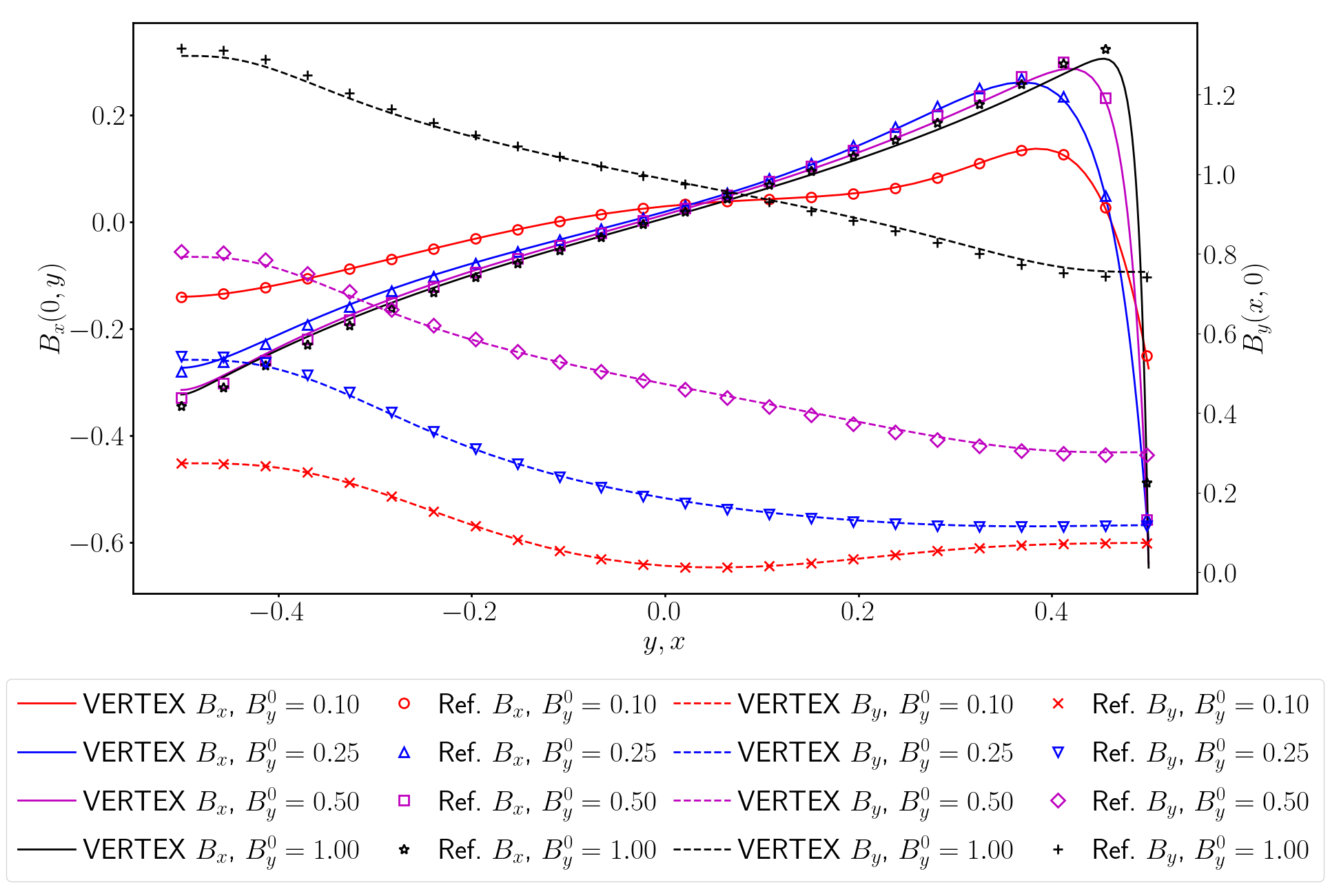}
        \caption{Magnetic field, $B^0_y$}
    \end{subfigure}
    \caption{Comparison of velocity (top) and magnetic field (bottom) component profiles along $x=0$ ($v_{x}$ and $B_{x}$) and $y=0$ ($v_{y}$ and $B_{y}$) from \vertex\ (lines) with reference data from \cite{fambri_etal_2023} (symbols) for the initially vertical magnetic field.}
    \label{fig:ldc_by_profiles}
\end{figure}

In Figure~\ref{fig:ldc_bx_vel}, increasing horizontal magnetic field strength compresses the primary rotation (compare velocity streamline configurations across panels) toward the lid and results in additional weaker zones of rotation. 
Figure~\ref{fig:ldc_bx_mag} shows the increasing impact of the magnetic field with increasing $B_{0}$, as the higher velocities driving the induced magnetic field are confined to the region near the cavity lid.  
In the case of the vertical initial magnetic field, a single rotating vortex is maintained at all strengths with the core flattening and compressing towards the lid as the field strength is increased, as shown in Figure~\ref{fig:ldc_by_vel}.  
As in the horizontal case, Figure~\ref{fig:ldc_by_mag} shows magnetic field lines increasingly aligned with the applied field, with deviations due to induction moving toward the higher-velocity regions at the cavity lid and side walls. 
Qualitatively, the results agree well with those presented by \cite[][see their Figures~18-21]{fambri_etal_2023}.  
The agreement is reinforced by a more quantitative comparisons of velocity and magnetic field profiles from \vertex\ with reference solution data provided by \cite{fambri_etal_2023} plotted in Figures~\ref{fig:ldc_bx_profiles} and \ref{fig:ldc_by_profiles}.
%
% Horizontal applied field velocity
%
%
% Horzontal applied field magnetic field
%
%
% Vertical applied field velocity
%
%
% Vertical applied field magnetic field
%
%
% LDC reference comparison plots
%

\section{Application to an Idealized Model of a Liquid Metal Fusion Blanket}
\label{sec:applications}

In this section, we consider the transient evolution of liquid metal flows in an idealized fusion blanket duct driven by the decay of an external magnetic field.  
The model was recently proposed and studied in detail by Smolentsev~\cite[][\citetalias{smolentsev_2025} from hereon]{smolentsev_2025}.  
To this end, the magnetic field is decomposed as
\begin{equation}
	\bB=\bB^{0} + \bb,
\end{equation}
where $\bB^{0}$ is the external, prescribed plasma magnetic field and $\bb$ is the blanket \emph{induced magnetic field}.  
The induced magnetic field is evolved by solving the induction equation in the form
\begin{equation}
	\pd{\bb}{t}+\nabla\cdot\big[\,\big(\bv\otimes\bB-\bB\otimes\bv\big)-\f{\eta}{\mu_{0}}\,\big(\,\nabla\bB-(\nabla\cdot\bB)\,\bI\,\big)\,\big] 
	= -\bv\,(\nabla\cdot\bB) - c_{h}\,\nabla\psi - \pd{\bB^{0}}{t},
	\label{eq:induction_equation_induced_magnetic_field}
\end{equation}
where the source due to the time rate of change in the external field, $-\pd{\bB^{0}}{t}$, is prescribed.  

\begin{figure}[!ht]
	\begin{center}
		\includegraphics[width=0.48\textwidth]{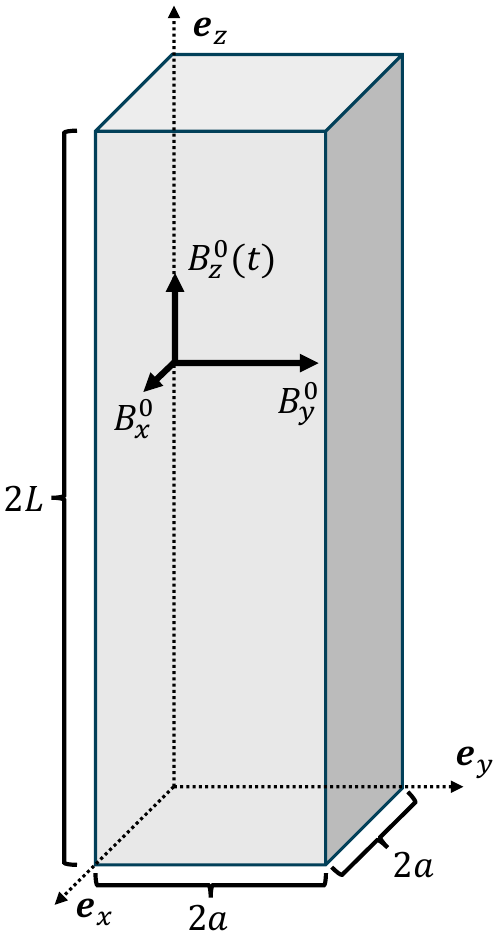}
	\end{center}
	\caption{Sketch of idealized blanket duct with cross sectional area $2a\times 2a$ and height $2L$, and applied plasma magnetic field $\bB^{0}=(B_{x}^{0},B_{y}^{0},B_{z}^{0})^{\intercal}$.  
	During the simulation of a plasma disruption, the \emph{poloidal magnetic field}, $B_{z}^{0}$ (initially set to $1$~T), decays at a constant rate $\pd{B_{z}^{0}}{t}=-10^{2}$~T~s$^{-1}$, while the \emph{toroidal magnetic field}, $B_{y}^{0}$, remains fixed.  
	In the simulations presented here, the \emph{radial magnetic field} $B_{x}^{0}=0$.}
	\label{fig:dbdt_flow_setup}
\end{figure}

The computational domain is a square duct $\Omega=[-a,a]^{2}\times[-L,L]$ with sides $2a$ and length $2L$, where $a\ll L$; see Figure~\ref{fig:dbdt_flow_setup}.  
Specifically, we set $L=1$~m and $a=0.1$~m.  
The components of the external plasma magnetic field threading the duct, $\bB^{0}=(B_{x}^{0},B_{y}^{0},B_{z}^{0})^{\intercal}$, is characterized by the toroidal ($B_{y}^{0}$) and poloidal ($B_{z}^{0}$) components.  
The radial component ($B_{x}^{0}$) is set to zero in the models considered here.  
We keep the toroidal component fixed in space and time, while the poloidal component is constant in space, but decays linearly in time with a constant time rate of change $\pd{B_{z}^{0}}{t}=-10^{2}$~T~s$^{-1}$.  
We take the initial poloidal field as $B_{z}^{0}(t=0)1$~T, while we will vary the toroidal component with $B_{y}^{0}\in\{\,0.2,0.5,1.0,4.0\,\}$~T, as in \citetalias{smolentsev_2025}.  
The liquid metal (PbLi) in the duct is characterized by the mass density $\rho_{0}=9486$~kg~m$^{-3}$, kinematic viscosity $\nu=1.0542\times10^{-7}$~m$^{2}$~s$^{-1}$, and electrical resistivity $\eta=1.4286\times10^{-6}$~$\Omega$~m.  
Under these conditions, the Reynolds number and magnetic Reynolds number are
\begin{equation}
    \Reynolds
    =\f{U\,\ell}{\nu}
    \sim 
    10^{7}\times\Big(\f{U}{5~{\rm m}~{\rm s}^{-1}}\Big)\,\Big(\f{\ell}{0.2~{\rm m}}\Big)
    \quad\text{and}\quad
    \ReynoldsM
    =\f{U\,\ell\,\mu_{0}}{\eta}
    \sim 
    \Big(\f{U}{5~{\rm m}~{\rm s}^{-1}}\Big)\,\Big(\f{\ell}{0.2~{\rm m}}\Big),
    \label{eq:reynoldsNumbers}
\end{equation}
respectively, where $U$ is a characteristic velocity and $\ell$ is a characteristic length scale.  
For characteristic velocities $U\sim5~{\rm m~s}^{-1}$ and length scales of order the width of the duct, $\Reynolds\gg1$, while $\ReynoldsM$ is of order unity.  
We shall obtain flow velocities that reach $\sim50$~m~s$^{-1}$, implying magnetic Reynolds numbers of $\cO(10)$.  
Another key non-dimensional number is the Hartmann number (e.g., \cite{huntShercliff_1971})
\begin{equation}
	\Ha
	=B\,\ell\,\sqrt{\f{1}{\eta\nu\rho_{0}}}
	\sim
	10^{4}\times\Big(\f{B}{2~{\rm T}}\Big)\,\Big(\f{\ell}{0.2~{\rm m}}\Big),
\end{equation}
where $B$ represents a characteristic magnetic field strength.  
The Hartmann number, which quantifies the ratio of electromagnetic to viscous forces, governs the thickness of the boundary layer that develops near the duct walls.  
Higher $\Ha$ values lead to thinner boundary layers (e.g., \cite{shercliff_1953}), necessitating finer mesh spacing for accurate resolution.  

We denote the boundaries in the $x$-dimension by $\Gamma_{x}=\{(x,y,z)~|~x=\pm a,\,y\in[-a,a],\,z\in[-L,L]\}$, with the $y$ and $z$ boundaries  defined similarly (denoted $\Gamma_{y}$ and $\Gamma_{z}$, respectively).  
On the duct side walls, $\Gamma_{xy}=\Gamma_{x}\cup\Gamma_{y}$, we impose no-slip boundary conditions for the velocity field; i.e., for boundary normal $\bn$, $\bn\cdot\bv=0$ and $\bn\times\bv=0$, which implies $\bv=0$ on $\Gamma_{xy}$.  
Similarly, we impose perfectly insulating wall boundary conditions on the induced magnetic field, $\bb\cdot\bn=0$ and $\bb\times\bn=0$, which implies $\bb=0$ on $\Gamma_{xy}$.  
On the top and bottom boundaries, $\Gamma_{z}$, we consider two cases: ({\it i}) periodic and ({\it ii}) no-slip conditions for $\bv$ and perfectly insulating conditions for $\bb$.  
The former is used exclusively for the two-dimensional models discussed in Section~\ref{sec:applications.2d}, where we solve the MHD equations using a single element in the $z$-dimension, while we compute models using both periodic and no-slip/insulating conditions for the three-dimensional models discussed in Section~\ref{sec:applications.3d}.  
(All two-dimensional models evolve the $z$-component of the velocity and the induced magnetic field.  
Such models are sometimes referred to as 2.5D MHD models.)

For initial conditions, both the velocity and induced magnetic field are set to zero throughout the computational domain.  
Although DCLL blankets exhibit a slow steady flow ($\sim0.1$~m~s$^{-1}$) under normal operating conditions, the flow induced by the decay of the external magnetic field is expected to far exceed this steady background.  
Therefore, initializing with zero velocity represents a reasonable simplification for the purposes of this study.  

\begin{figure}[ht!]
	\begin{center}
		\includegraphics[width=0.50\linewidth]{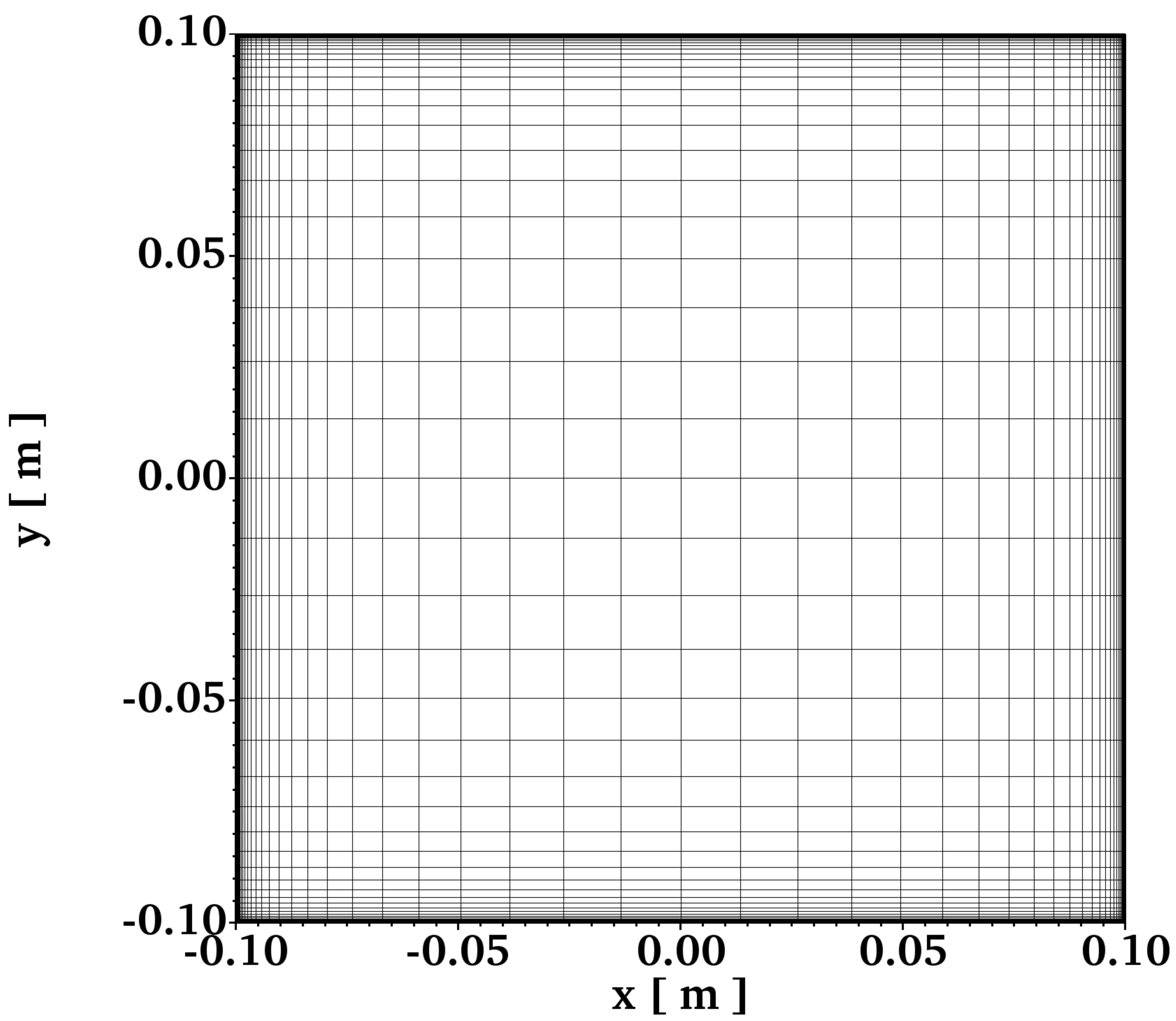}
		\hspace{12pt}
		\includegraphics[width=0.44\linewidth]{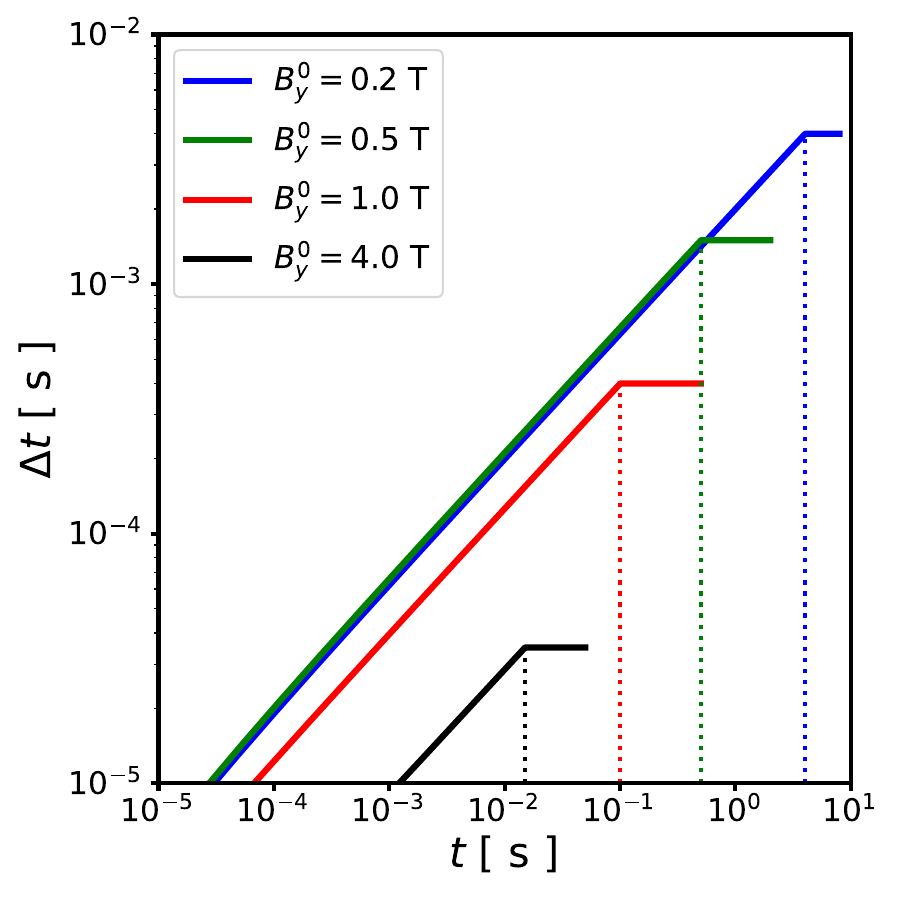}
		\caption{Coarse boundary layer mesh in the $xy$-plane (left panel) and time step versus time (right panel) for models with $B_{y}^{0}=0.2$~T (blue), $0.5$~T (green), $1.0$~T, and $4.0$~T (black).  
		The vertical, dotted lines in the right panel indicate the ``transition time'', $\tau$, for each model.}
		\label{fig:dbdt_mesh_dt}
	\end{center}
\end{figure}

The flows driven by the transient external magnetic field develop thin velocity shear layers near the duct boundaries at $y=\pm a$ (see Figure~\ref{fig:dbdt_By_4.0_F_composite} below).  
To resolve these ``Hartmann'' layers we employ a boundary layer mesh, where the mesh spacing is finest near the duct walls and increases towards the center of the duct as illustrated in the left panel of Figure~\ref{fig:dbdt_mesh_dt}.  
Specifically, we use the Pointwise \cite{pointwise} meshing tool to generate one-dimensional meshes based on the two-sided stretching function from \cite{vinokur_1983}, which is extended to two dimensions by tensorization.   
For the two-dimensional models discussed in Section~\ref{sec:applications.2d} we have created three meshes: coarse, medium, and fine.  
The coarse two-dimensional (2D) mesh displayed in Figure~\ref{fig:dbdt_mesh_dt} consists of $72\times72$ elements with a minimum element width of $\approx3.67\times10^{-6}$~m.  
The medium and fine 2D meshes consist of $108\times108$ and $162\times162$ elements and have minimum element widths of about $2.35\times10^{-6}$~m and $1.5\times10^{-6}$~m, respectively.  
By estimating the Hartmann layer thickness as $a/\Ha$, the selected meshes are expected to provide sufficient resolution for all two-dimensional models considered.  

The mesh for three-dimensional model with no-slip and electrically insulating conditions on all boundaries discussed in Section~\ref{sec:applications.3d} contains a total of $68\times68\times116$ hexahedral elements, with a mesh spacing in the $x$- and $y$-dimensions of $4\times10^{-6}$~m near $\Gamma_{xy}$, which increases geometrically towards the center with a geometric growth rate of $1.3$, reaching a maximum of approximately $2.3\times10^{-2}$~m.  
In the $z$-dimension, the mesh spacing is $6\times10^{-7}$~m near $\Gamma_{z}$, which increases geometrically towards the center with a growth rate of $1.25$, reaching a maximum of about $0.2$~m.

The time step is set to approximately capture the initial evolution with high temporal resolution, while taking relatively larger steps at later times.  
The external magnetic field drives a transient flow in the blanket duct, which eventually settles to a quasi-steady state, and the duration of the initial transient depends on the strength of the toroidal external magnetic field.  
Table~3 in \citetalias{smolentsev_2025} provides ``transition times'', $\tau$, as a function of $B_{y}^{0}$.  
Specifically, $\tau=4.0$~s for $B_{y}^{0}=0.2$~T, $\tau=0.5$~s for $B_{y}^{0}=0.5$~T, $\tau=0.1$~s for $B_{y}^{0}=1.0$~T, and $\tau=0.015$~s for $B_{y}^{0}=4.0$~T.  
For each model, we let the time step increase linearly with time, starting with $\dt=10^{-6}$~s, until it reaches the maximum value $\dt_{\max}$.  
The maximum time step is determined by requiring each model to complete the time interval from $\tau$ to $t_{\rm end}$ in $10^{3}$ step; i.e., $\dt_{\max}=(t_{\rm end}-\tau)/10^{3}$.  
For models with $B_{y}^{0}=0.2$, $0.5$, $1.0$, and $4.0$~T, we set $t_{\rm end}=7.0$, $2.0$, $0.1$, and $0.015$~s, so that $\dt_{\max}=4\times10^{-3}$, $1.5\times10^{-3}$, $4\times10^{-4}$, and $3.5\times10^{-5}$~s, respectively.  
The time step $\dt$ for each model we consider is plotted versus time in the right panel in Figure~\ref{fig:dbdt_mesh_dt}.  
For the medium mesh, each model starts out with CFL number of approximately $40$, as defined in Eq.~\eqref{eq:dtCFL}.  
As the simulations progress and reach their maximum step sizes ($\dt_{\max}$), the CFL number increases significantly.  
Specifically, for models with $B_{y}^{0}=0.2$, $0.5$, $1.0$, and $4.0$~T the CFL numbers at $\dt_{\max}$ are approximately $1.6\times10^{5}$, $6\times10^{4}$, $1.6\times10^{4}$, and $1.4\times10^{4}$, respectively.  
Clearly, implicit time stepping is advantageous for these simulations.  

In the three-dimensional simulation, a CFL-number schedule is used to set the time stepping behavior.  
$\CFL$ is initially set to $100$ ($\dt\approx10^{-6}$~s), and linearly ramped up to $1000$ over the first $200$ time steps, at which point it is held fixed for the remainder of the simulation, corresponding to a time step just below $\dt=1\times10^{-5}$~s.

The artificial compressibility parameter is set to $\beta=10^{4}$~J~m$^{-3}$ for all simulations in this section.  
With $\rho_{0}=9486$~kg~m$^{-3}$, this corresponds to an artificial sound speed of $c_{0}=\sqrt{\beta/\rho_{0}}\approx 1$~m~s$^{-1}$.  
In the two-dimensional models, the in-plane velocity components ($x$ and $y$) remain below $2\times10^{-2}$~m~s$^{-1}$, confirming that the low-Mach number assumption is satisfied for this choice of $\beta$.  
For the fully three-dimensional model, however, the $x$- and $y$-velocity components may grow larger near the duct end caps, and the low-Mach assumption may not hold everywhere, particularly at late times.  

The GMRES tolerance for the linear solves within Newton iterations is set to $\tolL=5\times10^{-9}$.  
For the nonlinear iterations, the absolute and relative residual tolerances are set to $\atolN=5\times10^{-9}$ and $\rtolN=1\times10^{-9}$, respectively.  

\subsection{Two-Dimensional Models}
\label{sec:applications.2d}

We begin by presenting results obtained by running two-dimensional (2.5D) models.  
First, we perform a comparison with results from \citetalias{smolentsev_2025}, specifically models discussed in Section~5 of that paper.  
This serves both to verify the MHD implementation in \vertex\ in an application-relevant setting, and to verify some model assumptions made in \citetalias{smolentsev_2025}.  
Specifically, \citetalias{smolentsev_2025} solves only for $v_{z}$ and $b_{z}$ in 2D, assuming the other components are zero, while we solve for all three components of the velocity and induced magnetic field (i.e., 2.5D).  
Next, we briefly investigate the impact of spatial resolution on some quantities of interest.  
Finally, we investigate the impact of various divergence cleaning approaches for one of the models.  
All 2.5D models were performed using linear, quadrilateral elements and the SDIRK22 time integrator.  

\begin{table}[!htbp]
	\begin{center}
		\begin{tabular}{lcccccc}
			\midrule
			Model Name & $B_{y}^{0}$ [T] & Mesh & GLM & $c_{h}$ [m s$^{-1}$] & $\alpha$ [s$^{-1}$] & GP \\
			\midrule\midrule
			\texttt{By\char`_4.0\char`_M} & 4.0 & Medium & \cmark & 50.0 & $c_{h}/0.18$ & \cmark \\
			\texttt{By\char`_1.0\char`_M} & 1.0 & Medium & \cmark & 50.0 & $c_{h}/0.18$ & \cmark \\
			\texttt{By\char`_0.5\char`_M} & 0.5 & Medium & \cmark & 50.0 & $c_{h}/0.18$ & \cmark \\
			\texttt{By\char`_0.2\char`_M} & 0.2 & Medium & \cmark & 50.0 & $c_{h}/0.18$ & \cmark \\
			\midrule
			\texttt{By\char`_4.0\char`_C} & 4.0 & Coarse & \cmark & 50.0 & $c_{h}/0.18$ & \cmark \\
			\texttt{By\char`_4.0\char`_F} & 4.0 & Fine & \cmark & 50.0 & $c_{h}/0.18$ & \cmark \\
			\midrule
			\texttt{By\char`_4.0\char`_F\char`_slow} & 4.0 & Fine & \cmark & 1 & $c_{h}/0.18$ & \cmark \\
			\texttt{By\char`_4.0\char`_F\char`_fast} & 4.0 & Fine & \cmark & 250 & $c_{h}/0.18$ & \cmark \\
			\texttt{By\char`_4.0\char`_F\char`_no\char`_damp} & 4.0 & Fine & \cmark & 50 & $0.0$ & \cmark \\
			\texttt{By\char`_4.0\char`_F\char`_GLM} & 4.0 & Fine & \cmark & 50 & $c_{h}/0.18$ & \xmark \\
			\texttt{By\char`_4.0\char`_F\char`_GP} & 4.0 & Fine & \xmark & N/A & N/A & \cmark \\
			\texttt{By\char`_4.0\char`_F\char`_no\char`_clean} & 4.0 & Fine & \xmark & N/A & N/A & \xmark \\
			\midrule\midrule
		\end{tabular}
		\caption{Overview of 2.5D models}
		\label{tab:2.5D_models}
	\end{center}
\end{table}

Table~\ref{tab:2.5D_models} provides an overview of the 2.5D models discussed in this section.  
The first four models, computed with the medium mesh, differ in the external toroidal magnetic field $B_{y}^{0}$.  
These models include full GLM divergence cleaning and the GP sources (indicated by the \cmark\ in the fourth and seventh column).  
The hyperbolic divergence cleaning speed is $c_{h}=50$~m~s$^{-1}$ and the damping parameter is set to the fiducial value $\alpha=c_{h}/0.18$.  
The next two models, \texttt{By\char`_4.0\char`_C} and \texttt{By\char`_4.0\char`_F}, are identical to model \texttt{By\char`_4.0\char`_M}, but are computed using the coarse and fine mesh, respectively.  
The last six models were also computed with the same physical parameters as model $\texttt{By\char`_4.0\char`_M}$, but differ in the approach to divergence cleaning.  
To minimize any finite spatial resolution effects, these models were all run with the fine mesh.  
Models \texttt{By\char`_4.0\char`_F\char`_slow} ($c_{h}=1$), \texttt{By\char`_4.0\char`_F\char`_fast} ($c_{h}=250$), and \texttt{By\char`_4.0\char`_F\char`_no\char`_damp} ($\alpha=0$) include GLM divergence cleaning and GP sources.  
Model \texttt{By\char`_4.0\char`_F\char`_GLM} includes only GLM divergence cleaning, while model \texttt{By\char`_4.0\char`_F\char`_GP} only includes the GP sources.  
Model \texttt{By\char`_4.0\char`_F\char`_no\char`_clean} does not include any mechanisms for divergence control.  

\begin{figure}[ht!]
	\centering
	\includegraphics[width=0.95\linewidth]{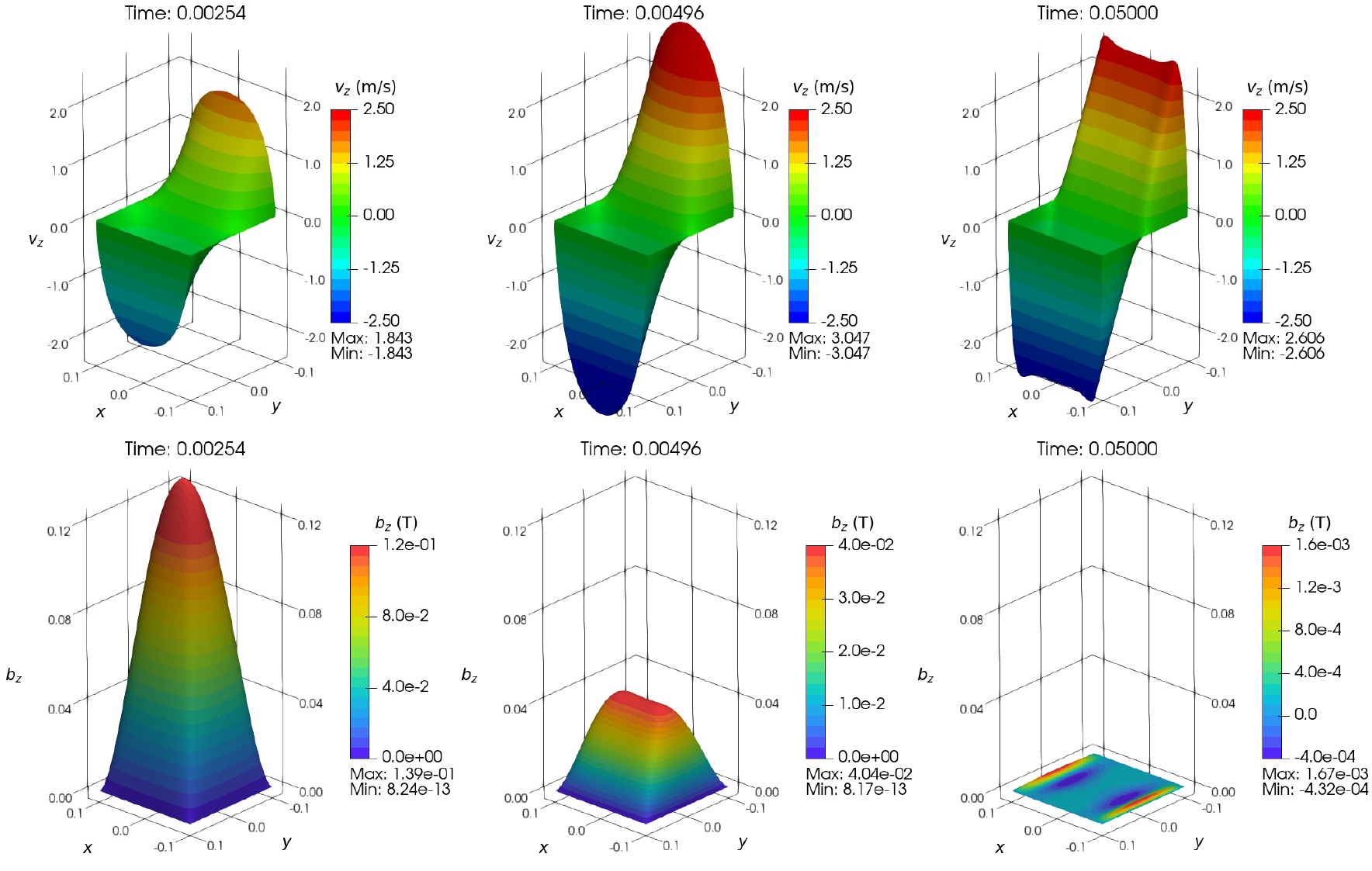}
	\caption{Surface plots illustrating the evolution of the $z$-component of velocity (top row) and induced magnetic field (bottom row) for model \texttt{By\char`_4.0\char`_F}.  Snapshots are shown $t\approx 2.5\times10^{-3}$~s (left panels), $t\approx 5\times10^{-3}$~s (middle panels), and $t=5\times10^{-2}$~s (right panels).}
	\label{fig:dbdt_By_4.0_F_composite}
\end{figure}

\subsubsection{Comparison with S25}

Figure~\ref{fig:dbdt_By_4.0_F_composite} illustrates the time evolution of $v_{z}$ (top panels) and $b_{z}$ (bottom panels) for model \texttt{By\char`_4.0\char`_F}, providing an overview of the dynamics driven by the decaying external field, $B_{z}^{0}$.  
As evident from the last term on the right-hand side of Eq.~\eqref{eq:induction_equation_induced_magnetic_field}, the decrease in $B_{z}^{0}$ leads to an increase in $b_{z}$.  
Due to the insulating wall boundary conditions, $b_{z}$ remains zero along the duct walls ($\Gamma_{xy}$), resulting in the induced magnetic field distribution shown in the lower left panel at $t\approx2.5$~ms.  
The associated current density $\bj=\nabla\times\bb/\mu_{0}$ forms closed loops in the $xy$-plane (see Figure~\ref{fig:dbdt_By_4.0_3D_composite} below), producing a Lorentz force $\bj\times\bB^{0}$ that drives upward flow near the inner $y$-boundary and a downward flow near the outer $y$-boundary, as seen in the top panels of Figure~\ref{fig:dbdt_By_4.0_F_composite}.  
Following an initial transient growth phase---peaking at approximately $0.14$~T around $2.5$~ms---the induced magnetic field begins to decay, falling below $2$~mT by the end of the simulation at $t=0.05$~s.  
The duct flow reaches a peak velocity of about $3$~m~s$^{-1}$ near $5$~ms and subsequently relaxes into a quasi-steady state as the Lorentz force diminishes.  
The middle and right top panels of Figure~\ref{fig:dbdt_By_4.0_F_composite} correspond closely with the upper and lower right panels of Figure~10 in \citetalias{smolentsev_2025}, showing qualitatively good agreement.  

\begin{figure}[ht!]
	\begin{center}
		\includegraphics[width=0.475\linewidth]{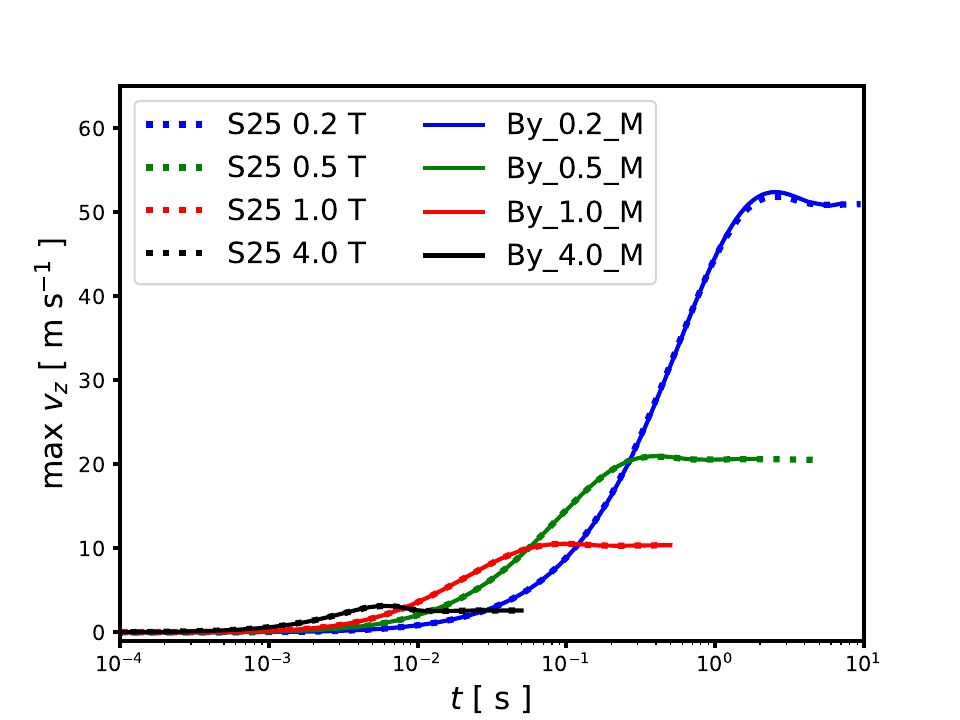}
		\hspace{12pt}
		\includegraphics[width=0.475\linewidth]{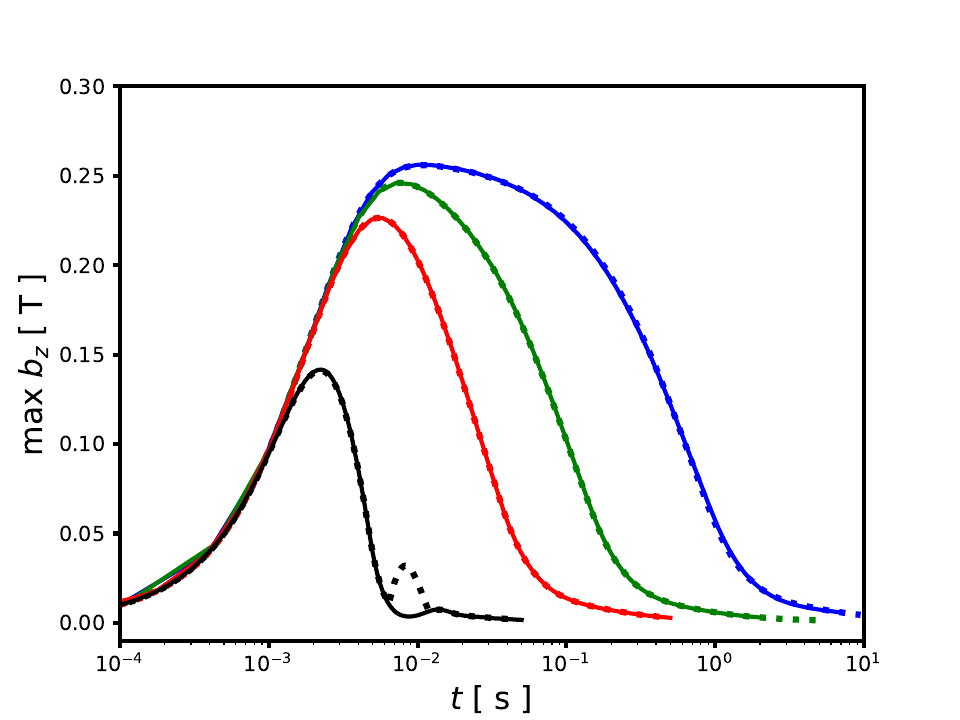}
		\caption{Plot of maximum $z$-component of velocity (left panel) and induced magnetic field (right panel) versus time for models \texttt{By\char`_0.2\char`_M} (solid blue), \texttt{By\char`_0.5\char`_M} (solid green), \texttt{By\char`_1.0\char`_M} (solid red), and \texttt{By\char`_4.0\char`_M} (solid black).  In each panel, the corresponding results from \citetalias{smolentsev_2025} are plotted with dotted lines.}
		\label{fig:dbdt_compare_S25}
	\end{center}
\end{figure}

\begin{figure}[ht!]
	\centering
	\includegraphics[width=0.9\linewidth]{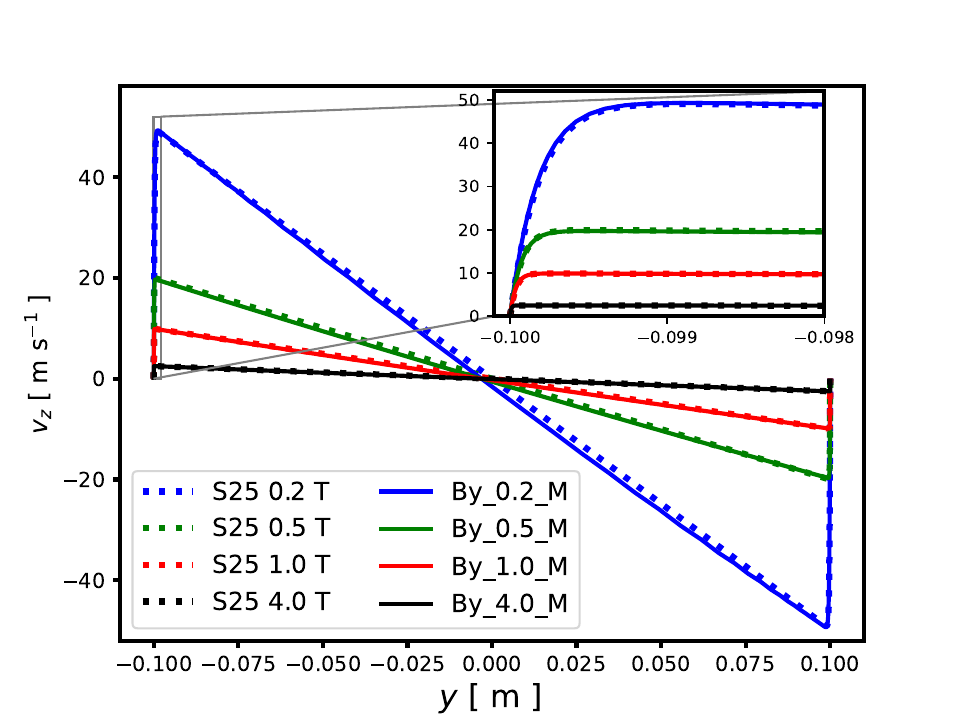}
	\caption{Plot of $v_{z}$ versus $y$ for $x=0$ at the final time for models \texttt{By\char`_0.2\char`_M} (solid blue; $t=8$~s), \texttt{By\char`_0.5\char`_M} (solid green; $t=2$~s), \texttt{By\char`_1.0\char`_M} (solid red; $t=0.5$~s), and \texttt{By\char`_4.0\char`_M} (solid black; $t=0.05$~s).  Corresponding results from \citetalias{smolentsev_2025} are plotted with dotted lines.  The inset shows the velocity profiles near the boundary layer for $y\in[-0.1,-0.098]$.}
	\label{fig:dbdt_compare_profiles_S25}
\end{figure}

Figures~\ref{fig:dbdt_compare_S25} and \ref{fig:dbdt_compare_profiles_S25} present a more quantitative comparison with the results from \citetalias{smolentsev_2025}.  
The left panel of Figure~\ref{fig:dbdt_compare_S25} shows excellent qualitative agreement in the time evolution of the maximum $v_{z}$ across a range of model parameter values $B_{y}^{0}\in\{0.2,0.5,1.0,4.0\}$~T.  
In general, both the duration of the transient phase and the peak and terminal velocities increase as $B_{y}^{0}$ decreases.  
This trend is well correlated with the general effect of the applied magnetic field to suppress convective flows due to Joule dissipation associated with Ohmic losses in the conducting fluid. 
Similarly, the right panel of Figure~\ref{fig:dbdt_compare_S25} demonstrates good agreement with \citetalias{smolentsev_2025} in the evolution of the maximum induced magnetic field $b_{z}$.  
A brief deviation from \citetalias{smolentsev_2025} is observed for the $B_{y}^{0}=4.0$~T model around $t=0.01$~s, though this behavior is not present in the other cases.  
The appearance of the local peak is explained in \citetalias{smolentsev_2025} by changes in the location of maximum velocity with time from the corner to the center line of the duct.  
This short-duration discrepancy with the current computations can indicate a deficiency of the 2D model compared to higher dimensionality models in this study.  
Figure~\ref{fig:dbdt_compare_profiles_S25} compares the $z$-component velocity profiles computed using \vertex\ (solid lines) with those from \citetalias{smolentsev_2025} (dotted lines) for the same models as in Figure~\ref{fig:dbdt_compare_S25}.  
The \vertex\ profiles are sampled at the final simulation time for each model.  
The qualitative agreement is again very good, particularly in the peak velocities near the duct walls and the thickness of the boundary (Hartmann) layers (see inset).  
For model \texttt{By\char`_0.2\char`_M}, the peak velocity reaches approximately $49.3$~m~s$^{-1}$, located around $10^{-3}$~m from the duct wall.  
For model \texttt{By\char`_4.0\char`_M}, the peak velocity is around $2.5$~m~s$^{-1}$, occurring approximately $6\times10^{-5}$~m from the wall.  
Recall that, for the medium mesh, the spacing closest to the wall is $2.35\times10^{-6}$~m.  

\begin{figure}[ht!]
	\centering
	\includegraphics[width=0.85\linewidth]{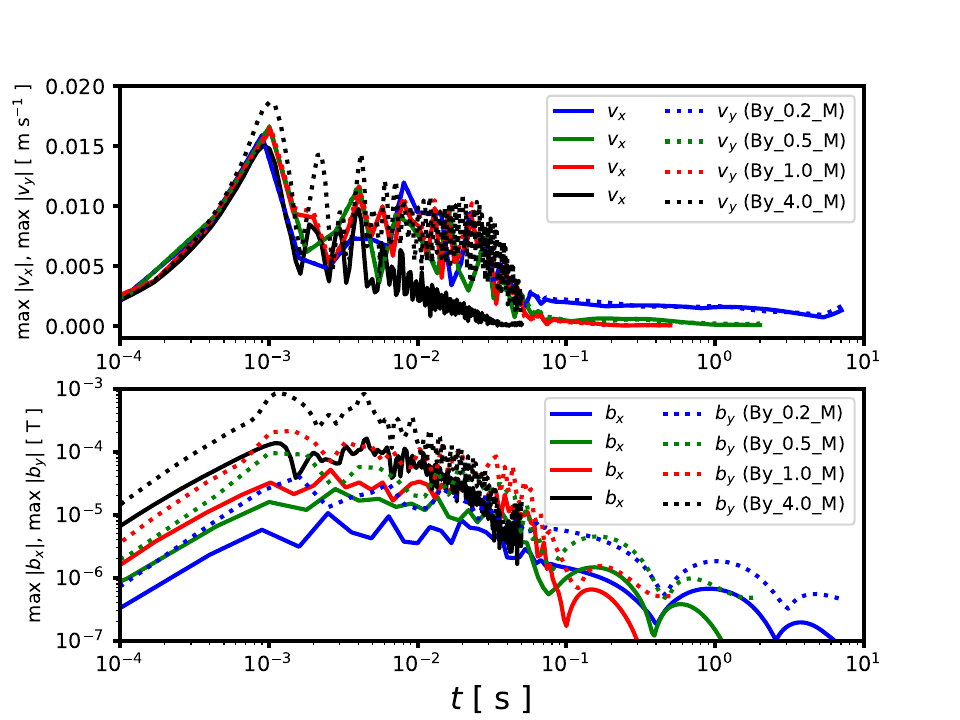}
	\caption{Plot of maximum in-plane (i.e., $x$ and $y$) components of velocity (upper panel) and induced magnetic field (lower panel) versus time for models \texttt{By\char`_0.2\char`_M} (blue), \texttt{By\char`_0.5\char`_M} (green), \texttt{By\char`_1.0\char`_M} (red), and \texttt{By\char`_4.0\char`_M} (black).  In each panel, $x$-components are plotted with solid lines while $y$-components are plotted with dotted lines.}
	\label{fig:dbdt_maxVxy_maxBxy}
\end{figure}

As shown in the top panel of Figure~\ref{fig:dbdt_maxVxy_maxBxy}, the $x$- and $y$-components of the velocity remain small throughout all simulations.  
These components reach peak amplitudes of approximately $0.015-0.02$~m~s$^{-1}$ around $t=1$~ms and remain lower at later times, with their time evolution exhibiting mild oscillatory behavior over the interval $t\in[10^{-3},10^{-1}]$~s.  
For comparison, at $t=1$~ms, the $z$-component of the velocity is approximately $0.04$~m~s$^{-1}$ for model \texttt{By\char`_0.2\char`_M} and $0.6$~m~s$^{-1}$ for model \texttt{By\char`_4.0\char`_M}.  
Similarly, the corresponding components of the induced magnetic field (bottom panel) remain small across all simulations, with peak values not exceeding $10^{-3}$~T (notably for model \texttt{By\char`_4.0\char`_M}) during $t\in[10^{-3},10^{-2}]$~s, and a general trend of decreasing amplitudes at later times.  
In contrast, the $z$-component of the induced magnetic field, although also decaying, maintains significantly larger amplitudes---ranging from $2\times10^{-3}$~T to $6\times10^{-3}$~T by the end of the simulations.  
These findings provide support for the assumptions made in the more simplified model considered by \citetalias{smolentsev_2025}.  

\subsubsection{Spatial Resolution}

\begin{figure}[ht!]
	\begin{center}
		\includegraphics[width=0.48\linewidth]{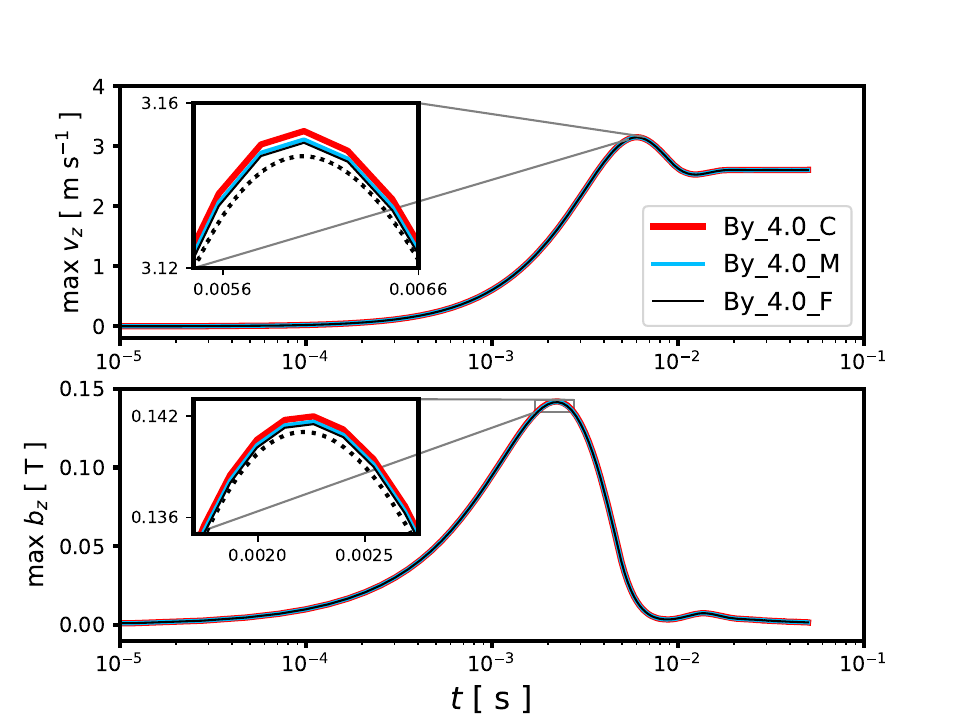}
		\hspace{8pt}
		\includegraphics[width=0.48\linewidth]{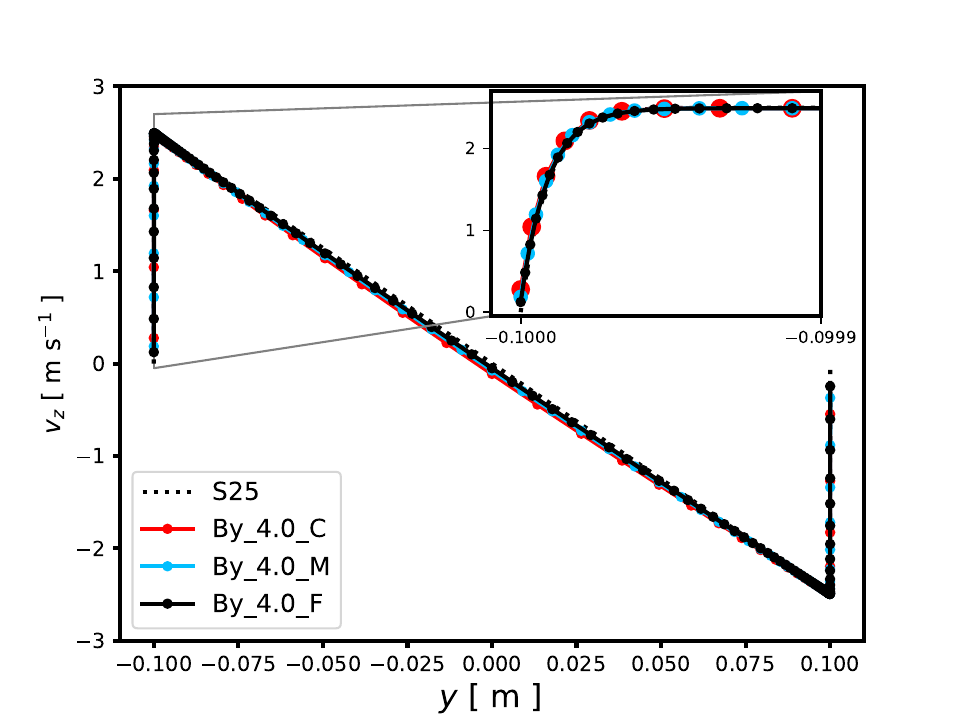}
		\caption{Plot of maximum $v_{z}$ and maximum $b_{z}$ versus time (top and bottom left panel, respectively) and $v_{z}$ versus $y$ for $x=0$ at $t=0.05$~s (right panel) for models \texttt{By\char`_4.0\char`_C} (red), \texttt{By\char`_4.0\char`_M} (blue), and \texttt{By\char`_4.0\char`_F} (black).  The insets in the left panels also include the corresponding result from \citetalias{smolentsev_2025} (dotted black).}
		\label{fig:dbdt_compare_resolutions}
	\end{center}
\end{figure}

In this section, we examine the impact of spatial resolution on key quantities of interest: the time evolution of the maximum $z$-component of velocity and induced magnetic field.  
Additionally, we assess the resolution of the sharp boundary layer that developes in $v_{z}$ near the duct walls ($y=\pm a$).  
These quantities are shown in Figure~\ref{fig:dbdt_compare_resolutions} for the model with $B_{y}^{0}=4.0$~T, using coarse, medium, and fine meshes (models \texttt{By\char`_4.0\char`_C}, \texttt{By\char`_4.0\char`_M}, and \texttt{By\char`_4.0\char`_F}, respectively).  
This particular model is selected because it produces the sharpest boundary layer among all cases considered.  

As shown in the left panels of Figure~\ref{fig:dbdt_compare_resolutions}, the overall time evolution of the maximum $v_{z}$ and $b_{z}$ is well captured across all mesh resolution, with the corresponding curves nearly overlapping.  
Insets in each panel zoom in on time intervals where these quantities reach their peak values and include reference results from \citetalias{smolentsev_2025}.  
These insets confirm good agreement with \citetalias{smolentsev_2025}, with a trend of improved accuracy as spatial resolution increases.  
At $t=6$~ms, the relative differences in $v_{z}$ compared to \citetalias{smolentsev_2025} are $2.0\times10^{-3}$, $1.3\times10^{-3}$, and $1.0\times10^{-3}$ for models \texttt{By\char`_4.0\char`_C}, \texttt{By\char`_4.0\char`_M}, and \texttt{By\char`_4.0\char`_F}, respectively.  
Similarly, near the peak of the induced magnetic field at $t\approx 2.26$~ms, the relative differences in $b_{z}$ are $6.7\times10^{-3}$ (coarse), $4.6\times10^{-3}$ (medium), and $3.4\times10^{-3}$ (fine).  
The right panel of Figure~\ref{fig:dbdt_compare_resolutions} shows that all the meshes are adequate for resolving the sharp boundary layer in $v_{z}$.  
The coarse mesh provides $8$ elements within the first $10^{-4}$~m from the duct wall, while the fine mesh provides approximately $18$.  
The inset reveals good agreement in the $v_{z}$ profiles across the different meshes and with the \citetalias{smolentsev_2025} results (dotted lines).  
Some deviations are observed around the domain center ($y=0$), where the mesh spacing is coarsest, but these discrepancies diminish with increasing mesh resolution.  

\subsubsection{Divergence Cleaning}

\begin{figure}[ht!]
	\centering
	\includegraphics[width=0.9\linewidth]{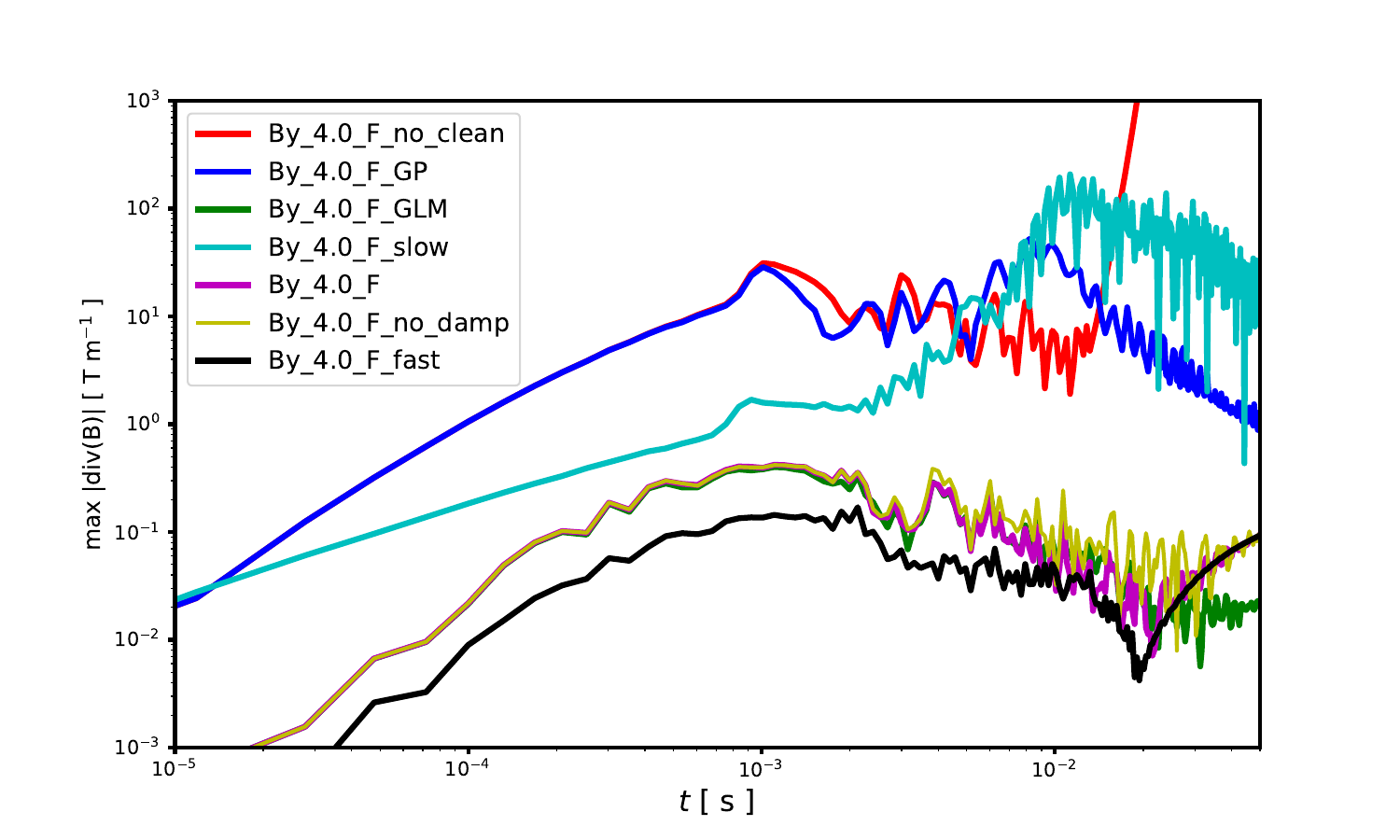}
	\caption{Plot of maximum absolute magnetic field divergence, as defined in Eq.~\eqref{eq:maxAbsDivB}, versus time for the last seven models of Table~\ref{tab:2.5D_models}: \texttt{By\char`_4.0\char`_F\char`_no\char`_clean} (red), \texttt{By\char`_4.0\char`_F\char`_GP} (blue), \texttt{By\char`_4.0\char`_F\char`_GLM} (green), \texttt{By\char`_4.0\char`_F\char`_slow} (cyan), \texttt{By\char`_4.0\char`_F} (magenta), \texttt{By\char`_4.0\char`_F\char`_no\char`_damp} (yellow), and \texttt{By\char`_4.0\char`_F\char`_fast} (black).}
	\label{fig:dbdt_divergence_cleaning}
\end{figure}

In this section, we examine the influence of various divergence control mechanisms on the evolution of the maximum absolute divergence of the magnetic field for the model with external toroidal magnetic field $B_{y}^{0}=4.0$~T.  
Figure~\ref{fig:dbdt_divergence_cleaning} presents this quantity for the last seven models listed in Table~\ref{tab:2.5D_models}.  
Notably, the model without explicit divergence control (\texttt{By\char`_4.0\char`_F\char`_no\char`_clean}) becomes unstable and diverges at approximately $t=0.021$~s---well before the intended end time of $0.05$~s (see solid red line in Figure~\ref{fig:dbdt_divergence_cleaning}).  
In contrast, all the other models successfully reach the final simulation time and exhibit consistent behavior.  
Over the interval $t\in[0,0.05]$~s, the relative deviations in the maximum $v_{z}$ and maximum $b_{z}$ from the corresponding average values across all stable models remain below $10^{-3}$ and $5\times10^{-3}$, respectively.  

Figure~\ref{fig:dbdt_divergence_cleaning} shows that the model using only GP sources for divergence control (\texttt{By\char`_4.0\char`_F\char`_GP}) closely follows the unstable model (\texttt{By\char`_4.0\char`_F\char`_no\char`_clean}) up to approximately $t=10^{-3}$~s, at which point the maximum magnetic field divergence reaches several tens of T~m$^{-1}$.  
Beyond this time, the divergence trajectories of the two models diverge: the no-cleaning model exhibits a brief period of reduced divergence before ultimately blowing up, while model \texttt{By\char`_4.0\char`_F\char`_GP} demonstrates a sustained overall decrease in divergence through the end of the simulation.  
Among the stable models, model \texttt{By\char`_4.0\char`_F\char`_slow}, which uses a reduced hyperbolic cleaning speed of $c_{h}=1$~m~s$^{-1}$, reaches the largest divergence values---exceeding $10^{2}$~T~m$^{-1}$ around $t=0.01$~s.  
In contrast, models employing hyperbolic divergence cleaning with $c_{h}\ge50$~m~s$^{-1}$ maintain significantly smaller divergence errors, consistently below $5\times10^{-1}$~T~m$^{-1}$, with only modest variation across models.  
Notably, model \texttt{By\char`_4.0\char`_F\char`_fast} ($c_{h}=250$~m~s$^{-1}$) shows the lowest divergence violations up to about $0.01$~s, after which its trajectory aligns with that of models \texttt{By\char`_4.0\char`_F} and \texttt{By\char`_4.0\char`_F\char`_no\char`_damp} towards the end of the simulation.  
The model with only hyperbolic divergence cleaning (\texttt{By\char`_4.0\char`_F\char`_GLM}) exhibits the smallest divergence violations overall at the final simulation time.  
These results indicate that some form of divergence control is necessary for stable evolution.  
However, among the models that incorporate divergence control, there is no significant variation in key physical quantities of interest, such as maximum $v_{z}$ and $b_{z}$.  
Models employing hyperbolic divergence cleaning with hyperbolic cleaning speeds significantly higher than typical flow speeds ($c_{h}\ge50$~m~s$^{-1}$) offer the best performance in maintaining the solenoidal condition of the induced magnetic field.  

\begin{figure}[ht!]
	\centering
	\includegraphics[width=0.9\linewidth]{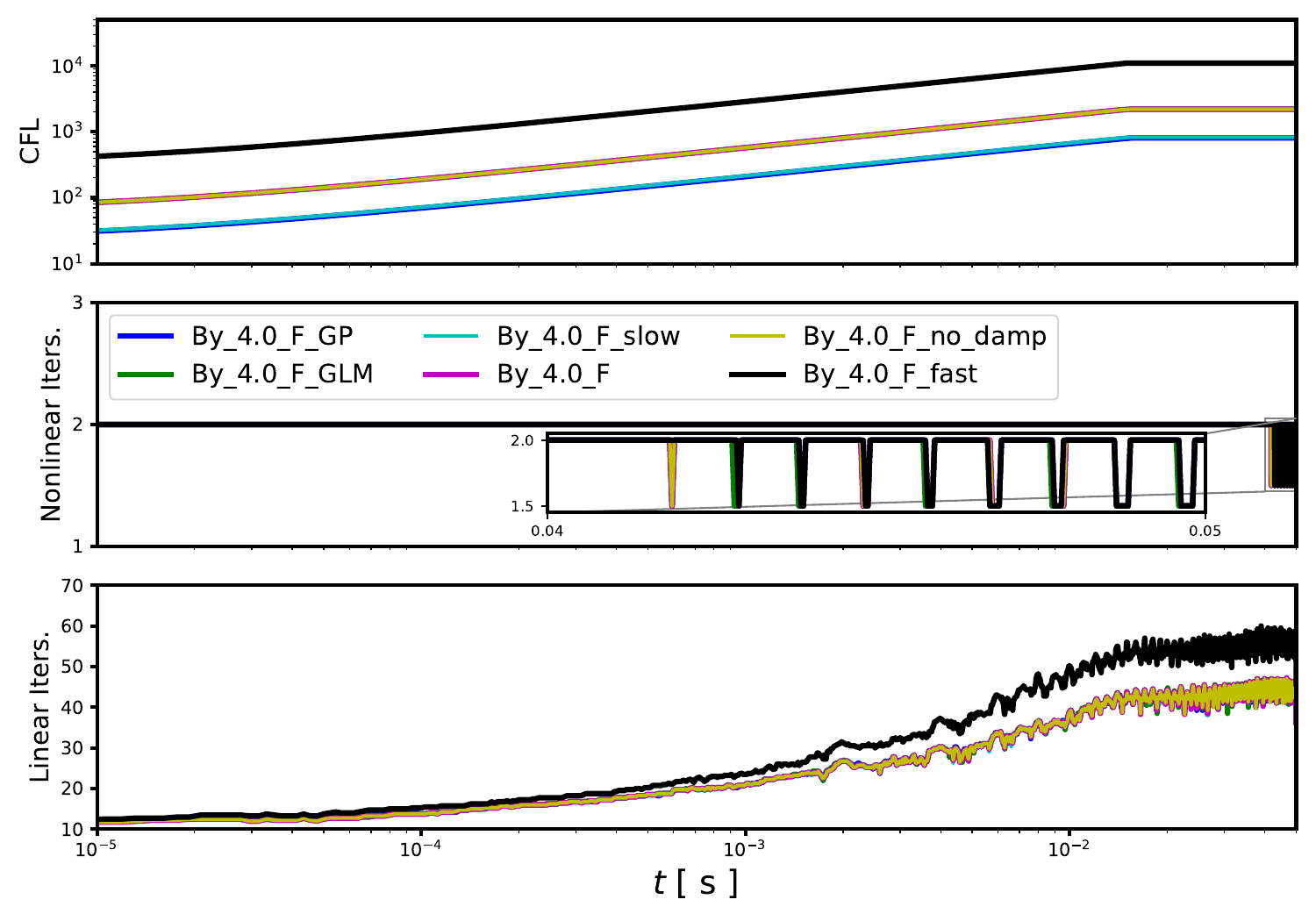}
	\caption{Plot of CFL number (top panel), average number of nonlinear iterations per SDIRK stage per time step (middle panel), and average number of linear iterations per nonlinear iteration per SDIRK stage per time step (bottom panel) versus time for the convergent models plotted in Figure~\ref{fig:dbdt_divergence_cleaning} (i.e., all but model \texttt{By\char`_4.0\char`_F\char`_no\char`_clean}).}
	\label{fig:dbdt_divergence_cleaning_solver_data}
\end{figure}

The hyperbolic cleaning speed directly influences the CFL number for a given time step and spatial resolution (see Eq.~\eqref{eq:dtCFL}), and can therefore affect the performance of the implicit solver.  
We investigate this in Figure~\ref{fig:dbdt_divergence_cleaning_solver_data}, which plots the CFL number as a function of time, along with the number of nonlinear and linear iterations for the converging models from Figure~\ref{fig:dbdt_divergence_cleaning}.  
All models follow the same time step progression, shown by the black line in the right panel of Figure~\ref{fig:dbdt_mesh_dt}: beginning with $\dt=10^{-6}$~s and increasing linearly to a maximum of $\dt_{\max}=3.5\times10^{-5}$~s at $t=0.015$~s.  
The evolution of $\CFL$ for models \texttt{By\char`_4.0\char`_F\char`_GP} and \texttt{By\char`_4.0\char`_F\char`_slow} is nearly identical, indicating that $c_{h}=1$~m~s$^{-1}$ is smaller than other characteristic wave speeds for this problem.  
For these models, $\CFL$ starts at approximately $20$ and increases to a maximum of around $800$.  
Models with $c_{h}=50$~m~s$^{-1}$ (\texttt{By\char`_4.0\char`_F\char`_GLM}, \texttt{By\char`_4.0\char`_F}, and \texttt{By\char`_4.0\char`_F\char`_no\char`_damp}) reach higher $\CFL$, beginning at about $63$ and peaking near $2200$.  
Model \texttt{By\char`_4.0\char`_F\char`_fast} ($c_{h}=250$~m~s$^{-1}$) evolves with the largest $\CFL$, staring at $314$ and reaching $\CFL=1.1\times10^{4}$ by the end of the simulation.  
Despite this wide range in $\CFL$, the number of nonlinear iterations per Runge--Kutta stage remains largely unaffected.  
On average, all models require two nonlinear iterations per stage.  
Toward the end of the simulations, some time steps require only one nonlinear iteration in the first stage, reducing the average to $1.5$ (see inset, middle panel of Figure~\ref{fig:dbdt_divergence_cleaning_solver_data}).  
In contrast, the number of linear iterations per nonlinear iteration (bottom panel) shows greater sensitivity to $\CFL$, but only the model with $c_{h}=250$~m~s$^{-1}$ stands out as requiring more iterations than the other models.  
Most models start out requiring an average of $10$-$12$ linear iterations, which increases to $40$-$45$ when the time step reaches its maximum; i.e., about a fourfold increase in linear iterations after a $35$-fold increase in $\dt$.  
Model \texttt{By\char`_4.0\char`_F\char`_fast} also starts with an average of $10$-$12$ linear iterations, but ultimately requires $50$-$60$ iterations when the time step is maximal.  

\subsection{Three-Dimensional Models}
\label{sec:applications.3d}

We conclude this section by discussing results from three-dimensional (3D) models.  
Due to the larger computational cost associated with fully 3D models---particularly from resolving the thin boundary layers and capturing the initial transient dynamics---the number of 3D models included in this study is limited.  
Our 3D models were performed using resources provided by NERSC (Perlmutter) and a dedicated cluster at ORNL.  
On Perlmutter, the simulations were executed on 6 CPU nodes, with 64 MPI tasks per node.  
Each task utilized 4 CPU cores, corresponding to a total of 384 MPI ranks and 1536 CPU cores.  
On the ORNL cluster, the simulations were run using 8 CPU nodes with 31 tasks per node and 4 CPUs per task (248 MPI ranks and 992 CPU cores).  
The primary objectives are to demonstrate the computational capability of \vertex\ and to provide a point of comparison against predictions obtained from the reduced models of \citetalias{smolentsev_2025} and the 2.5D models discussed in the previous section.  
Similar to the 2.5D models, these were performed using linear, hexahedral elements and the SDIRK22 time integrator.  

\begin{figure}[ht!]
	\centering
  \begin{subfigure}[t]{0.495\textwidth}
    \centering
	  \includegraphics[width=0.9\linewidth]{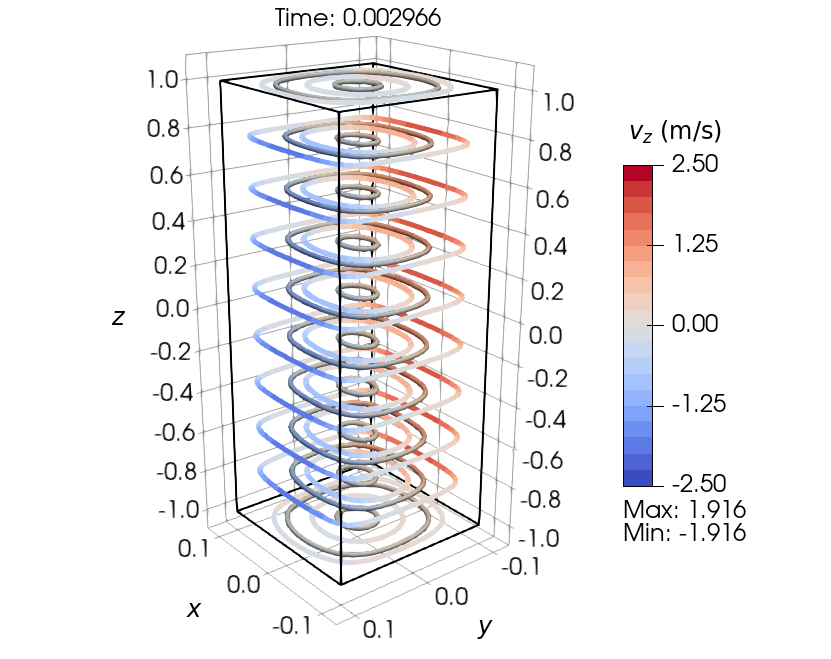}
    \caption{$\nabla\times\bb$, $t=3.0$~ms}
  \end{subfigure}
  \hfill
  \begin{subfigure}[t]{0.495\textwidth}
    \centering
	  \includegraphics[width=0.9\linewidth]{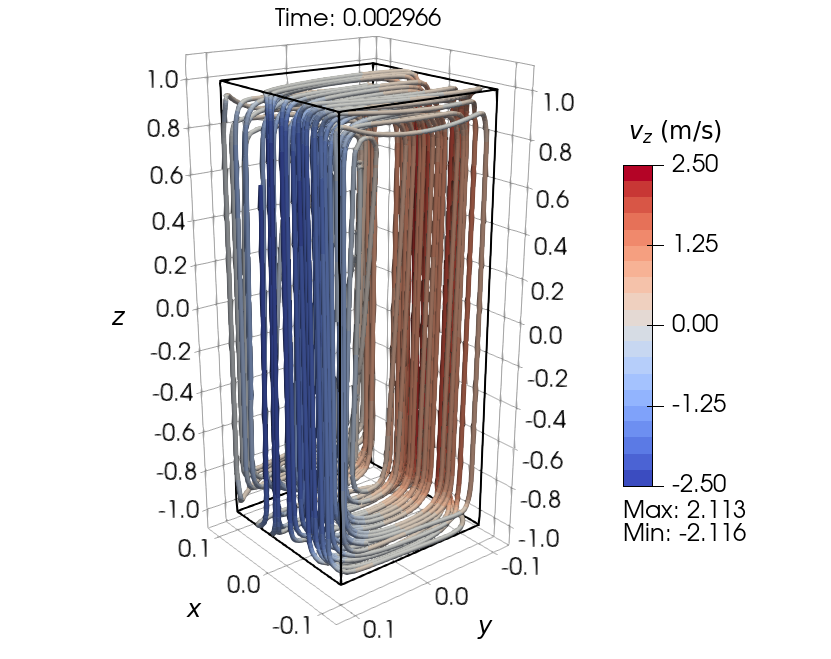}
    \caption{Velocity, $t=3.0$~ms}
  \end{subfigure}
  \\
  \vspace{2.5mm}
  \begin{subfigure}[t]{0.495\textwidth}
    \centering
	  \includegraphics[width=0.9\linewidth]{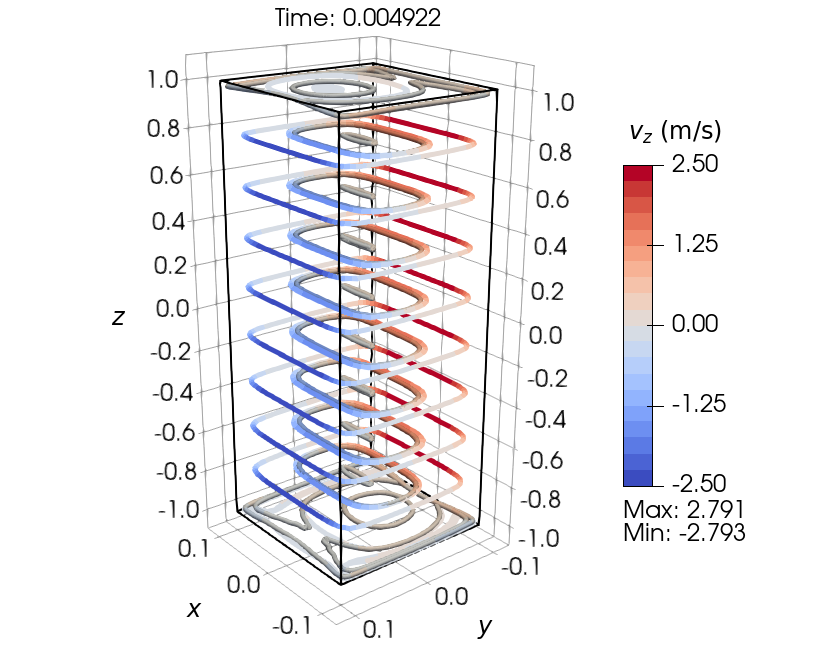}
    \caption{$\nabla\times\bb$, $t=4.9$~ms}
  \end{subfigure}
  \hfill
  \begin{subfigure}[t]{0.495\textwidth}
    \centering
	  \includegraphics[width=0.9\linewidth]{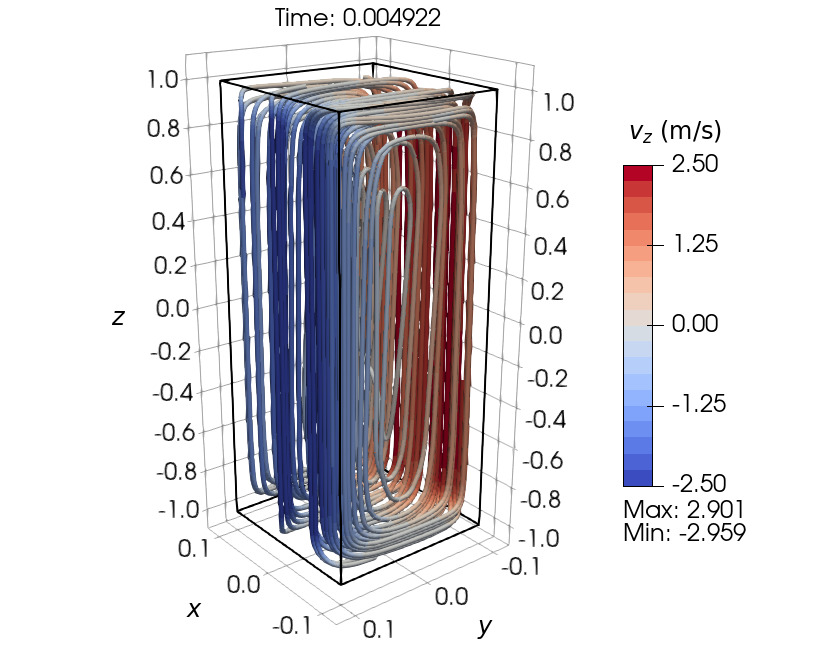}
    \caption{Velocity, $t=4.9$~ms}
  \end{subfigure}
  \\
  \vspace{2.5mm}
  \begin{subfigure}[t]{0.495\textwidth}
    \centering
	  \includegraphics[width=0.9\linewidth]{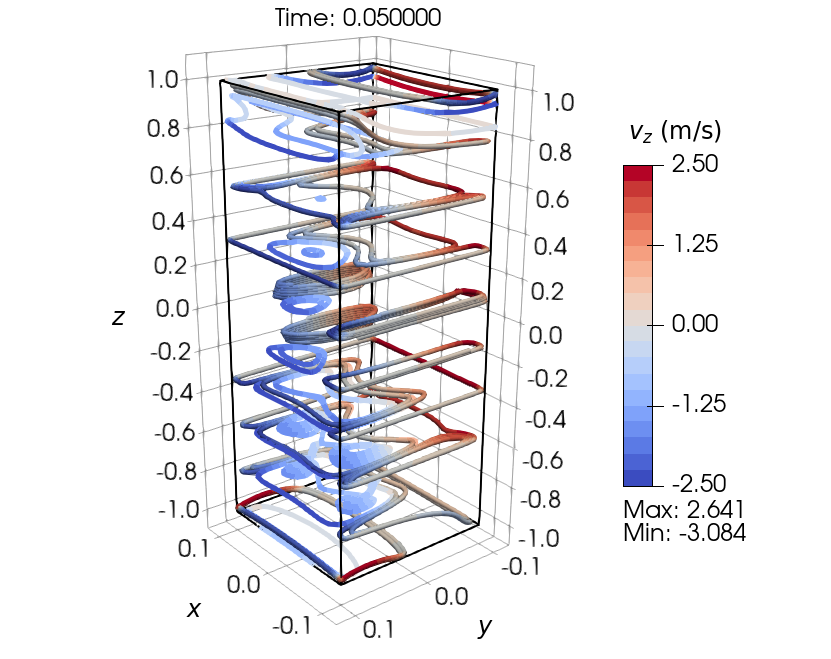}
    \caption{$\nabla\times\bb$, $t=50$~ms}
  \end{subfigure}
  \hfill
  \begin{subfigure}[t]{0.495\textwidth}
    \centering
	  \includegraphics[width=0.9\linewidth]{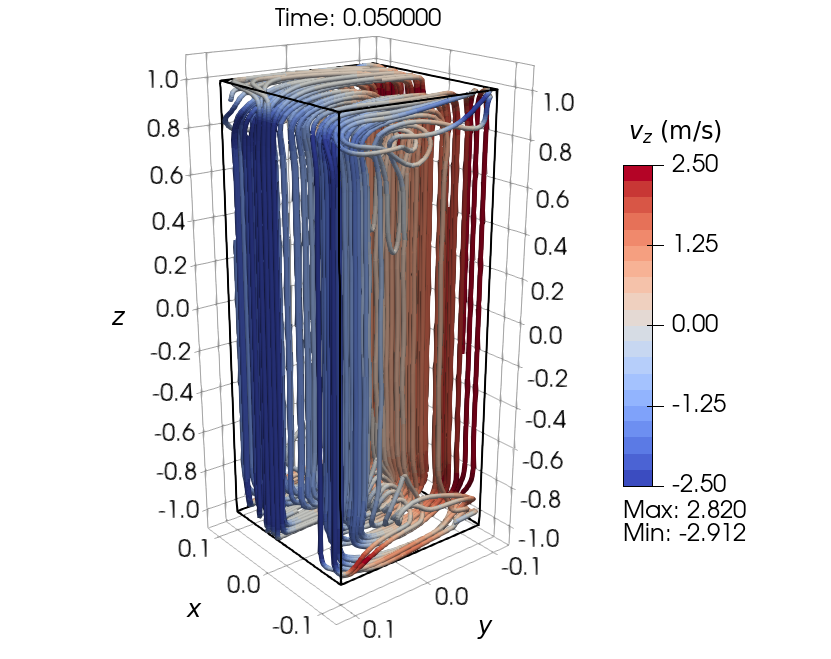}
    \caption{Velocity, $t=50$~ms}
  \end{subfigure}
  \caption{Streamlines of the curl of the induced magnetic field $\nabla\times\bb$ (left panels) and velocity (right panels) at several times for the 3D model with no-slip and insulating wall boundary conditions, and $B_{y}^{0}=4.0$~T. Note, $x$ and $y$ coordinates have been scaled by a factor of four for visibility.}
	\label{fig:dbdt_By_4.0_3D_composite}
\end{figure}

\begin{figure}[ht!]
	\begin{center}
		\includegraphics[width=0.48\linewidth]{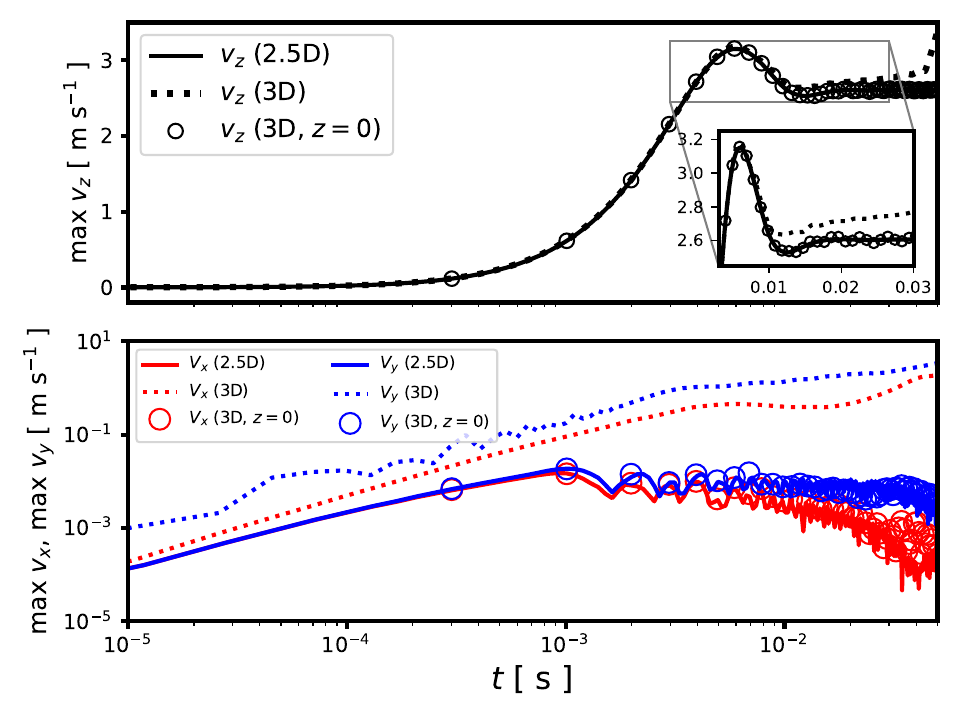}
		\hspace{8pt}
		\includegraphics[width=0.48\linewidth]{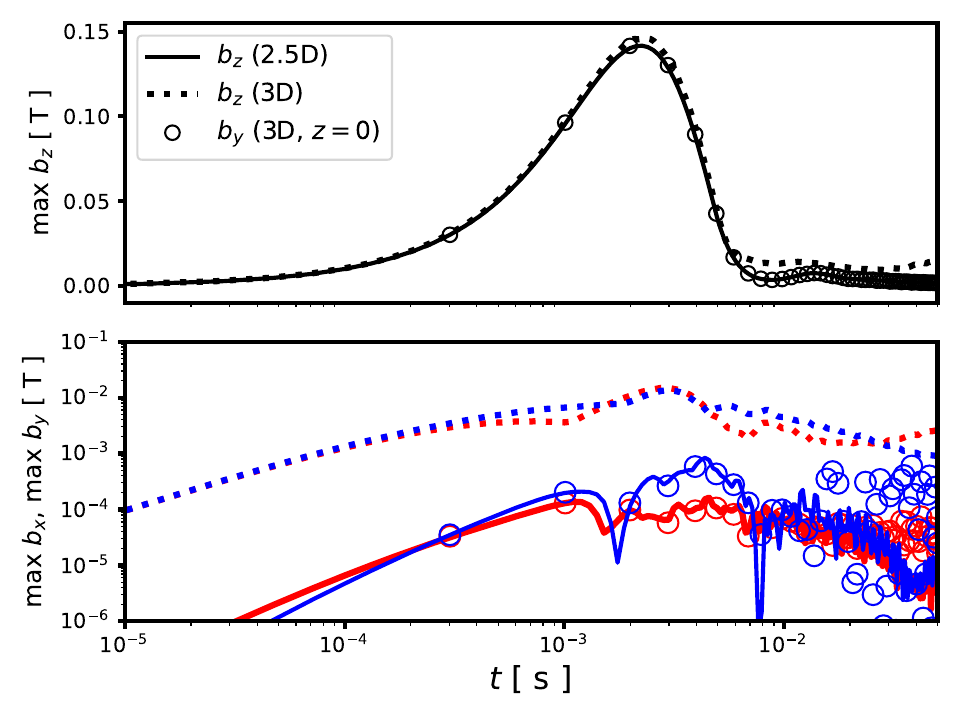}
		\caption{Plot of maximum velocity (left panels) and induced magnetic field (right panels) components versus time for the genuinely three-dimensional model from Figure~\ref{fig:dbdt_By_4.0_3D_composite}, compared with the corresponding 2.5D model \texttt{By\char`_4.0\char`_M}.  
		The upper panels plot the maximum $z$-component of velocity and induced magnetic field, where the maximum is over the full computational domain (dotted lines), in the $xy$-plane with $z=0$ (circles), and from the 2.5D model (solid lines).  
		The lower panels plot the corresponding maximum $x$- and $y$-components of the velocity (left) and induced magnetic field (right).}
		\label{fig:dbdt_compare_S25_3D}
	\end{center}
\end{figure}

Figure~\ref{fig:dbdt_By_4.0_3D_composite} highlights key aspects of the flow evolution in the three-dimensional (3D) model with no-slip, electrically insulating boundary conditions at $z=\pm1$~m.  
(We note that the genuinely 3D model with periodic boundary conditions in the $z$-dimension reproduces the results of the corresponding 2.5D model uniformly across all $z$, and will therefore not be discussed further.)  
The left panels of Figure~\ref{fig:dbdt_By_4.0_3D_composite} shows streamlines of the current density associated with the induced magnetic field, specifically $\mu_{0}\bj=\nabla\times\bb$, at various times during the evolution.  
Specifically, the middle left panel shows the current density at $t\approx4.9$~ms, shortly before the vertical velocity component $v_{z}$ reaches its peak (see Figure~\ref{fig:dbdt_compare_S25_3D}).  
The current circulates in closed counterclockwise loops within the $xy$-plane, remaining relative uniform along the $z$-direction, except near the top and bottom boundaries where 3D effects become significant.  
The interaction of this induced current with the external magnetic field, particularly the toroidal component $B_{y}^{0}$, produces the Lorentz force $\bj\times\bB^{0}$ that drives flow in the vertical direction.  
The streamlines are colored-coded by $v_{z}$, revealing a general upward flow for $y<0$ and downward flow for $y>0$.  
The resulting flow circulation pattern is illustrated in the right panels of Figure~\ref{fig:dbdt_By_4.0_3D_composite}.  
The lower right panel displays velocity streamlines at the final time ($t=50$~ms), after the duct flow in the midplane ($z=0$) has settled into a quasi-steady state.  
At this stage of the simulation, the $z$-component of the induced magnetic field has decayed significantly (see Figure~\ref{fig:dbdt_compare_S25_3D}), and the current no longer maintains the coherent loop structure observed at earlier times.  
As in the 2.5D models, the flow is predominately vertical, while near the top and bottom boundaries it turns laterally to complete closed-loop circulations within the duct.  

Figure~\ref{fig:dbdt_compare_S25_3D} presents the time evolution of the maximum velocity and induced magnetic field components from the 3D simulation.  
The top panels compare the evolution of the maximum $v_{z}$ and $b_{z}$ in the midplane (black circles) with corresponding results from the 2.5D model (solid black lines), showing very good agreement for all times.  
Consequently, the agreement with the solution from \citetalias{smolentsev_2025} is also very good.  

The bottom panels show the evolution of the $x$- and $y$-components of both velocity and induced magnetic field.  
Here too, the 3D midplane results (red and blue open circles) align closely with the 2.5D model predictions (red and blue solid lines).   
This agreement holds consistently for the velocity components throughout the simulation and for the magnetic field components up to $t\lesssim10$~ms.  
At later times, however, oscillatory behavior emerges, especially in $\max b_{x}$, whose amplitude becomes comparable to that of $\max b_{z}$.  
These increased fluctuations may contribute to the more disordered structure of the current density seen in the lower left panel of Figure~\ref{fig:dbdt_By_4.0_3D_composite}.  

The maximum $v_{z}$ and $b_{z}$ values computed across the entire 3D domain (dotted black lines) track the corresponding midplane values well during the early evolution, but $v_{z}$ begins to diverge after $t\approx40$~ms, when a notable upward trend is observed.  
The maximum values of the $x$- and $y$-components values of both velocity and magnetic field (dotted lines in the lower panels) exceed their midplane counterparts, consistent with the genuinely three-dimensional flow structure.  

At present, the cause of the late-time growth observed in maximum $v_{z}$ in the 3D model remains unclear.  
It may be linked to hydromagnetic instabilities in the quasi-steady flow, which warrants further investigation but lies beyond the scope of this study.  
Alternatively, the growth could be of numerical origin.  
The finite element method employed is not stabilized and may be susceptible to numerical instabilities, particularly in under-resolved regions.  
Evidence of such behavior is visible in the lower right panel in Figure~\ref{fig:dbdt_By_4.0_3D_composite}, where a significant upflow develops in the lower corned of the domain, suggesting the presence of steep gradients.  
Future work will explore this behavior using higher spatial resolution and/or stabilization techniques in the numerical method.  

\section{Summary, Conclusions, and Outlook}
\label{sec:conclusions}

We have implemented and tested a numerical method for the incompressible, full-induction MHD equations within the open-source \vertex\ framework, targeting future applications to liquid metal flows in fusion blankets, including transient scenarios driven by plasma disruptions.  
The solenoidal constraints on the velocity and magnetic field are relaxed using the artificial compressibility method \cite{chorin_1967} and the generalized Lagrange multiplier formulation \cite{dedner_etal_2002}, respectively.  
The AC-GLM-MHD model is further augmented with the Godunov--Powell sources, which enhance numerical orthogonality between the Lorentz force and the magnetic field in the presence of divergence errors \cite{brackbillBarnes_1980}, and transports these errors with the flow \cite{powell_etal_1999}.  
The GP sources also symmetrize the MHD equations \cite{godunov_1972}, though this aspect has not been analyzed here in the context of the AC-GLM-MHD model.  
The governing hyperbolic-parabolic PDEs are discretized in space using the FEM, yielding a system of ODEs in time that is advanced with SDIRK methods.  
The resulting nonlinear system is solved with Newton's method, using a preconditioned iterative solver for the associated linear systems.  
The \vertex\ implementation leverages Trilinos packages to enable a portable, HPC-ready solver framework, and has been verified against several benchmarks with known (analytical or numerical) solutions, where we focused on solver accuracy and efficacy of divergence control mechanisms.  
We also applied the full-induction MHD solver to an idealized blanket model with a decaying external magnetic field in 2.5D and full 3D, and performed a detailed comparison with the recent quasi-2D simulations by Smolentsev~\cite{smolentsev_2025}.  

Benchmark simulations with analytic solutions demonstrate that the implementation exhibits convergent behavior.  
Using linear elements with the SDIRK22 time integrator, we verified second-order accuracy in space and time for both the magnetic advection-diffusion test (Section~\ref{sec:magnetic_advection_diffusion}) and the circularly polarized Alfv{\'e}n wave (Section~\ref{sec:cpAlfvenWave}).  
With quadratic elements and the SDIRK54 integrator for the advection-diffusion test at $\eta=(2\pi)^{-2}$, third-order accuracy for all magnetic field components was achieved only with $\nabla\cdot\bB$-terms in the MHD equations disabled; otherwise, the $x$- and $y$-components showed second-order accuracy, while the $z$-component retained third-order accuracy.  
This indicates that relaxing the divergence-free constraint can impact asymptotic convergence.  
In the advection limit ($\eta=0$), all components converged at second order, with $B_{z}$ errors largely independent of $\nabla\cdot\bB$-terms, and $B_{x}$ and $B_{y}$ errors reduced with these terms included.  
For the circularly polarized Alfv{\'e}n wave, quadratic elements and SDIRK54 also yielded sub-optimal second-order convergence.  
We suspect the observed order reduction in both the advection-diffusion (with $\eta=0$) and Alfv{\'e}n wave problems is due in part to the lack of dissipation, a behavior previously noted in analyses of wave-propagation methods (e.g., \cite{cheng_etal_2017,garcia-archilla_etal_2021}).  

The divergence cleaning test (Section~\ref{sec:divergenceCleaningTest}) shows good qualitative agreement with previously published results (e.g., \cite{dedner_etal_2002,TRICCO20127214}).  
While both the GP source term and undamped GLM cleaning disperse divergence errors, adding damping reduces divergence errors markedly.  
With implicit time stepping, stable evolution is achieved with $c_{h}=10^{3}$ without time step reduction, and the damping factor of $\alpha=c_{h}/0.18$ recommended by \cite{dedner_etal_2002} is confirmed to be effective.  

For the lid-driven cavity, which lacks an analytic solution, we compared \vertex\ results with those published by Fambri et al. \cite{fambri_etal_2023}, obtained using their well-balanced, exactly divergence-free scheme.  
These comparisons are comprehensive in that they probe both incompressibility and divergence-free properties in a viscous, resistive setting.  
Simulations with vertical and horizontal initial fields, across a range of field strengths, show good qualitative agreement with the steady-state structures of $|\bv|$ and $|\bB|$ published in \cite{fambri_etal_2023}.  
Direct comparisons of solution profiles further confirm close quantitative agreement, supporting the conclusion that the \vertex\ implementation of the AC-GLM-MHD model is consistent and robust.  

Application of the full-induction AC-GLM-MHD model in 2.5D and 3D \vertex\ simulations of the idealized blanket configuration proposed by \citetalias{smolentsev_2025} (Section~\ref{sec:applications}) leads to two main conclusions: ({\it i}) the \vertex\ implementation is capable of handling transient MHD flows that develop thin Hartmann boundary layers with $\Ha\sim\cO(10^{4})$, a key prerequisite for modeling disruption-driven events in more realistic blanket geometries; and ({\it ii}) the quasi-2D modeling assumption --- evolving only $v_{z}$ and $b_{z}$ --- remains a reasonable approximation, with our results providing further evidence in support of \citetalias{smolentsev_2025}.  
Across a range of external toroidal field strengths, we found excellent agreement in the time evolution of maximum $v_{z}$ and $b_{z}$, as well as in the quasi-steady $v_{z}$ profiles, particularly in the structure of the Hartmann layers.  

The successful implementation and verification of the full-induction model in \vertex\ provides a foundation for extending its capabilities and performance.  
Continuing development will focus on incorporating multi-material domain capabilities and far-field or pseudo-vacuum boundary conditions to enable studies of the coupled behavior of liquid metals, structural materials, and the plasma.  
Realistic modeling of liquid metal flows driven by plasma disruptions will ultimately require coupling to plasma MHD, e.g., as provided by M3D-C1 \cite{breslau_etal_2009,ferraroJardin_2009}.  
We also plan a systematic study of solver scalability on HPC systems, including both CPU-only and GPU-accelerated platforms, to identify any existing performance bottlenecks and guide improvements to enhance efficiency at application-relevant resolutions.
From a numerical perspective, future work will include incorporating finite-element stabilization to improve robustness in under-resolved regions, e.g., via the streamline-upwind Petrov--Galerkin, variational multiscale, or entropy viscosity method \cite{hughesMallet_1986,hughes_etal_1998,kirkCarey_2008,guermond_etal_2011}.  
Supporting this, we plan to conduct a continuous entropy analysis and develop suitable viscous regularization, e.g., as in \cite{guermondPopov_2014,dao_etal_2024}, for the AC-GLM-MHD system.  
In parallel, we will design preconditioners tailored to the system's characteristic coupling and stiffness properties.  
These developments will strengthen \vertex\ as a versatile platform for modeling transient MHD phenomena to support fusion blanket designs.

\section*{Acknowledgements}
\noindent
Research at Oak Ridge National Laboratory is supported under contract DE-AC05-00OR22725 from the U.S. Department of Energy (DOE) to UT-Battelle, LLC.
This work was performed under the SciDAC project “Center for Simulation of Plasma-Liquid Metal Interactions in Plasma Facing Components and Breeding Blankets of a Fusion Power Reactor” under contract DE-AC05-00OR22725 with the US DOE.  
This research used resources of the National Energy Research Scientific Computing Center (NERSC), a DOE User Facility using NERSC award FES-ERCAP~32482.

\section*{Declaration of generative AI and AI-assisted technologies in the writing process}
\noindent
During the preparation of this manuscript the authors used ChatGPT to provide editing suggestions for several sentences. 
After using this tool, the authors reviewed and edited the content as needed and take full responsibility for the content of the published article.

\bibliographystyle{elsarticle-num}
\bibliography{references}

@BOOK{kwakKiris_2013,
title="{Computation of Viscous Incompressible Flows}",
author={{Kwak}, D. and {Kiris}, C.C.},
series={Scientific Computation},
publisher={Springer Dordrecht},
year={2013}}

@BOOK{hairerWanner_1996,
title="{Solving Ordinary Differential Eqiations II.  Stiff and Differential-Algebraic Problems.}",
author={{Hairer}, E. and {Wanner}, G.},
series={Springer Series in Computational Mathematics},
publisher={Springer Berlin, Heidelberg},
year={1996}}

@article{abdou_etal_2015,
title = {Blanket/first wall challenges and required R\&D on the pathway to DEMO},
journal = {Fusion Engineering and Design},
volume = {100},
pages = {2-43},
year = {2015},
doi = {https://doi.org/10.1016/j.fusengdes.2015.07.021},
author = {Mohamed {Abdou} and Neil B. {Morley} and Sergey {Smolentsev} and Alice {Ying} and Siegfried {Malang} and Arthur {Rowcliffe} and Mike {Ulrickson}}}

@article{boozer_2012,
    author = {Boozer, Allen H.},
    title = "{Theory of tokamak disruptions)}",
    journal = {Physics of Plasmas},
    volume = {19},
    pages = {058101},
    year = {2012},
    doi = {10.1063/1.3703327}}

@article{brackbillBarnes_1980,
title = "{The Effect of Nonzero $\nabla\cdot \bB$ on the Numerical Solution of the Magnetohydrodynamic Equations}",
journal = {Journal of Computational Physics},
volume = {35},
number = {3},
pages = {426-430},
year = {1980},
issn = {0021-9991},
doi = {https://doi.org/10.1016/0021-9991(80)90079-0},
author = {J.U Brackbill and D.C Barnes}}

@article{breslau_etal_2009,
    author = {Breslau, J. and Ferraro, N. and Jardin, S.},
    title = "{Some properties of the M3D-C1 form of the three-dimensional magnetohydrodynamics equations}",
    journal = {Physics of Plasmas},
    volume = {16},
    number = {9},
    pages = {092503},
    year = {2009},
    month = {09},
    issn = {1070-664X},
    doi = {10.1063/1.3224035}}

@article{cheng_etal_2017,
  title="{$L^2$ stable discontinuous Galerkin methods for one-dimensional two-way wave equations}",
  author={{Cheng}, Yingda and {Chou}, Ching-Shan and {Li}, Fengyan and {Xing}, Yulong},
  journal={Mathematics of Computation},
  volume={86},
  number={303},
  pages={121--155},
  year={2017}}

@article{chorin_1967,
title = "{A numerical method for solving incompressible viscous flow problems}",
journal = {Journal of Computational Physics},
volume = {2},
number = {1},
pages = {12-26},
year = {1967},
doi = {https://doi.org/10.1016/0021-9991(67)90037-X},
author = {Alexandre Joel {Chorin}}}

@article{clausen_2013,
  title = "{Entropically damped form of artificial compressibility for explicit simulation of incompressible flow}",
  author = {Clausen, Jonathan R.},
  journal = {Phys. Rev. E},
  volume = {87},
  issue = {1},
  pages = {013309},
  numpages = {12},
  year = {2013},
  month = {Jan},
  publisher = {American Physical Society},
  doi = {10.1103/PhysRevE.87.013309}}

@article{cyr_etal_2016,
author = {Cyr, Eric C. and Shadid, John N. and Tuminaro, Raymond S.},
title = {Teko: A Block Preconditioning Capability with Concrete Example Applications in Navier--Stokes and MHD},
journal = {SIAM Journal on Scientific Computing},
volume = {38},
number = {5},
pages = {S307-S331},
year = {2016},
doi = {10.1137/15M1017946}}

@article{dao_etal_2024,
author = {{Dao}, Tuan Anh and {Lundgren}, Lukas and {Nazarov}, Murtazo},
title = {Viscous Regularization of the MHD Equations},
journal = {SIAM Journal on Applied Mathematics},
volume = {84},
number = {4},
pages = {1439-1459},
year = {2024},
doi = {10.1137/23M1564274}}

@ARTICLE{dedner_etal_2002,
       author = {{Dedner}, A. and {Kemm}, F. and {Kr{\"o}ner}, D. and {Munz}, C. -D. and {Schnitzer}, T. and {Wesenberg}, M.},
        title = "{Hyperbolic Divergence Cleaning for the MHD Equations}",
      journal = {Journal of Computational Physics},
         year = 2002,
       volume = {175},
       number = {2},
        pages = {645-673},
          doi = {10.1006/jcph.2001.6961}}

@ARTICLE{derigs_etal_2018,
       author = {{Derigs}, Dominik and {Winters}, Andrew R. and {Gassner}, Gregor J. and {Walch}, Stefanie and {Bohm}, Marvin},
        title = "{Ideal GLM-MHD: About the entropy consistent nine-wave magnetic field divergence diminishing ideal magnetohydrodynamics equations}",
      journal = {Journal of Computational Physics},
         year = 2018,
        month = jul,
       volume = {364},
        pages = {420-467},
          doi = {10.1016/j.jcp.2018.03.002}}

@article{donatelli_2013,
author = {{Donatelli}, Donatella},
title = {The Artificial Compressibility Approximation for MHD Equations in Unbounded Domain},
journal = {Journal of Hyperbolic Differential Equations},
volume = {10},
pages = {181-198},
year = {2013},
doi = {10.1142/S0219891613500082}}

@article{eardley-Brunt_etal_2024,
doi = {10.1088/1361-6587/ad100a},
year = {2023},
month = {dec},
publisher = {IOP Publishing},
volume = {66},
number = {1},
pages = {015015},
author = {Eardley-Brunt, R W and Dubas, A J and Davis, A},
title = {On scalable liquid-metal MHD solvers for fusion breeder blanket multiphysics applications},
journal = {Plasma Physics and Controlled Fusion}}

@article{ferraroJardin_2009,
title = {Calculations of two-fluid magnetohydrodynamic axisymmetric steady-states},
journal = {Journal of Computational Physics},
volume = {228},
number = {20},
pages = {7742-7770},
year = {2009},
doi = {https://doi.org/10.1016/j.jcp.2009.07.015},
author = {N.M. Ferraro and S.C. Jardin}}

@article{garcia-archilla_etal_2021,
title = {On the convergence order of the finite element error in the kinetic energy for high Reynolds number incompressible flows},
journal = {Computer Methods in Applied Mechanics and Engineering},
volume = {385},
pages = {114032},
year = {2021},
doi = {https://doi.org/10.1016/j.cma.2021.114032},
author = {Bosco García-Archilla and Volker John and Julia Novo}}

@article{gardinerStone_2008,
title = {An unsplit Godunov method for ideal MHD via constrained transport in three dimensions},
journal = {Journal of Computational Physics},
volume = {227},
pages = {4123-4141},
year = {2008},
doi = {https://doi.org/10.1016/j.jcp.2007.12.017},
author = {Thomas A. Gardiner and James M. Stone}}

@article{ghia_etal_1982,
title = {High-Re solutions for incompressible flow using the Navier-Stokes equations and a multigrid method},
journal = {Journal of Computational Physics},
volume = {48},
number = {3},
pages = {387-411},
year = {1982},
doi = {https://doi.org/10.1016/0021-9991(82)90058-4},
author = {U Ghia and K.N Ghia and C.T Shin}}

@article{godunov_1972,
author = {{Godunov}, S.K.},
title = "{Symmetric form of the equations of magnetohydrodynamics}",
journal = {Numer. Methods Mech. Contin. Medium},
year = {1972},
volume = {1},
pages = {26-34}}

@article{guermond_etal_2011,
title = {Entropy viscosity method for nonlinear conservation laws},
journal = {Journal of Computational Physics},
volume = {230},
pages = {4248-4267},
year = {2011},
doi = {https://doi.org/10.1016/j.jcp.2010.11.043},
author = {Jean-Luc {Guermond} and Richard {Pasquetti} and Bojan {Popov}}}

@article{guermondPopov_2014,
author = {{Guermond}, Jean-Luc and {Popov}, Bojan},
title = "{Viscous Regularization of the Euler Equations and Entropy Principles}",
journal = {SIAM Journal on Applied Mathematics},
volume = {74},
number = {2},
pages = {284-305},
year = {2014},
doi = {10.1137/120903312}}

@article{hartmannHouston_2006,
  title={Symmetric interior penalty DG methods for the compressible Navier--Stokes equations II: Goal--oriented a posteriori error estimation},
  author={Hartmann, Ralf and Houston, Paul},
  journal={Int. J. Num. Anal. Model},
  volume={3},
  number={2},
  pages={141--162},
  year={2006}}

@article{he_etal_2015,
title = "{Acceleration of the OpenFOAM-based MHD solver using graphics processing units}",
journal = {Fusion Engineering and Design},
volume = {101},
pages = {88-93},
year = {2015},
doi = {https://doi.org/10.1016/j.fusengdes.2015.09.017},
author = {Qingyun {He} and Hongli {Chen} and Jingchao {Feng}}}

@article{hender_etal_2007,
        author = {T.~C.~Hender and J.~C.~Wesley and J.~Bialek and A.~Bondeson and A.~H.~Boozer and R.~J.~Buttery and A.~Garofalo and T.~P.~Goodman and R.~S.~Granetz and Y.~Gribov and O.~Gruber and M.~Gryaznevich and G.~Giruzzi and S.~G{"u}nter and N.~Hayashi and P.~Helander and C.~C.~Hegna and D.~F.~Howell and D.~A.~Humphreys and G.~T.~A.~Huysmans and A.~W.~Hyatt and A.~Isayama and S.~C.~Jardin and Y.~Kawano and A.~Kellman and C.~Kessel and H.~R.~Koslowski and R.~J.~La Haye and E.~Lazzaro and Y.~Q.~Liu and V.~Lukash and J.~Manickam and S.~Medvedev and V.~Mertens and S.~V.~Mirnov and Y.~Nakamura and G.~Navratil and M.~Okabayashi and T.~Ozeki and R.~Paccagnella and G.~Pautasso and F.~Porcelli and V.~D.~Pustovitov and V.~Riccardo and M.~Sato and O.~Sauter and M.~J.~Schaffer and M.~Shimada and P.~Sonato and E.~J.~Strait and M.~Sugihara and M.~Takechi and A.~D.~Turnbull and E.~Westerhof and D.~G.~Whyte and R.~Yoshino and H.~ Zohm and the ITPA MHD, Disruption and Magnetic Control Topical Group},
        journal = {Nucl.~Fusion},
        pages = {S128},
        title = {Chapter 3: {MHD} stability, operational limits and disruptions},
        volume = 47,
        year = 2007}

@article{hu_etal_2021,
title = {Helicity-conservative finite element discretization for incompressible MHD systems},
journal = {Journal of Computational Physics},
volume = {436},
pages = {110284},
year = {2021},
issn = {0021-9991},
doi = {https://doi.org/10.1016/j.jcp.2021.110284},
url = {https://www.sciencedirect.com/science/article/pii/S0021999121001790},
author = {Kaibo Hu and Young-Ju Lee and Jinchao Xu}}

@techreport{huang_etal_2017,
  author       = {{Huang}, Peter and {Chhabra}, Rupanshi and {Munipalli}, Ramakanth and {Pulugundla}, Gautam and {Kawczynski}, Charlie and {Smolentsev}, Sergey},
  title        = {A Comprehensive High Performance Predictive Tool for Fusion Liquid Metal Hydromagnetics},
  institution  = {HyPerComp Inc., Westlake Village, CA (United States)},
  url          = {https://www.osti.gov/biblio/1402053},
  place        = {United States},
  year         = {2017},
  month        = {10}}

@article{hughesMallet_1986,
title = {A new finite element formulation for computational fluid dynamics: III. The generalized streamline operator for multidimensional advective-diffusive systems},
journal = {Computer Methods in Applied Mechanics and Engineering},
volume = {58},
pages = {305-328},
year = {1986},
doi = {https://doi.org/10.1016/0045-7825(86)90152-0},
author = {Thomas J.R. {Hughes} and Michel {Mallet}}}

@article{hughes_etal_1998,
title = {The variational multiscale method—a paradigm for computational mechanics},
journal = {Computer Methods in Applied Mechanics and Engineering},
volume = {166},
pages = {3-24},
year = {1998},
doi = {https://doi.org/10.1016/S0045-7825(98)00079-6},
author = {Thomas J.R. {Hughes} and Gonzalo R. {Feijóo} and Luca {Mazzei} and Jean-Baptiste {Quincy}}}

@article{humrickhouseMerrill_2018,
title = "{An equation of state for liquid Pb83Li17}",
journal = {Fusion Engineering and Design},
volume = {127},
pages = {10-16},
year = {2018},
doi = {https://doi.org/10.1016/j.fusengdes.2017.11.023},
author = {P.W. {Humrickhouse} and B.J. {Merrill}}}

@article{huntShercliff_1971,
   author = {{Hunt}, J C R and {Shercliff}, J A},
   title = "{Magnetohydrodynamics at High Hartmann Number}",
   journal= {Annual Review of Fluid Mechanics},
   year = {1971},
   volume = {3},
   pages = {37-62},
   doi = {https://doi.org/10.1146/annurev.fl.03.010171.000345}}

@BOOK{jackson_ClassicalElectroDynamics,
       author = {{Jackson}, John David},
        title = "{Classical Electrodynamics, 2nd Edition}",
         year = 1975,
    publisher = {McGraw-Hill}}

@phdthesis{kawczynski_thesis,
   author = {{Kawczynski}, C.N.},
    title = "{A Numerical Investigation of Moderate Magnetic Reynolds Number Fusion Liquid Metal Magnetohydrodynamic Flows}",
   school = {University of California, Los Angeles},
publisher = {ProQuest},
  address = {Ann Arbor, MI},
     year = {2018}}

@article{kawczynski_etal_2018,
    author = {{Kawczynski}, Charles and {Smolentsev}, Sergey and {Abdou}, Mohamed},
    title = {Characterization of the lid-driven cavity magnetohydrodynamic flow at finite magnetic Reynolds numbers using far-field magnetic boundary conditions},
    journal = {Physics of Fluids},
    volume = {30},
    number = {6},
    pages = {067103},
    year = {2018},
    doi = {10.1063/1.5036775}}

@techreport{kennedyCarpenter_2016,
  author={Christopher A. Kennedy and Mark H. Carpenter},
  title="{Diagonally Implicit Runge-Kutta Methods for Ordinary Differential Equations. A Review}",
  year = 2016,
  number = {NASA/TM-2016-219173},
  institution = {Langley Research Center}}

@article{kessel_etal_2018,
title = {Overview of the fusion nuclear science facility, a credible break-in step on the path to fusion energy},
journal = {Fusion Engineering and Design},
volume = {135},
pages = {236-270},
year = {2018},
doi = {https://doi.org/10.1016/j.fusengdes.2017.05.081},
author = {C.E. Kessel and J.P. Blanchard and A. Davis and L. El-Guebaly and L.M. Garrison and N.M. Ghoniem and P.W. Humrickhouse and Y. Huang and Y. Katoh and A. Khodak and E.P. Marriott and S. Malang and N.B. Morley and G.H. Neilson and J. Rapp and M.E. Rensink and T.D. Rognlien and A.F. Rowcliffe and S. Smolentsev and L.L. Snead and M.S. Tillack and P. Titus and L.M. Waganer and G.M. Wallace and S.J. Wukitch and A. Ying and K. Young and Y. Zhai}}

@article{keyes_etal_2013,
author = {David E Keyes and Lois C McInnes and Carol Woodward and William Gropp and Eric Myra and Michael Pernice and John Bell and Jed Brown and Alain Clo and Jeffrey Connors and Emil Constantinescu and Don Estep and Kate Evans and Charbel Farhat and Ammar Hakim and Glenn Hammond and Glen Hansen and Judith Hill and Tobin Isaac and Xiangmin Jiao and Kirk Jordan and Dinesh Kaushik and Efthimios Kaxiras and Alice Koniges and Kihwan Lee and Aaron Lott and Qiming Lu and John Magerlein and Reed Maxwell and Michael McCourt and Miriam Mehl and Roger Pawlowski and Amanda P Randles and Daniel Reynolds and Beatrice Rivière and Ulrich Rüde and Tim Scheibe and John Shadid and Brendan Sheehan and Mark Shephard and Andrew Siegel and Barry Smith and Xianzhu Tang and Cian Wilson and Barbara Wohlmuth},
title ={Multiphysics simulations: Challenges and opportunities},
journal = {The International Journal of High Performance Computing Applications},
volume = {27},
number = {1},
pages = {4-83},
year = {2013},
doi = {10.1177/1094342012468181}}

@article{kirkCarey_2008,
author = {{Kirk}, Benjamin S. and {Carey}, Graham F.},
title = {Development and validation of a SUPG finite element scheme for the compressible Navier–Stokes equations using a modified inviscid flux discretization},
journal = {International Journal for Numerical Methods in Fluids},
volume = {57},
pages = {265-293},
doi = {https://doi.org/10.1002/fld.1635},
year = {2008}}

@article{knollKeyes_2004,
title = {Jacobian-free Newton–Krylov methods: a survey of approaches and applications},
journal = {Journal of Computational Physics},
volume = {193},
number = {2},
pages = {357-397},
year = {2004},
doi = {https://doi.org/10.1016/j.jcp.2003.08.010},
author = {D.A. Knoll and D.E. Keyes}}

@article{liuOsher_1996,
     title = {Nonoscillatory High Order Accurate Self-Similar Maximum Principle Satisfying Shock Capturing Schemes {I}},
     author = {{Liu}, X. D. and {Osher}, S.},
     journal = {SIAM J. Numer. Anal.},
     volume = {33},
     number = {2},
     pages = {760-779},
     year = {1996}}

@article{martelli_etal_2019,
title = "{Literature review of lead-lithium thermophysical properties}",
journal = {Fusion Engineering and Design},
volume = {138},
pages = {183-195},
year = {2019},
doi = {https://doi.org/10.1016/j.fusengdes.2018.11.028},
author = {D. {Martelli} and A. {Venturini} and M. {Utili}}}

@article{ni_etal_2007,
author = {Ming-Jiu {Ni} and Ramakanth {Munipalli} and Neil B. {Morley} and Peter {Huang} and Mohamed A. {Abdou}},
title = "{Validation Case Results for 2D and 3D MHD Simulations}",
journal = {Fusion Science and Technology},
volume = {52},
pages = {587--594},
year = {2007},
publisher = {American Nuclear Society},
doi = {10.13182/FST07-A1552}}

@article{ohm_etal_2024,
author = {Ohm, Peter and Bonilla, Jesus and Phillips, Edward and Shadid, John N. and Crockatt, Michael and Tuminaro, Ray S. and Hu, Jonathan and Tang, Xian-Zhu},
title = {Scalable Multiphysics Block Preconditioning for Low Mach Number Compressible Resistive MHD with Application to Magnetic Confinement Fusion},
journal = {SIAM Journal on Scientific Computing},
volume = {46},
number = {5},
pages = {S170-S201},
year = {2024},
doi = {10.1137/23M1582667}}

@article{planas_etal_2011,
title = "{Approximation of the inductionless MHD problem using a stabilized finite element method}",
journal = {Journal of Computational Physics},
volume = {230},
number = {8},
pages = {2977-2996},
year = {2011},
doi = {https://doi.org/10.1016/j.jcp.2010.12.046},
author = {Ramon {Planas} and Santiago {Badia} and Ramon {Codina}}}

@misc{pointwise,
  title = {Fidelity Pointwise for CFD Meshing},
  howpublished = {\url{www.pointwise.com}},
  note = {Accessed: 2025-07-11},
  year = {2025}}

@ARTICLE{powell_etal_1999,
       author = {{Powell}, Kenneth G. and {Roe}, Philip L. and {Linde}, Timur J. and {Gombosi}, Tamas I. and {De Zeeuw}, Darren L.},
        title = "{A Solution-Adaptive Upwind Scheme for Ideal Magnetohydrodynamics}",
      journal = {Journal of Computational Physics},
         year = 1999,
        month = sep,
       volume = {154},
       number = {2},
        pages = {284-309},
          doi = {10.1006/jcph.1999.6299}}

@article{rogers_1987,
title = "{On the accuracy of the pseudocompressibility method in solving the incompressible Navier-Stokes equations}",
journal = {Applied Mathematical Modelling},
volume = {11},
number = {1},
pages = {35-44},
year = {1987},
doi = {https://doi.org/10.1016/0307-904X(87)90182-X},
author = {S.E. {Rogers} and D. {Kwak} and U. {Kaul}}}

@article{shercliff_1953,
    title = "{Steady motion of conducting fluids in pipes under transverse magnetic fields}",
   volume = {49},
      doi = {10.1017/S0305004100028139},
   number = {1},
  journal = {Mathematical Proceedings of the Cambridge Philosophical Society},
   author = {Shercliff, J. A.},
     year = {1953},
    pages = {136–144}}

@article{smolentsev_etal_2008,
title = {Characterization of key magnetohydrodynamic phenomena in PbLi flows for the US DCLL blanket},
journal = {Fusion Engineering and Design},
volume = {83},
number = {5},
pages = {771-783},
year = {2008},
doi = {https://doi.org/10.1016/j.fusengdes.2008.07.023},
author = {Sergey Smolentsev and René Moreau and Mohamed Abdou}}

@article{smolentsev_etal_2015,
	title = {Dual-coolant lead–lithium (DCLL) blanket status and R\&D needs},
	journal = {Fusion Engineering and Design},
	volume = {100},
	pages = {44-54},
	year = {2015},
	issn = {0920-3796},
	doi = {https://doi.org/10.1016/j.fusengdes.2014.12.031},
	author = {Sergey Smolentsev and Neil B. Morley and Mohamed A. Abdou and Siegfried Malang}}

@article{smolentsev_etal_2015b,
	title = {An approach to verification and validation of MHD codes for fusion applications},
	journal = {Fusion Engineering and Design},
	volume = {100},
	pages = {65-72},
	year = {2015},
	issn = {0920-3796},
	doi = {https://doi.org/10.1016/j.fusengdes.2014.04.049},
	author = {S. {Smolentsev} and S. {Badia} and R. {Bhattacharyay} and L. {Bühler} and L. {Chen} and Q. {Huang} and H.-G. {Jin} and D. {Krasnov} and D.-W. {Lee} and E. Mas {de les Valls} and C. {Mistrangelo} and R. {Munipalli} and M.-J. {Ni} and D. {Pashkevich} and A. {Patel} and G. {Pulugundla} and P. {Satyamurthy} and A. {Snegirev} and V. {Sviridov} and P. {Swain} and T. {Zhou} and O. {Zikanov}}}

@article{smolentsev_etal_2020,
	author = {S. {Smolentsev} and T. {Rhodes} and Y. {Yan} and A. {Tassone} and C. {Mistrangelo} and L. {Bühler} and F. R. {Urgorri}},
	title = {Code-to-Code Comparison for a PbLi Mixed-Convection MHD Flow},
	journal = {Fusion Science and Technology},
	volume = {76},
	number = {5},
	pages = {653--669},
	year = {2020},
	publisher = {Taylor \& Francis},
	doi = {10.1080/15361055.2020.1751378}}

@Article{smolentsev_2021,
AUTHOR = {{Smolentsev}, Sergey},
TITLE = "{Physical Background, Computations and Practical Issues of the Magnetohydrodynamic Pressure Drop in a Fusion Liquid Metal Blanket}",
JOURNAL = {Fluids},
VOLUME = {6},
YEAR = {2021},
NUMBER = {3},
ARTICLE-NUMBER = {110},
DOI = {10.3390/fluids6030110}}

@article{smolentsev_2025,
	title = {A model and assessments of the effect of transient plasma on liquid metal flows in a fusion blanket},
	journal = {Fusion Engineering and Design},
	volume = {213},
	pages = {114854},
	year = {2025},
	issn = {0920-3796},
	doi = {https://doi.org/10.1016/j.fusengdes.2025.114854},
	url = {https://www.sciencedirect.com/science/article/pii/S0920379625000572},
	author = {Sergey Smolentsev}}

@article{stone_etal_2020,
doi = {10.3847/1538-4365/ab929b},
year = {2020},
volume = {249},
pages = {4},
author = {Stone, James M. and Tomida, Kengo and White, Christopher J. and Felker, Kyle G.},
title = {The Athena++ Adaptive Mesh Refinement Framework: Design and Magnetohydrodynamic Solvers},
journal = {The Astrophysical Journal Supplement Series}}

@ARTICLE{toth_2000,
       author = {{T{\'o}th}, G{\'a}bor},
        title = "{The {\ensuremath{\nabla}}{\textperiodcentered} B=0 Constraint in Shock-Capturing Magnetohydrodynamics Codes}",
      journal = {Journal of Computational Physics},
         year = 2000,
        month = jul,
       volume = {161},
       number = {2},
        pages = {605-652},
          doi = {10.1006/jcph.2000.6519}}

@misc{VERTEX-CFD-v100,
  author       = {Delchini, Marc Olivier and
                  Furkan Oz and
                  Kincaid, Kellis and
                  Erwin, Jon Taylor and
                  Slattery, Stuart and
                  Curtis, Franklin and
                  Glasby, Ryan and
                  Brandao, Filipe and
                  Gottiparthi, Kalyana and
                  DeGraw, Jason and
                  Ryan Savery},
  title        = {ORNL/VERTEX-CFD: vertex-cfd-v1.0.0-alpha},
  month        = feb,
  year         = 2025,
  publisher    = {Zenodo},
  version      = {v1.0.0},
  doi          = {10.5281/zenodo.14907174},
  url          = {https://doi.org/10.5281/zenodo.14907174},
  swhid        = {swh:1:dir:3ea1469d6974d2e4972bc56707b48e67267354dd
                   ;origin=https://doi.org/10.5281/zenodo.14907173;vi
                   sit=swh:1:snp:f5cf93739bbdcf134c41ed8f6db4c3a694dc
                   972a;anchor=swh:1:rel:4d02f164be05b6b66380bcae1ea5
                   1cf54d291275;path=ORNL-VERTEX-CFD-1ff10fb
                  },
}

@ARTICLE{vinokur_1983,
       author = {{Vinokur}, Marcel},
        title = "{On One-Dimensional Stretching Functions for Finite-Difference Calculations}",
      journal = {Journal of Computational Physics},
         year = 1983,
       volume = {50},
        pages = {215-234}}

@article{zhangShu_2011,
author = {{Zhang}, Xiangxiong  and {Shu}, Chi-Wang },
title = {Maximum-principle-satisfying and positivity-preserving high-order schemes for conservation laws: survey and new developments},
journal = {Proceedings of the Royal Society A: Mathematical, Physical and Engineering Sciences},
volume = {467},
number = {2134},
pages = {2752-2776},
year = {2011},
doi = {10.1098/rspa.2011.0153}}

@article{zhou_2010,
title = "{Code development and validation for analyzing liquid metal MHD flow in rectangular ducts}",
journal = {Fusion Engineering and Design},
volume = {85},
number = {10},
pages = {1736-1741},
year = {2010},
doi = {https://doi.org/10.1016/j.fusengdes.2010.05.034},
author = {Tao {Zhou} and Zhiyi {Yang} and Mingjiu {Ni} and Hongli {Chen}}}

@article{fambri_etal_2023,
title = {A well-balanced and exactly divergence-free staggered semi-implicit hybrid finite volume / finite element scheme for the incompressible MHD equations},
journal = {Journal of Computational Physics},
volume = {493},
pages = {112493},
year = {2023},
doi = {https://doi.org/10.1016/j.jcp.2023.112493},
author = {F. Fambri and E. Zampa and S. Busto and L. Río-Martín and F. Hindenlang and E. Sonnendrücker and M. Dumbser}}

@Manual{trilinos-website,
title = {The {T}rilinos {P}roject {W}ebsite},
author = {The {T}rilinos {P}roject {T}eam},
year = {2025 (acccessed March 13, 2025)},
url = {https://trilinos.github.io}
}

@Manual{panzer-website,
title = {The {P}anzer {P}roject {W}ebsite},
author = {The {P}anzer {P}roject {T}eam},
year = {2025 (acccessed March 13, 2025)},
url = {https://trilinos.github.io/panzer.html}
}

@article{TRICCO20127214,
title = {Constrained hyperbolic divergence cleaning for smoothed particle magnetohydrodynamics},
journal = {Journal of Computational Physics},
volume = {231},
number = {21},
pages = {7214-7236},
year = {2012},
issn = {0021-9991},
doi = {https://doi.org/10.1016/j.jcp.2012.06.039},
url = {https://www.sciencedirect.com/science/article/pii/S0021999112003737},
author = {Terrence S. Tricco and Daniel J. Price}
}

@article{phipps2022automatic,
  title={Automatic differentiation of C++ codes on emerging manycore architectures with sacado},
  author={Phipps, Eric and Pawlowski, Roger and Trott, Christian},
  journal={ACM Transactions on Mathematical Software},
  volume={48},
  number={4},
  pages={1--29},
  year={2022},
  publisher={ACM New York, NY}
}

@article{bochev2012solving,
  title={Solving pdes with intrepid},
  author={Bochev, Pavel and Edwards, H Carter and Kirby, Robert C and Peterson, Kara and Ridzal, Denis},
  journal={Scientific Programming},
  volume={20},
  number={2},
  pages={151--180},
  year={2012},
  publisher={Wiley Online Library}
}

@article{pawlowski2012automating,
  title={Automating embedded analysis capabilities and managing software complexity in multiphysics simulation, Part I: Template-based generic programming},
  author={Pawlowski, Roger P and Phipps, Eric T and Salinger, Andrew G},
  journal={Scientific Programming},
  volume={20},
  number={2},
  pages={197--219},
  year={2012},
  publisher={Wiley Online Library}
}

@article{pawlowski2012automating2,
  title={Automating embedded analysis capabilities and managing software complexity in multiphysics simulation, Part II: Application to partial differential equations},
  author={Pawlowski, Roger P and Phipps, Eric T and Salinger, Andrew G and Owen, Steven J and Siefert, Christopher M and Staten, Matthew L},
  journal={Scientific Programming},
  volume={20},
  number={3},
  pages={327--345},
  year={2012},
  publisher={Wiley Online Library}
}

@article{devine2002zoltan,
  title={Zoltan data management services for parallel dynamic applications},
  author={Devine, Karen and Boman, Erik and Heaphy, Robert and Hendrickson, Bruce and Vaughan, Courtenay},
  journal={Computing in Science \& Engineering},
  volume={4},
  number={2},
  pages={90--96},
  year={2002},
  publisher={IEEE}
}

@incollection{karypis2011metis,
  title={METIS and ParMETIS},
  author={Karypis, George},
  booktitle={Encyclopedia of parallel computing},
  pages={1117--1124},
  year={2011},
  publisher={Springer}
}

@techreport{edwards2011stk,
  title={STK-mesh tutorial minisymposium.},
  author={Edwards, Harold C and Coffey, Todd S and Sunderland, Daniel and Williams, Alan B},
  year={2011},
  institution={Sandia National Lab.(SNL-NM), Albuquerque, NM (United States)}
}

@article{sjaardema2006exodus,
  title={EXODUS II: A finite element data model},
  author={Sjaardema, Gregory D and Schoof, Larry A and Yarberry, Victor R},
  journal={Computational Mechanics and Visualization Department, Sandia National Laboratories, Albuquerque, NM},
  volume={87185},
  year={2006}
}

@techreport{blacker1994cubit,
  title={CUBIT mesh generation environment. Volume 1: Users manual},
  author={Blacker, Ted D and Bohnhoff, William J and Edwards, Tony L},
  year={1994},
  institution={Sandia National Lab.(SNL-NM), Albuquerque, NM (United States)}
}

@techreport{bader2005robust,
  title={Robust large-scale parallel nonlinear solvers for simulations.},
  author={Bader, Brett William and Pawlowski, Roger Patrick and Kolda, Tamara Gibson},
  year={2005},
  institution={Sandia National Laboratories (SNL), Albuquerque, NM, and Livermore, CA~…}
}

@article{bavier2012amesos2,
  title={Amesos2 and Belos: Direct and iterative solvers for large sparse linear systems},
  author={Bavier, Eric and Hoemmen, Mark and Rajamanickam, Sivasankaran and Thornquist, Heidi},
  journal={Scientific Programming},
  volume={20},
  number={3},
  pages={241--255},
  year={2012},
  publisher={Wiley Online Library}
}

@incollection{schenk2011pardiso,
  title={Pardiso},
  author={Schenk, Olaf and G{\"a}rtner, Klaus},
  booktitle={Encyclopedia of parallel computing},
  pages={1458--1464},
  year={2011},
  publisher={Springer}
}

@article{prokopenko2016ifpack2,
  title={Ifpack2 User’s Guide 1.0 (Trilinos version 12.6)},
  author={Prokopenko, Andrey and Siefert, Christopher M and Hu, Jonathan J and Hoemmen, Mark and Klinvex, Alicia},
  journal={Sandia Report, SAND2016-5338},
  year={2016}
}

@article{yamazaki2025shylu,
  title={ShyLU node: On-node Scalable Solvers and Preconditioners Recent Progresses and Current Performance},
  author={Yamazaki, Ichitaro and Ellingwood, Nathan and Rajamanickam, Sivasankaran},
  journal={arXiv preprint arXiv:2506.05793},
  year={2025}
}

@techreport{obert2017tempus,
  title={Tempus v. 1.0},
  author={Ober, Curtis C and Pawlowski, Roger and Cyr, Eric C and Conde, Sidafa},
  year={2017},
  institution={Sandia National Lab.(SNL-NM), Albuquerque, NM (United States)}
}

@article{trott2021kokkos,
  title={Kokkos 3: Programming model extensions for the exascale era},
  author={Trott, Christian R and Lebrun-Grandi{\'e}, Damien and Arndt, Daniel and Ciesko, Jan and Dang, Vinh and Ellingwood, Nathan and Gayatri, Rahulkumar and Harvey, Evan and Hollman, Daisy S and Ibanez, Dan and others},
  journal={IEEE Transactions on Parallel and Distributed Systems},
  volume={33},
  number={4},
  pages={805--817},
  year={2021},
  publisher={IEEE}
}

@book{kelley2003solving,
  title={Solving nonlinear equations with Newton's method},
  author={Kelley, Carl T},
  year={2003},
  publisher={SIAM}
}

@article{saad1986gmres,
  title={GMRES: A generalized minimal residual algorithm for solving nonsymmetric linear systems},
  author={Saad, Youcef and Schultz, Martin H},
  journal={SIAM Journal on scientific and statistical computing},
  volume={7},
  number={3},
  pages={856--869},
  year={1986},
  publisher={SIAM}
}

@article{phillips2014block,
  title={A block preconditioner for an exact penalty formulation for stationary MHD},
  author={Phillips, Edward G and Elman, Howard C and Cyr, Eric C and Shadid, John N and Pawlowski, Roger P},
  journal={SIAM Journal on Scientific Computing},
  volume={36},
  number={6},
  pages={B930--B951},
  year={2014},
  publisher={SIAM}
}

@article{klawonn2000comparison,
  title={A comparison of overlapping Schwarz methods and block preconditioners for saddle point problems},
  author={Klawonn, Axel and Pavarino, Luca F},
  journal={Numerical linear algebra with applications},
  volume={7},
  number={1},
  pages={1--25},
  year={2000},
  publisher={Wiley Online Library}
}

@article{sip-wheeler,
author = {Wheeler, M. F.},
title = {An Elliptic Collocation-Finite Element Method with Interior Penalties},
journal = {SIAM Journal on Numerical Analysis},
volume = {15},
number = {1},
pages = {152--161},
year = {1978},
doi = {10.1137/0715010},
URL = {https://doi.org/10.1137/0715010},
eprint = {https://doi.org/10.1137/0715010},
}

@article{HARTMANN20089670,
title = {An optimal order interior penalty discontinuous Galerkin discretization of the compressible Navier–Stokes equations},
journal = {Journal of Computational Physics},
volume = {227},
number = {22},
pages = {9670-9685},
year = {2008},
issn = {0021-9991},
doi = {https://doi.org/10.1016/j.jcp.2008.07.015},
url = {https://www.sciencedirect.com/science/article/pii/S0021999108003975},
author = {Ralf Hartmann and Paul Houston}
}

\appendix

\section{Wave Speeds of the AC-GLM-MHD System}
\label{sec:wavespeeds}

Here we consider the wave speeds associated with the MHD system in the inviscid and ideal limit of Eq.~\eqref{eq:mhd.compact}; i.e.,
\begin{equation}
    \pd{\bU}{t}+\nabla\cdot\bF=\bS,
    \label{eq:mhd.compact.inviscid}
\end{equation}
which is obtained from Eq.~\eqref{eq:mhd.compact} by setting $\nu=\eta=0$.  
We also set $\alpha=0$ in the source term in Eq.~\eqref{eq:mhd.compact.sources}.  
Defining primitive variables $\bV=[\bv^{\intercal},P,\bB^{\intercal},\psi]^{\intercal}$, we can write Eq.~\eqref{eq:mhd.compact.inviscid} in quasilinear form
\begin{equation}
    \pd{\bV}{t}
    +A^{x}(\bV)\pd{\bV}{x}
    +A^{y}(\bV)\pd{\bV}{y}
    +A^{z}(\bV)\pd{\bV}{z}
    =0,
\end{equation}
where 
\begin{equation}
    A^{x}=\Big(\pderiv{\bU}{\bV}\Big)^{-1}\Big(\pderiv{\bF^{x}}{\bV}\Big)
    =
    \left[\begin{array}{cccccccc}
        2v_{x} & 0 & 0 & 1/\rho_{0} & 0 & \f{B_{y}}{\mu_{0}\rho_{0}} & \f{B_{z}}{\mu_{0}\rho_{0}} & 0 \\
         v_{y} & v_{x} & 0 & 0 & 0 & -\f{B_{x}}{\mu_{0}\rho_{0}} & 0 & 0 \\
         v_{z} & 0 & v_{x} & 0 & 0 & 0 & -\f{B_{x}}{\mu_{0}\rho_{0}} & 0 \\
         \rho_{0}c_{0}^{2} & 0 & 0 & 0 & 0 & 0 & 0 & 0 \\
         0 & 0 & 0 & 0 & v_{x} & 0 & 0 & c_{h} \\
         B_{y} & -B_{x} & 0 & 0 & 0 & v_{x} & 0 & 0 \\
         B_{z} & 0 & -B_{x} & 0 & 0 & 0 & v_{x} & 0 \\
         0 & 0 & 0 & 0 & c_{h} & 0 & 0 & 0
    \end{array}\right],
\end{equation}
with $A^{y}$ and $A^{z}$ defined similarly.  
We let $\tilde{A}^{x}$ denote the upper left $4\times4$ block of $A^{x}$, corresponding to the inviscid part of the NS subsystem in Eq.~\eqref{eq:navierStokes} with $\bB=0$.  

Computing the characteristic polynomial of $\tilde{A}^{x}$, we obtain
\begin{equation}
	\det(\tilde{A}^{x}-\lambda^{x}\,I) = \big(\,\lambda^{x}-v_{x}\,\big)^{2}\,\big(\,\lambda^{x}-(v_{x}-c)\,\big)\,\big(\,\lambda^{x}-(v_{x}+c)\,\big),
\end{equation}
where $c^{2}=v_{x}^{2}+c_{0}^{2}$.  
Thus, the eigenvalues of the NS system correspond to entropy ($\lambda^{x}=v_{x}$) and acoustic ($\lambda^{x}=v_{x}\pm c$) modes.  
Since, for $c_{0}>0$, 
\begin{equation}
	0\le |v_{x}|/\max(|v_{x}+c|,|v_{x}-c|)<1
\end{equation}
it follows that the flow is always subsonic.  

Computing the characteristic polynomial of $A^{x}$ for the case with $\bv=0$, we obtain
\begin{align}
	&\det(A^{x}-\lambda^{x}\,I) \\
	&\stackrel{\bv=0}{=}
	\big(\,\lambda^{x}-c_{{\rm s},x}\,\big)\,
	\big(\,\lambda^{x}+c_{{\rm s},x}\,\big)\,
	\big(\,\lambda^{x}-b_{x}\,\big)\,
	\big(\,\lambda^{x}+b_{x}\,\big)\,
	\big(\,\lambda^{x}-c_{{\rm f},x}\,\big)\,
	\big(\,\lambda^{x}+c_{{\rm f},x}\,\big)\,
	\big(\,\lambda^{x}-c_{h}\,\big)\,
	\big(\,\lambda^{x}+c_{h}\,\big), \nonumber
\end{align}
where $b_{x}=B_{x}/\sqrt{\mu_{0}\rho_{0}}$, the fast and slow magnetosonic speeds are
\begin{subequations}\label{eq:magnetosonic_speed}
	\begin{align}
		c_{{\rm f},x}^{2} 
		&=\f{1}{2}\,\big[\,(c_{0}^{2}+|\bb|^{2})+\sqrt{(c_{0}^{2}+|\bb|^{2})^{2}-4\,c_{0}^{2}\,b_{x}^{2}}\,\big], \label{eq:fast_magnetosonic_speed} \\
		c_{{\rm s},x}^{2}
		&=\f{1}{2}\,\big[\,(c_{0}^{2}+|\bb|^{2})-\sqrt{(c_{0}^{2}+|\bb|^{2})^{2}-4\,c_{0}^{2}\,b_{x}^{2}}\,\big], \label{eq:slow_magnetosonic_speed}
	\end{align}
\end{subequations}
and $|\bb|=|\bB|/\sqrt{\mu_{0}\rho_{0}}$ is the Alfv{\'e}n speed.  

We do not have closed-form expressions for the eigenvalues of $A^{x}$ for the general case with $\bv\ne0$ and $\bB\ne0$.  
Instead, we use the approximate expressions
\begin{equation}
	\big\{\,
		v_{x}-c_{h},\,
		v_{x}-c_{{\rm f},x},\,
		v_{x}-b_{x},\,
		v_{x}-c_{{\rm s},x},\,
		v_{x}+c_{{\rm s},x},\,
		v_{x}+b_{x},\,
		v_{x}+c_{{\rm f},x},\,
		v_{x}+c_{h}
	\,\big\},
	\label{eq:eigenvalues}
\end{equation}
with $c_{0}$ replaced with $c=\sqrt{v_{x}^{2}+c_{0}^{2}}$ in Eq.~\eqref{eq:magnetosonic_speed}.  
(Similar expressions can be found for $A^{y}$ and $A^{z}$.)  
Our FEM with implicit time stepping does not require exact expressions for the eigenvalues, only estimates for time step selection.  
We believe the estimates in Eq.~\eqref{eq:eigenvalues} are sufficient for this purpose.  

\section{Additional Implementation Details}
\label{sec:implementations}

In this appendix, we provide implementation details of the methods outlined in Section~\ref{sec:methods}, covering mesh representation and partitioning, operator assembly, time integrators, and the subsequent linear solvers and preconditioners that emerge from linearization of the nonlinear problem.  
Our overview is intentionally brief, as \vertex\ heavily leverages the technology and formalisms described in \cite{pawlowski2012automating,pawlowski2012automating2} and other references, to which we refer the reader to for additional detail.  

\paragraph{Time Integrators}
As described in Section~\ref{sec:methods}, \vertex\ uses implicit time integration methods.  
The implementation of these methods is provided by the Tempus library in Trilinos \cite{obert2017tempus}.  
Tempus orchestrates the construction of the nonlinear problem solved at each time step for a given time integrator, which in turn builds the linear solver and preconditioner that drive the operator assembly through evaluation of physics terms; i.e., the individual terms in Eq.~\eqref{eq:mhd.compact.fem}.  
Time step selection is controlled automatically through a specified CFL number.  
Physics kernels compute local wave speeds and element lengths, which are combined to build a local time step sizes via Eq.~\eqref{eq:dtCFL}; a global minimum then sets the time step.  
In general, early stages of a simulation may use small CFL values due to initial transients resulting from initial conditions that are not equilibrium solutions to Eq.~\eqref{eq:mhd.compact.fem}, and therefore make challenging nonlinear problems.  
We have implemented a CFL-based time step controller within the Tempus framework to smoothly ramp the CFL in time, and we use this capability for the simulations discussed in Section~\ref{sec:applications}.  

\paragraph{Nonlinear Solvers}
The solution to Eq.~\eqref{eq:mhd.compact.ode} within the time integrator requires the solution of a nonlinear problem, for which we use the NOX library in Trilinos \cite{bader2005robust}.  
While many algorithms are available in NOX for the solution of nonlinear problems, we use inexact Newton methods \cite{kelley2003solving}, which rely on linearizations of the nonlinear residual in terms of the Jacobian matrix in Eq.~\eqref{eq:nonlinear.jacobian}, and an iterative method is used to approximately solve for the update vector in Eq.\eqref{eq:nonlinear.iteration}.  
We obtain both the residual vectors and the Jacobian matrices through coordination of NOX and the finite element assembly tools in Trilinos, which will be further outlined below.  
Subsequent matrix and vector operations needed for the implementation of Newton's method, such as vector norms and dot products, can be executed on GPUs using the Trilinos Tpetra library.  
To improve robustness, many globalization methods (e.g., line search methods) are available in NOX for use with Newton's method.  
We generally choose a simple backtracking scheme to modify $\delta\bY$ in Eq.~\eqref{eq:nonlinear.iteration} to ensure reduction of the residual norm for a given nonlinear iteration.  
In the simulations documented in Sections~\ref{sec:tests} and \ref{sec:applications}, we report nonlinear solver tolerances $\tolN$ based on the norm of the nonlinear residual.  

\paragraph{Linear Solvers and Preconditioners}
An inexact Newton method requires the inversion of the Jacobian matrix to compute the update vector in Eq.\eqref{eq:nonlinear.iteration}.  
This is generally among the most expensive parts of a simulation as it is repeated in each nonlinear iteration of which there may be several per time step, depending on the initial guess and time stepping method.  
The discretization in Eq.~\eqref{eq:mhd.compact.fem} results in Jacobian matrices that are generally neither symmetric nor positive definite.  
Furthermore, the systems are large, sparse, and distributed in an HPC context.  
We use GMRES \cite{saad1986gmres} for the solution of linear problems with the Belos library in Trilinos \cite{bavier2012amesos2}.  
In Sections~\ref{sec:tests} and \ref{sec:applications}, reported linear solver convergence tolerances $\tolL$ apply to the GMRES residual norm convergence criteria, which is distinct from the nonlinear solver convergence criteria discussed above.  

The implicit FEM discretization tends to produce particularly stiff matrices when large time steps are taken (see for example CFL numbers in Figure~\ref{fig:dbdt_divergence_cleaning_solver_data}).  
As a result, a preconditioner is typically required with GMRES for efficient convergence of the nonlinear problem.  
For our initial efforts, we consider algebraic preconditioners (e.g., preconditioners that only consider Jacobian matrix entries, and no additional information about the PDE being solved) because we have easy access to the matrix elements through the finite element assembly procedure via automatic differentiation as outlined below, and we avoid, at least initially, the challenge of developing specific physics-based approaches (e.g., \cite{phillips2014block}).  
Unless small time steps are used to resolve dynamics in a time accurate way, general transient simulations with larger time steps or steady state problems do not result in diagonally dominant Jacobian matrices that make schemes based on diagonal preconditioning or incomplete factorizations effective.  
Furthermore, due to the hyperbolic nature of the AC-GLM-MHD model, off-the-shelf algebraic multigrid methods also do not work effectively.  
We have found sparse-direct solvers within an overlapping Schwarz method to be the most effective and algorithmically scalable \cite{klawonn2000comparison}.
The subdomain and overlap construction is handled by the Ipack2 library in Trilinos \cite{prokopenko2016ifpack2}.  
Within each local subdomain, we apply a sparse direct solve.  
For multi-threaded CPU architectures we have found the Pardiso solver \cite{schenk2011pardiso} to be effective as well as the Trilinos Tacho solver \cite{yamazaki2025shylu}, both of which are available through a common Ifpack2 programming interface.  
On GPUs, the performance portable Tacho solver in Trilinos is also generally used. 
We used the Tacho solver for the simulations presented in this paper.  

\begin{figure}[ht!]
	\centering  
	\includegraphics[width=0.6\linewidth]{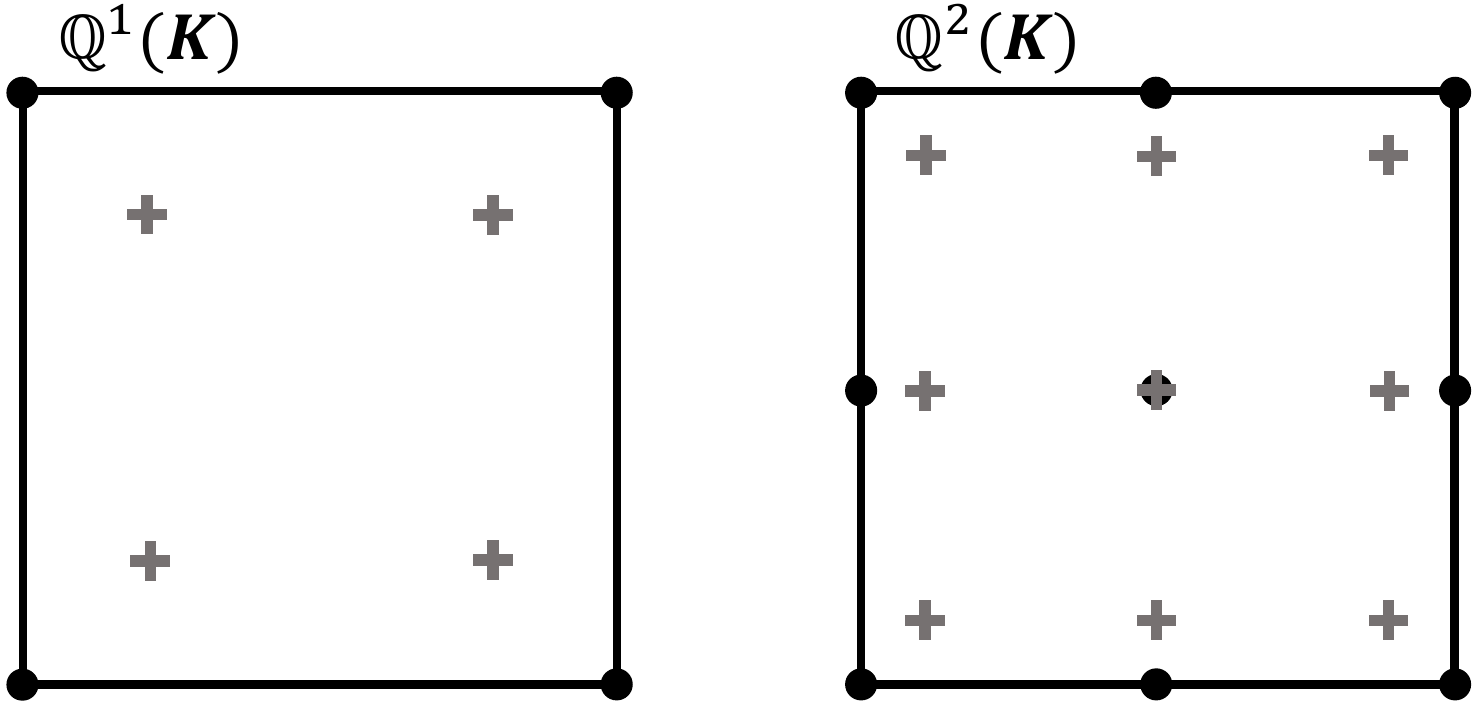}
	\caption{Illustration of linear (left) and quadratic (right) quadrilateral elements commonly used in \vertex, including many of the simulations in this paper.
	For linear elements, $\bbM^{k}(\bK)\vcentcolon=\bbQ^{1}(\bK)$ in Eq.~\eqref{eq:fem.approximation_space}; that is, solution fields on the element $\bK$ are approximated using tensor products of one-dimensional Lagrange polynomials of degree $k=1$, defined at two-point Legendre--Gauss--Lobatto (LGL) \emph{nodes} (solid circles).  
	The gray plusses mark the \emph{quadrature points}, obtained from tensorization of the one-dimensional two-point Legendre--Gauss (LG) rule, which is used to evaluate the volume integrals in Eq.~\eqref{eq:mhd.compact.fem}.
	For quadratic elements, $\bbM^{k}(\bK)\vcentcolon=\bbQ^{2}(\bK)$, fields are approximated by tensor products of degree $k=2$ Lagrange polynomials at three-point LGL nodes (solid circles).  
	Quadrature is performed using the three-point LG rule, whose tensorization provides the interior quadrature points (gray plusses).  
	The $N$-point LG quadrature rule integrates polynomials of degree $2N-1$ exactly.}
	\label{fig:finiteElements}
\end{figure}

\paragraph{Physics Terms}
The assembly of matrices and vectors needed to implement the time integration and solver schemes described above begins with the evaluation of physics terms --- i.e., the individual terms in the PDEs that comprise the AC-GLM-MHD model.  
Integrals in Eq.~\eqref{eq:mhd.compact.fem} are evaluated with quadrature rules over individual elements; see Figure~\ref{fig:finiteElements} for examples of \emph{nodes} and \emph{quadrature points} on an element.  
Quadrature rules specify discrete points within the element and associated weights such that evaluation of the integrand at the quadrature points can later be used to assemble the global integral.  
Physics terms are implemented at the quadrature point level using the Phalanx library in Trilinos \cite{pawlowski2012automating}.  
In \vertex, developers describe the components contributing to the global residual by indicating the fields required to perform the evaluation and the fields that are generated as a product of the evaluation.  
Phalanx uses these field data descriptions to generate multidimensional arrays over groups of elements, called worksets, allowing for element-local loops over the nodes and quadrature points of the workset elements to be implemented for the evaluation of field terms.  
These multidimensional arrays are represented using the Kokkos library \cite{trott2021kokkos} allowing for a single array definition to contain data in CPU or GPU memory, depending on compilation parameters.  
Generally, the number of elements in a workset is chosen either to effectively use CPU cache or to enable as many threads as possible for throughput on a GPU.  

\paragraph{Kernel Performance Portability} 
The evaluation of physics terms is performed within function evaluation kernels.  
These kernels use C++ templates to determine, at compile time, both the target computing architecture (e.g., CPU versus GPU) and the mode of function evaluation.  
In \emph{residual mode}, field data is represented as standard floating-point numbers, and the kernels compute residual function terms, which are then assembled into the global residual vector.  
In \emph{Jacobian mode}, used to construct the linearization of residuals for Newton’s method, field data is represented by forward automatic differentiation (FAD) types, provided by the Sacado library \cite{phipps2022automatic}.  
The number of derivatives in this mode equals the number of element nodes and degrees of freedom contributing to the local kernel evaluation.  
With Kokkos, array looping schemes are optimized for the data type and evaluation mode.  
For all modes, quadrature point loops are threaded over elements and points in a multidimensional layout.  
However, in Jacobian mode, additional threading is applied over derivative dimensions.  
By integrating Kokkos and Sacado within the Phalanx framework, a single unified physics kernel, implemented through C++ templates, can seamlessly support both CPUs and GPUs. 
This kernel can operate in basic function evaluation mode or in a mode where derivatives are automatically computed.  

\paragraph{Operator Assembly}
Each local physics term described above includes information about its dependencies on other terms (i.e., the fields on which it depends and are therefore created by others and the fields it creates which are therefore depended upon by others).  
These dependencies are used by the Phalanx library to construct a Directed Acyclic Graph (DAG) which can represent a general physics operator describing the entire problem \cite{pawlowski2012automating2}.  
Examples of operators include a global residual vector, global Jacobian matrix, and auxiliary integrated quantities for analysis. 
Translating the Phalanx DAG into operators is achieved through the Panzer library in Trilinos \cite{panzer-website}.  
Panzer provides a FEM assembly abstraction through which the DAG is used to generate all the needed physics terms.  
In addition, it provides access to finite element basis functions and quadrature rules from the Trilinos Intrepid2 library \cite{bochev2012solving} and comes with a variety of graph kernels needed for implementing FEMs.  
Examples include gathering data from a global state vector into element-local arrays for kernel evaluation (i.e., the map between between $\bU_{h}$ in Eq.~\eqref{eq:mhd.compact.fem} and $\bY$ in Eq.~\eqref{eq:mhd.compact.ode}), projecting fields and their gradients from nodes to quadrature points (i.e., the discrete values of $\bU_{h}$ and their gradients), evaluating common residual components such as the nodal divergence of a flux computed at a quadrature point (i.e., basis functions and their gradients in Eq.~\eqref{eq:mhd.compact.fem}), and completing the global integration by scattering values back to the global residual vector, all while considering domain decomposition and MPI communication.  
Jacobian assembly is similar where FAD types are used to evaluate and integrate, and derivatives are unpacked into sparse matrices for physics operations.  
Every time a new matrix or vector is needed in Newton's method, the operator assembly routines are triggered which perform DAG execution of kernels.  
If the matrices and vectors are needed on the GPU, the kernels launch on the GPU and assembly occurs in GPU memory.  
The Trilinos Tpetra library is used to represent the distributed sparse linear algebra objects and also includes performance portability via Kokkos.  

\paragraph{Computational Mesh}
To represent elements $\bK$ comprising $\Omega$, \vertex\ uses general unstructured grids containing a single element topology type (e.g., only 8-node hexahedrons or 4-node tetrahedra).
This is implemented by the STK Mesh library \cite{edwards2011stk} in Trilinos and the associated Shards element topology library.  
STK Mesh creates and stores the element descriptions including node connectivity, subdivisions of the domain into contiguous mesh blocks, and subdivision of domain boundaries into mesh sidesets.  
The Shards library applies nodal layouts to the elements to enable higher order basis functions (e.g., 8-node linear hexahedra versus 27-node quadratic hexahedra).  
In this work, we restrict ourselves to using meshes of linear and quadratic element topologies although the finite element formulation is equally applicable to higher order element types.  
STK reads Exodus files and Exodus-formatted data \cite{sjaardema2006exodus} which can be created through conversion from other file types or directly created by several mesh generation tools including Cubit \cite{blacker1994cubit} and Pointwise \cite{pointwise}. 

\paragraph{Load Balancing and Partitioning}
To enable use of distributed memory parallel computers, which are required for large meshes to speed up computations and ensure that the problem fits in memory, we subdivide $\Omega$ by assigning a contiguous subset of elements $\bK$ to each parallel process via a partitioning algorithm.  
Mesh partitioning is achieved in \vertex\ through the use of the Zoltan2 library \cite{devine2002zoltan} which can read and redistribute mesh data structures in a parallel setting.  
Many algorithms are available in Zoltan2 including geometric-based schemes such as Recursive Coordinate Bisectioning (RCB) as well as graph-based algorithms through the use of the ParMETIS library \cite{karypis2011metis}.  
Optimal partitioning seeks to balance the computational load assigned to each parallel process with the communication needed between processes to complete discretization stencils, compute matrix-vector products, and perform other linear algebra operations. 
For most simulations presented in this paper, we use the ``geom k-way'' algorithm provided by ParMETIS to partition our grids. 
In our testing, though not exhaustive, we have found this particular algorithm to be performant in its combination of a multilevel algorithm and a space-filling curve algorithm compared to purely geometric algorithms such as RCB.  

\end{document}